\documentclass[12pt]{article}
\newcommand{\Comment}[1]{{}}
\usepackage[textwidth = 450 pt, textheight = 630 pt]{geometry}
\usepackage[parfill]{parskip}
\usepackage{amssymb,euscript,amsmath,amsfonts}
\usepackage{tikz,tikz-cd,url}
\usetikzlibrary{decorations.pathreplacing,calligraphy}
\usepackage{xcolor}
\usepackage{graphicx}
\usepackage{color}
\usepackage{bbm} 
\usepackage{tensor}
\usepackage{slashed}
\usepackage{bigints}
\usepackage{mathrsfs}

\renewcommand{\frak}[1]{\mathfrak{#1}}
\renewcommand{\cal}[1]{\mathcal{#1}}

\newcommand{\inv}[1]{\frac{1}{#1}}

\newcommand{\brac}[1]{\left(#1\right)}

\newcommand{\mn}{\mu\nu}
\newcommand{\bepsilon}{\bar{\epsilon}}

\newcommand{\bbm}[1]{\mathbbm{#1}}

\newcommand{\bpsi}{\Bar{\psi}}

\newcommand{\brho}{\Bar{\rho}}

\newcommand{\bD}{\Tilde{D}}
\newcommand{\bX}{\Tilde{X}}
\newcommand{\bA}{\Tilde{A}}
\newcommand{\bF}{\Tilde{F}}
\newcommand{\bcalX}{\Tilde{\cal{X}}}
\newcommand{\dalpha}{\dot{\alpha}}
\newcommand{\dbeta}{\dot{\beta}}
\newcommand{\bY}{\Tilde{Y}}

\allowdisplaybreaks

\definecolor{MyDarkBlue}{rgb}{0.15,0.15,0.45}
\usepackage[linktocpage=true]{hyperref}
\hypersetup{
colorlinks=true,
citecolor=MyDarkBlue,
linkcolor=MyDarkBlue,
urlcolor=MyDarkBlue,
pdfauthor={Neil Lambert },
pdftitle={ },
pdfsubject={hep-th}
}

\usepackage{hypernat}
\usepackage{physics}
\usepackage{accents}
\usepackage{bm}
\usepackage[sorting=none, style=numeric-comp]{biblatex}
\begin{document}

 \centerline{\Huge  {  Quantum Effects in Supersymmetric  }} \vskip 12pt
 \centerline{\Huge  {   Galilean Yang-Mills}}

\centerline{\LARGE \bf {\sc  }} \vspace{2truecm} \thispagestyle{empty} \centerline{
    {\large {{\sc Neil~Lambert${}^{\,a \, }$}}}\footnote{E-mail address: \href{mailto:neil.lambert@kcl.ac.uk}{\tt neil.lambert@kcl.ac.uk} } and  {\large {{\sc Joseph~Smith${}^{\,b \, }$}}}\footnote{E-mail address: \href{mailto:joseph.m.smith@kcl.ac.uk}{\tt j.smith.24@bham.ac.uk} }}

%\vspace{1cm}
\centerline{${}^a${\it Department of Mathematics}}
\centerline{{\it King's College London }} 
\centerline{{\it The Strand }} 
\centerline{{\it  WC2R 2LS, UK}} 

\vspace{1cm}
\centerline{${}^b${\it School of Mathematics}}
\centerline{{\it University of Birmingham}} 
\centerline{{\it Edgbaston, Birmingham}} 
\centerline{{\it  B15 2TT, UK }} 

\vspace{1.0truecm}

\thispagestyle{empty}

\centerline{\sc Abstract}
\vspace{0.4truecm}
\begin{center}
\begin{minipage}[c]{360pt}{
    \noindent}

We study the quantum dynamics  of  supersymmetric Galilean Yang-Mills (SGYM), with a  focus on three dimensions where there is a classical scale symmetry. This is believed to be the worldvolume theory describing D2-branes within type IIA non-relativistic string theory and arises as a non-relativistic limit of maximally supersymmetric (relativistic) Yang-Mills. Unlike the relativistic case, we find that SGYM only admits perturbative dynamical excitations on the Coulomb branch. We show that in three dimensions the classical scale symmetry is broken at one-loop due to logarithmic divergences which cannot be absorbed by renormalization of existing couplings. As such we find that SGYM is non-renormalizable. However, we argue that by including   additional marginal deformations the original coupling can be renormalized and rendered asymptotically free. Finally, we propose that three-dimensional SGYM flows to a  strongly coupled non-relativistic M2-brane theory.

\end{minipage}
\end{center}

\newpage
\tableofcontents

\section{Introduction} \label{sect: intro}

Field theories with Galilean invariance\footnote{See \cite{Baiguera:2023fus} for a recent review of the topic.} are ubiquitous in descriptions of low-energy systems, with the paradigmatic example being the low-velocity dynamics of a system of particles. However, there has been recent interest in generalisations of Galilean symmetry based on brane-like foliations of spacetime  within string theory, which arise in various near-BPS limits associated with extended objects \cite{Andringa:2012uz, Bergshoeff:2019pij, Bergshoeff:2018yvt, Blair:2023noj}. Such limits are naturally implemented by non-Lorentzian geometries and define decoupled non-Lorentzian corners of string theory, leading to novel perspectives on well-studied examples of holography \cite{Blair:2024aqz, Guijosa:2025mwh} and motivating the construction of consistent supersymmetry-preserving non-Lorentzian limits of supergravity and field theories \cite{Blair:2021waq, Bergshoeff:2023ogz, Lambert:2024uue, Lambert:2024yjk, Bergshoeff:2025grj, Blair:2025ewa, Blair:2025prd, Maskalaniec:2026vlk}. Additionally, it has been proposed \cite{Fontanella:2024rvn, Blair:2024aqz, Lambert:2024ncn, Fontanella:2026gaq} that taking additional near-BPS limits of existing holographic dualities enables us to construct top-down examples of holography in which the QFT is non-relativistic. It is clearly a worthwhile endeavour to try and make statements of this form precise, for which we require an understanding of quantum effects in non-relativistic supersymmetric field theory\footnote{Alongside this, see \cite{Auzzi:2026gan} for recent progress in understanding quantum effects in a non-supersymmetric two-dimensional theory arising from a spin matrix theory \cite{Harmark:2014mpa} limit of $\cal{N}=4$ SYM.}.

The class of theories that we shall concern ourselves with in this work are those that contain a gauge field $A_{\mu}$ for the gauge algebra $\frak{g}$ and a singled-out $\frak{g}$-valued scalar $X$. The couplings of these fields to each other takes the form
\begin{equation} \label{eq: GYM action}
    S_{GYM} = \inv{2g^2} \tr \int dt d^d x \bigg( D_0 X D_0 X - 2 D_i X F_{0i} - \inv{2} F_{ij} F_{ij} \bigg) \ ,
\end{equation}
which was first constructed for the gauge group $U(1)$ in \cite{Santos:2004pq}, with a construction for arbitrary gauge group using the null reduction of $(d+2)$-dimensional Yang-Mills and Maxwell theory given in \cite{Bagchi:2022twx} following general work on Galilean gauge theories in \cite{LeBellac:1973unm, Bagchi:2014ysa, Bagchi:2015qcw, Bagchi:2017yvj}. Previous work has mainly focused on Abelian theories, with their symmetries discussed in \cite{Festuccia:2016caf, Fontanella:2024hgv,Hernandez:2026stx}, the one-loop renormalisation of the theory coupled to a Schr\"odinger scalar and to the null reduction of a Dirac fermion performed in \cite{Chapman:2020vtn} and \cite{Banerjee:2022uqj} respectively, and extensions to supersymmetric theories constructed by null reduction studied in \cite{Baiguera:2022cbp} following similar work on supersymmetric non-relativistic non-gauged models in \cite{Auzzi:2019kdd}.

Our understanding of non-Abelian Galilean gauge theories is less developed than their Lorentzian counterparts. A supersymmetric extension of \eqref{eq: GYM action} with non-Abelian gauge group was constructed in \cite{Fontanella:2024rvn} in three spatial dimensions using the machinery of non-relativistic string theory and was studied further in \cite{Kim:2026hpe}, but to our knowledge there has been no work on quantization of such theories. Our aim here is to rectify this by studying perturbative quantum effects in the supersymmetric extension of three-dimensional Galilean Yang-Mills that is obtained from either the null reduction of four-dimensional $\mathcal{N}=4$ supersymmetric Yang-Mills (SYM) or from a non-relativistic limit of three-dimensional $\cal{N}=8$ SYM. As four-dimensional $\cal{N}=4$ SYM is conformally invariant one may expect the null reduction of its action to be invariant under the Schr\"odinger algebra, including the $z=2$ Lifshitz scaling symmetry \cite{Son:2008ye}
\begin{equation}
    t \to \lambda^2 t \ , \ x^i \to \lambda x^i \ ,
\end{equation}
which is indeed the case. However, null reduction is an incredibly formal procedure, and it is unclear if one would expect this symmetry (and the larger Schr\"odinger superalgebra \cite{Duval:1993hs} it is part of) to be preserved once quantum effects are included. This is further compounded by the theory's relation to three-dimensional MSYM, which is not conformal. The fate of the classical symmetries under quantum corrections is therefore an important puzzle to explore.

The relationship between maximally supersymmetric (Lorentzian) Yang-Mills and string theory suggests we should look for a similar embedding of supersymmetric Galilean Yang-Mills (SGYM) within the framework of non-relativistic string theory. This is made somewhat more concrete by noting that the non-relativistic limit that must be taken of the three-dimensional theory is induced by the non-relativistic string limit of type IIA string theory \cite{Blair:2021waq}. As we review in appendix \ref{sect: IIA NRST}, the same limit of the relativistic D2-brane solution also leads to a backreacting solution of the non-relativistic string limit of type IIA supergravity with finite R-R three-form charge. The most obvious proposal for the role of the QFT is then as the worldvolume theory of a non-relativistic D2-brane stack. There is, however, a subtlety we've brushed under the rug here: in order to take the supergravity limit while preserving the solution's charge we must smear it along a circle parametrised by a transverse coordinate $X$. This coordinate is associated with the scalar field $X$ in the non-Lorentzian D2-brane worldvolume theory that forms part of the GYM action's gauge sector. One would then imagine that matching this set-up with the field theory would require both coupling the theory to an infinite number of fields to accommodate the transverse circle \cite{Taylor:1996ik}, and moving onto the Coulomb branch to smear the branes. While we shall not explore the former, we will see that the latter is forced on us; a non-trivial VEV for $X$ is required for fields to have non-vanishing kinetic terms, meaning we only find perturbative dynamics in this case. From the perspective of the non-relativistic limit the mass given to the perturbative BPS particle states by this VEV creates the hierarchy of scales between rest and excitation energies needed for a non-relativistic limit to be well-defined.

The remainder of this work is structured as follows. In section \ref{sect: GYM review} we review SGYM and its symmetries in all spatial dimensions $d\leq 8$, and discuss its perturbative dynamics on the Coulomb branch. In section \ref{sect: quantization} we move on to the quantization of the three-dimensional theory using the background field method, where we compute correlation functions of gauge and scalar fields. In section \ref{sect: deformation} we discuss the theory's classically marginal deformations that descend from the $d=8$ theory and their effect on correlation functions, before presenting our conclusions in section \ref{sect: conclusion}. We also include two appendices: in appendix \ref{sect: IIA NRST} we review the non-relativistic string limit of type IIA supergravity and its D2-brane solution, and in appendix \ref{sect: deformed quant} we include details about quantum effects once marginal deformations are turned on.

\section{Supersymmetric Galilean Yang-Mills} \label{sect: GYM review}

Let us begin with a brief review of supersymmetric Galilean Yang-Mills in general spatial dimension $d\leq 8$, the bosonic sector of which was first written down in \cite{Fontanella:2024rvn} for the four-dimensional theory following previous work on general non-Abelian Galilean gauge theories in \cite{Bagchi:2015qcw, Bagchi:2022twx}. For simplicity we shall restrict ourselves to a simple Lie algebra $\frak{g}$. The theory contains a $\frak{g}$-valued gauge-field $A$, $(9-d)$ adjoint-valued real scalars $\{ X, Y^A \}$ and an $SO(1,9)$ Majorana-Weyl spinor $\psi$, for which it will be necessary to define the projections
\begin{equation}
    \psi_{\pm} = \inv{2} \brac{\bbm{1}_{16} \pm \Gamma_{0X} } \psi 
\end{equation}
in terms of the Clifford algebra $\{ \Gamma_0 , \Gamma_i , \Gamma_X , \Gamma_A \}$. In this paper we will always take this representation to be real. The action for the theory is
\begin{align} \nonumber
    S_{\text{SGYM}} = \inv{2g^2} \tr \int dt d^d x \bigg(& 
    D_0 X D_0 X - 2 D_i X F_{0i} - \inv{2} F_{ij} F_{ij} - 2i
    [X, Y^A] D_0 Y^A \\ \nonumber
    &- D_i Y^A D_i Y^A + \inv{2} [Y^A , Y^B]^2 - \sqrt{2} i \psi_-^T D_0 \psi_- \\ \label{eq: sgym action}
    &- 2 i \psi_-^T \Gamma_{0i} D_i \psi_+ - \sqrt{2} \psi_+^T [X,\psi_+] - 2 \psi_-^T \Gamma_{0A} [Y^A , \psi_+] \bigg) \ ,
\end{align}
where we define the covariant derivative of $X$ by
\begin{equation}
    D_{\mu} X = \partial_{\mu} X - i [A_{\mu}, X] \ ,
\end{equation}
and similarly for the other fields. 

There are two paths that we know of that reach this from a relativistic theory. The first, as in \cite{Fontanella:2024rvn}, is to view the gauge theory as arising on the worldvolume of a D-brane within the non-relativistic string limit of the relevant type II string theory. In practical terms this can be achieved by starting with maximally supersymmetric Yang-Mills in $(d+1)$-dimensions and performing the rescaling
\begin{subequations} \label{eq: NRST limit}
\begin{align}
    (t , x^i) &\to (ct , x^i) \ , \\
    X &\to c X \ , \\
    Y^A &\to Y^A \ , \\
    \psi_+ &\to 2^{-1/4} c^{-1/2} \psi_+ \ , \\
    \psi_- &\to 2^{1/4} c^{1/2} \psi_- \ ,
\end{align}
\end{subequations}
of the fields and coordinates (with the gauge fields scaling tensorially with respect to the coordinate transformation), as well as further redefining the temporal gauge-field component post-rescaling to
\begin{equation}
    A_0 \to A_0 + c^2 X \ ,
\end{equation}
and the Yang-Mills coupling to
\begin{equation}
    g^2 \to c g^2 \ ,
\end{equation}
before taking the $c\to\infty$ limit. The second approach is to take the null reduction of maximally supersymmetric Yang-Mills in $(d+2)$-spacetime dimensions, as was first done in \cite{Bagchi:2022twx} to obtain the non-supersymmetric theory and in \cite{Lambert:2024yjk} to construct \eqref{eq: sgym action}. The lightcone components of the gauge-field transverse to and along the null-reduced direction become $A_0$ and $X$ respectively, with the fermionic projections corresponding to chirality in the lightcone coordinates. It is worth noting that if one performs the transformation
\begin{subequations} \label{eq: coupling scaling transformation}
\begin{align}
    t &\to \lambda t \ , \\
    X &\to \lambda X \ , \\
    \psi_+ &\to \lambda^{-1/2} \psi_+ \ , \\
    \psi_- &\to \lambda^{1/2} \psi_- \ ,
\end{align}
\end{subequations}
in \eqref{eq: sgym action} the only effect is to rescale the coupling to $\lambda g^2$, which can therefore be used to fix the coupling to any value; we shall revisit this point momentarily. Such a transformation is just the avatar of a finite rescaling of $c$ in \eqref{eq: NRST limit}. The energy one obtains from \eqref{eq: GYM action} as the Noether charge for time-translational symmetry is, after including an improvement term to restore gauge invariance, 
\begin{align} \nonumber
    E = \inv{2g^2} \tr \bigintssss d^d x \bigg(& (D_0X)^2 + \inv{2} F_{ij} F_{ij} + D_i Y^A D_i Y^A - \inv{2} [Y^A , Y^B]^2 \\
    &+ 2i \psi_-^T \Gamma_{0i} D_i \psi_+ + \sqrt{2} \psi_+^T [X , \psi_+] + 2 \psi_-^T \Gamma_{0A} [Y^A , \psi_+] \bigg) \ .
\end{align}
Crucially, the bosonic terms are positive-definite, and so despite the unconventional structure of the action the theory is classically well-defined.

We also observe that, in the absence of fermions, the Gauss Law constraint is
\begin{align}
D_iD_iX +i[X,D_0X] +[Y^A,[X,Y^A]]=0	\ ,
\end{align}
which leads to \begin{align}
\tr \big(XD_iD_iX+ [X,Y^A][X,Y^A]\big) =0	\ .
\end{align}
Integrating this over  space we find
\begin{align}
\tr \int  d^d x \big(D_iXD_iX - [X,Y^A][X,Y^A]\big) = \tr\oint dS_i \,XD_iX \ .
\end{align}
Both terms on the left hand side are positive definite. If  $X$ vanishes on the boundary we must then have $D_iX=0$ and $[X,Y^A]=0$ everywhere, effectively decoupling $X$. Therefore we are most interested in the Coulomb branch where $X$ has a non-zero vacuum expectation value.  

It is still not immediately clear what the dynamics of \eqref{eq: sgym action} are. To get a handle on this we should first examine the free theory obtained by rescaling all fields by a factor of $g$ before taking the $g\to0$ limit. As we've seen we should perform this limit on the Coulomb branch of the theory so we expand the scalar field $X$ as
\begin{equation} \label{eq: X expansion}
    X = v \sigma + \Phi \ ,
\end{equation}
where $\sigma\in\frak{g}$ is a semisimple element, and rescale $\Phi$ in order to take the free-field limit. After integrating by parts the result of this is
\begin{align} \nonumber
    S_{\text{free}} = \inv{2} \tr \int dt d^d x \bigg(&
    (\partial_0 \Phi)^2 + 2 i v [\sigma, A_0] \partial_0 \Phi - v^2 [\sigma, A_0]^2 - 2 i v [\sigma, A_i] \partial_0 A_i \\ \nonumber
    &+ 2 i v [\sigma, A_0] \partial_i A_i - 2 \partial_0 \Phi \partial_i A_i + 2 \partial_i \Phi \partial_i A_0 \\ \nonumber
    &- (\partial_i A_j)^2 + (\partial_i A_i)^2 - 2 i v [\sigma, Y^A] \partial_0 Y^A - \partial_i Y^A \partial_i Y^A \\
    &- \sqrt{2} i \psi_-^T \partial_0 \psi_- - 2 i \psi_-^T \Gamma_{0i} \partial_i \psi_+ - \sqrt{2} v \psi_+^T [\sigma, \psi_+] 
    \bigg) \ .
\end{align}
We shall choose to gauge-fix by setting
\begin{equation} \label{eq: gauge-fixing 1}
    \partial_0 \Phi - i v [\sigma, A_0] - \partial_i A_i = 0 \ ,
\end{equation}
for which the action is
\begin{align} \nonumber
    S_{\text{free}} = \inv{2} \tr \int dt d^d x \bigg( &
    4 i v [\sigma, A_0] \partial_0 \Phi + 2 \partial_i \Phi \partial_i A_0 - 2 i v [\sigma, A_i] \partial_0 A_i - (\partial_i A_j)^2 \\ \nonumber
    &- 2 i v [\sigma, Y^A] \partial_0 Y^A - \partial_i Y^A \partial_i Y^A - \sqrt{2} i \psi_-^T \partial_0 \psi_- \\ \label{eq: free action}
    &- 2 i \psi_-^T \Gamma_{0i} \partial_i \psi_+ - \sqrt{2} v \psi_+^T [\sigma, \psi_+]  
    \bigg) \ .
\end{align}
The structure of the derivative terms depends on whether the fields' components commute with $\sigma$. This is most easily illustrated with a simple example. Let us consider the case $\frak{g}=\frak{su}(2)$. We may then work with the basis 
\begin{align}
    \sigma = \inv{2}\begin{pmatrix}
        1 & 0 \\
        0 & -1
    \end{pmatrix} \ , \ \sigma^+ = \begin{pmatrix}
        0 & 1 \\
        0 & 0
    \end{pmatrix} \ , \ \sigma^- = \begin{pmatrix}
        0 & 0 \\
        1 & 0
    \end{pmatrix} \ ,
\end{align}
and expand the fields as
\begin{subequations} \label{eq: matrix expansions}
\begin{align}
    A_{\mu} &= \sqrt{2} A^{(\sigma)}_{\mu} \sigma + B_{\mu} \sigma^+ + \bar{B}_{\mu} \sigma^- \ , \\
    \Phi &= \sqrt{2} \Phi^{(\sigma)} \sigma + \cal{X} \sigma^+ + \bar{\cal{X}} \sigma^- \ , \\
    Y^A &= \sqrt{2} Y^A_{(\sigma)} \sigma + \Omega^A \sigma^+ + \bar{\Omega}^A \sigma^- \ , \\
    \psi_{\pm} &= \sqrt{2} \psi_{\pm}^{(\sigma)} \sigma + \rho_{\pm} \sigma^{+} + \bar{\rho}_{\pm} \sigma^- \ ,
\end{align}
\end{subequations}
to reach the action
\begin{align} \nonumber
    S_{\text{free}} = \int dt d^d x \bigg(& 2 i v B_0 \partial_0 \bar{\cal{X}} -2 i v \bar{B}_0 \partial_0 \cal{X} + \partial_i \cal{X} \partial_i \bar{B}_0 + \partial_i \bar{\cal{X}} \partial_i B_0 \\ \nonumber
    &+
    2i v \bar{B}_i \partial_0 B_i - \partial_i B_j \partial_i \bar{B}_j + 2i v \bar{\Omega}^A \partial_0 \Omega^A - \partial_i \Omega^A \partial_i \bar{\Omega}^A \\ \nonumber
    &+ \partial_i A_0^{(\sigma)} \partial_i \Phi^{(\sigma)} - (\partial_i A_j^{(\sigma)})^2 - ( \partial_i Y^A_{(\sigma)})^2 
    - \sqrt{2} \bar{\rho}_-^T \partial_0 \rho_- \\ \nonumber
    &- i \bar{\rho}_-^T \Gamma_{0i} \partial_i \rho_+ - i \bar{\rho}_+^T \Gamma_{0i} \partial_i \rho_- - \sqrt{2} v \bar{\rho}_+^T \rho_+ \\
    &- \inv{\sqrt{2}} \psi_{-}^{(\sigma) \, T} \partial_0 \psi_{-}^{(\sigma)} - i \psi_{-}^{(\sigma) \, T} \Gamma_{0i} \partial_i \psi_{+}^{(\sigma)} \bigg) \ .
\end{align}
Focusing first on the bosonic fields we see that the components along $\sigma$ are non-dynamical, whereas the others form Schr\"odinger-like complex pairs of mass $M = v$. It is now clear why we had to work on the Coulomb branch; the only way kinetic terms can arise is through interactions with $X$, leading to a coefficient in the free theory proportional to the field's VEV. This is somewhat analogous to the structure of fermionic terms in the Green-Schwarz string, which only have a standard quadratic kinetic term when the bosonic fields are expanded around a background field. One may worry that the mixed terms involving $B_0$ and $\cal{X}$ do not possess a definite sign and lead to a theory that is unbounded from below. We will see later that, as in the relativistic theory, this is solved via the inclusion of Faddeev-Popov ghosts for the gauge-fixing condition \eqref{eq: gauge-fixing 1}. 

Turning our attention to the fermions, we see something similar. As both $\psi_+^{(\sigma)}$ and $\rho_+$ are non-dynamical we are free to impose their equations of motion
\begin{subequations}
\begin{align}
    \Gamma_{0i} \partial_i \psi_-^{(\sigma)} &= 0 \ , \\
    i \Gamma_{0i} \partial_i \rho_- + \sqrt{2} v \rho_+ &= 0 \ .
\end{align}
\end{subequations}
The first of these renders $\psi_-^{(\sigma)}$ dynamically trivial in the free theory, while the second gives $\rho_-$ the same Schr\"odinger-like derivative terms as the complex bosonic fields. The structure arising in the free-field limit is perhaps not surprising. Taking the non-relativistic limit \eqref{eq: NRST limit} of the three-dimensional gauge theory on its Coulomb branch reduces the dynamics to infinitesimal excitations above the BPS bound associated with W-bosons of mass equal to the Coulomb branch VEV; this makes the appearance of a non-relativistic supermultiplet of the same mass in \eqref{eq: sgym action} incredibly natural. The non-dynamical fields emerge from the non-relativistic limit of the (gauge-fixed) gauge sector of the Abelian Coulomb branch EFT, forming the equivalent sector of the supersymmetric extension of Galilean electrodynamics (GED) \cite{Chapman:2020vtn, Baiguera:2022cbp}. While it appears that we have two parameters ($v$ and $g^2$) that define the Coulomb branch theory, we recall that using the transformation \eqref{eq: coupling scaling transformation} we can rescale both simultaneously. The true invariant coupling of the theory is therefore
\begin{equation} \label{eq: true coupling}
    e^2 = g^2 v^{-1}  \ ,
\end{equation}
which we shall see naturally emerging in perturbation theory later on.

We can now generalise to an arbitrary simple Lie algebra, focussing on the scalar fields to get a feel for the dynamics. Let $\{H^m\}$ be a basis for a Cartan subalgebra of $\frak{g}$ containing $\sigma$, with the remaining generators taken to be a Cartan-Weyl basis $\{ E^{\alpha}, E^{-\alpha} \}$, where we work in terms of the positive roots $\alpha$ defined by
\begin{subequations}
\begin{align}
    [H^m , E^{\alpha}] &= \alpha^m E^{\alpha} \ , \\
    [H^m , E^{-\alpha}] &= - \alpha^m E^{-\alpha} \ .
\end{align}
\end{subequations}
We shall take a positive root to be defined by the condition $\alpha^{\sigma}>0$ when this component is non-zero, with an arbitrary choice made if it vanishes. With this choice, the trace defined by the Killing form is schematically given by
\begin{subequations}
\begin{align}
    \tr \brac{ H^m H^n } &= C_{mn} \ , \\
    \tr \brac{E^{\alpha} E^{-\alpha} } &= k_{\alpha} \ ,
\end{align}
\end{subequations}
with all other combinations vanishing. Expanding $Y^A$ in this basis as
\begin{equation}
    Y^A = \sum_m Y^A_m H^m + \sum_{\alpha} \brac{\omega^A_{\alpha} E^{\alpha} + \bar{\omega}^A_{\alpha} E^{-\alpha} } k^{-1/2}_{\alpha} \ ,
\end{equation}
the free scalar action is
\begin{align}
    S_{\text{free},Y} = \,&\sum_{\alpha} \int dt d^d x \bigg( 
    2i v \alpha^{\sigma} \bar{\omega}^A_{\alpha} \partial_0 \omega^A_{\alpha} - \partial_i \bar{\omega}^A_{\alpha} \partial_i \omega^A_{\alpha} \bigg) \\
    &- \inv{2} \sum_{m,n} \int dt d^d x \, C_{mn} \partial_i Y^A_m \partial_i Y^A_n \ .
\end{align}
For each root with $\alpha^{\sigma} \neq0$ we get an $SO(8-d)$ vector of Schr\"odinger scalars of mass $M_{\sigma} = v\alpha^{\sigma}$; these are the dynamical non-relativistic particles of the theory. The rest of the components lack kinetic terms, and in the full theory mediate instantaneous interactions between the particles. While it would be interesting to study the interacting theory for a generic gauge group, particularly its structure at large-$N$, we shall restrict our attention to $\frak{su}(2)$ when we look at the theory's one-loop properties in section \ref{sect: quantization} for simplicity.

The bosonic symmetries of the bosonic sector were studied in \cite{Bagchi:2022twx, Lambert:2024yjk, Fontanella:2024hgv} for the four-dimensional theory; let us briefly review this, extending to a general number of dimensions. The first set of transformations to consider are those transforming the time coordinate, which are (working infinitesimally)
\begin{subequations}
\begin{align}
    \hat{t} &= t + a + (4-d) bt + (4-d) ct^2 \ , \\
    \hat{x}^i &= x^i \brac{1 + b + 2 ct} \ ,
\end{align}
\end{subequations}
and correspond to translations, dilatations, and special conformal transformations (SCTs) of time. It is straightforward to show that the field transformations
\begin{subequations}
\begin{align}
    \hat{A}_0 (\hat{t}, \hat{x}) &= \big(1 - (4-d) b - 2(4-d) ct\big) A_0(t,x) - 2 cx^i A_i(t,x) \ , \\
    \hat{A}_i(\hat{t}, \hat{x}) &= (1 - b - 2 ct)A_i(t,x) - 2 cx^i X(t,x) \ , \\
    \hat{X}(\hat{t}, \hat{x}) &= \big(1 + (2-d) ( b + 2 ct)\big) X(t,x) \ , \\
    \hat{Y}^A (\hat{t}, \hat{x}) &= \brac{1 - b - 2 ct} Y^A(t,x) \ , \\
    \hat{\psi}_-(\hat{t}, \hat{x}) &= \brac{ 1 - \frac{d}{2}(b+2ct) } \psi_-(t,x) \ , \\
    \hat{\psi}_+(\hat{t}, \hat{x}) &= \brac{ 1 - \brac{ 3 - \frac{d}{2}}(b + 2 c t) } \psi_+(t,x) + \sqrt{2} c x^i \Gamma_{0i} \psi_-(t,x) \ ,
\end{align}
\end{subequations}
leave the action invariant up to total derivative terms. However, as we are working on the Coulomb branch we must impose the asymptotic boundary condition
\begin{equation}
    \lim_{r\to\infty} X(t,x) = v \sigma \ ,
\end{equation}
and demand that any symmetry of the theory leaves this invariant. This requires us to discard the dilatation and SCT when $d\neq2$ as the transformations act non-trivially on $X$. We note that the inhomogeneous term in the transformation of $A_i$ under a SCT does not alter the asymptotic values of any physical quantities provided that the components of all fields that do not commute with $\sigma$ decay sufficiently quickly as we take $r\to\infty$; we shall assume this condition is met from here onwards, and so SCTs remain a symmetry in $d=2$. The three-dimensional theory is therefore special, and shall be our focus in the next section. The dilatation is a $z=2$ Lifshitz scaling symmetry, with time scaling twice as fast as space.

The remaining spacetime symmetries of the action are those which only act on space,
\begin{equation}
    \hat{x}^i = x^i + \xi^i(t) + \tensor{r}{^i_j} x^j \ ,
\end{equation}
with the infinitesimal field transformations
\begin{subequations}
\begin{align}
    \hat{A}_0 (\hat{t}, \hat{x}) &= A_0(t,x) - \dot{\xi}^i(t) A_i(t,x) \ , \\
    \hat{A}_i(\hat{t}, \hat{x}) &= \brac{\delta_{i}^j - \tensor{r}{_{i}^j}}A_j(t,x) - \dot{\xi}^i(t) X(t,x) \ , \\
    \hat{X}(\hat{t}, \hat{x}) &= X(t,x) \ , \\
    \hat{Y}^A (\hat{t}, \hat{x}) &= Y^A(t,x) \ , \\
    \hat{\psi}_-(\hat{t}, \hat{x}) &= \brac{\bbm{1} + \inv{4} r^{ij} \Gamma_{ij}} \psi_-(t,x) \ , \\
    \hat{\psi}_+(\hat{t}, \hat{x}) &= \brac{\bbm{1} + \inv{4} r^{ij} \Gamma_{ij}} \psi_+(t,x) + \inv{\sqrt{2}} \Gamma_{0i} \dot{\xi}^i(t) \psi_-(t,x) \ ,
\end{align}
\end{subequations}
that leave the action invariant up to total derivatives. Here $r_{ij}$ is antisymmetric. However, unlike the previous set of transformations, the inhomogeneous term in the transformation of $A_i$ changes the asymptotic value of $F_{0i}$ by an amount proportional to $\Ddot{\xi}^i$ and the VEV of $X$. The symmetries of the theory are therefore those with constant $\dot{\xi}^i$, i.e. spatial translations and Galilean boosts. The last set of bosonic symmetries to consider are the action's internal symmetries, for which the non-trivial field transformations are
\begin{subequations}
\begin{align}
    \hat{A}_0 &= A_0 + f_A(t) Y^A \ , \\
    \hat{Y}^A &= \brac{ \delta^A_B + \tensor{L}{^A_B} }Y^B + f^A(t) X \ , \\
    \psi_- &= \brac{\bbm{1} + \inv{4} L^{AB} \Gamma_{AB} }\psi_- \ , \\
    \psi_+ &= \brac{ \bbm{1} + \inv{4} L^{AB} \Gamma_{AB} }\psi_+ + \inv{\sqrt{2}} \Gamma_{0A} f^A(t) \psi_- \ ,
\end{align}
\end{subequations}
with $L_{AB}$ antisymmetric. As above, the inhomogeneous term in the transformation of $Y^A$ alters its boundary condition and forces us to set $f^A(t)=0$ in order to get a symmetry of the theory, leaving us with only the R-symmetry rotations. The full bosonic symmetry algebra on the Coulomb branch is
\begin{equation}
    \frak{g}_{d\neq2} = \frak{gal}(d) \oplus \frak{so}(8-d)
\end{equation}
for $d\neq 2$, and
\begin{equation}
    \frak{g}_{d=2} = \frak{schr}(2) \oplus \frak{so}(6) \ ,
\end{equation}
when $d=2$. As some of the more exotic symmetries  of the action are broken by the non-trivial VEV of $X$ (for instance, arbitrarily time-dependent spatial translations), one may wonder if the theory is well-defined if we set the VEV to zero and recover the broken symmetries. From our analysis above the theory in this regime possesses no perturbative dynamics as the kinetic terms for all components now vanish. However, it may be the case that the theory recovers its dynamics non-perturbatively: as the rescaling-invariant coupling \eqref{eq: true coupling} diverges in this limit this would necessarily require an understanding of the theory at strong coupling. We shall not consider this question further, though it is obviously an interesting avenue to pursue.

We now turn our attention to the fermionic symmetries of \eqref{eq: sgym action}. The simplest way to obtain these is to use the connection between the theory and the relativistic theory it arises from via null reduction. This is most straightforward to see when $d=8$. The relativistic theory of interest is then ten-dimensional $\cal{N}=1$ SYM,
\begin{equation} \label{eq: 10d N=1 action}
    S_{10d} = \inv{2g^2} \tr \int d^{10} x \brac{
    -\inv{2} F_{\mn} F^{\mn} + i \bpsi \Gamma^{\mu} D_{\mu} \psi } \ ,
\end{equation}
which is invariant under the supersymmetry transformations
\begin{subequations} \label{eq: 10d N=1 susy}
\begin{align}
    \delta A_{\mu} &= - i \bepsilon \Gamma_{\mu} \psi \ , \\
    \delta \psi &= - \inv{2} F_{\mn} \Gamma^{\mn} \epsilon \ .
\end{align}
\end{subequations}
By rewriting \eqref{eq: 10d N=1 action} in terms of the lightcone coordinates
\begin{equation}
    x^{\pm} = \inv{\sqrt{2}} \brac{x^0 \pm x^9} \ ,
\end{equation}
followed by a reduction along $x^+$ and the identifications $x^-\Rightarrow t$, $(A_- , A_+) \Rightarrow (A_0 , X)$ we obtain $d=8$ SGYM, where the fermions are projected onto the eigenspaces of $\Gamma_{09}$. Reducing the supersymmetry transformations in the same way, we see that the non-relativistic theory is invariant under
\begin{subequations} \label{eq: 8d SGYM susy}
\begin{align}
    \delta X &= -\sqrt{2} i \epsilon_-^T \psi_- \ , \\
    \delta A_0 &= -\sqrt{2} i \epsilon_+^T \psi_+ \ , \\
    \delta A_i &= i \brac{ \epsilon_+^T \Gamma_{0i} \psi_- + \epsilon_-^T \Gamma_{0i} \psi_+} \ , \\
    \delta \psi_+^T &= \sqrt{2}\epsilon_-^T \Gamma_{0i} F_{0i} + \epsilon_+^T \brac{ \inv{2} F_{ij} \Gamma_{ij} + D_0 X} \ , \\
    \delta \psi_-^T &= - \sqrt{2}\epsilon_+^T \Gamma_{0i} D_i X + \epsilon_-^T \brac{\inv{2} F_{ij} \Gamma_{ij} - D_0 X} \ ,
\end{align}
\end{subequations}
as can be checked explicitly. A further spatial dimensional reduction of this gives us the rigid supersymmetry for SGYM in arbitrary dimensions. By examining the transformations \eqref{eq: 8d SGYM susy} (or through a reduction of the relativistic superalgebra) we see that the non-relativistic supersymmetry obeys an algebra of the schematic form
\begin{equation}
    \{\cal{Q}_-, \cal{Q}_- \} \sim H \ , \ \{ \cal{Q}_- , \cal{Q}_+ \} \sim P_i \ , \ \{ \cal{Q}_+ , \cal{Q}_+ \} \sim 0 \ ,
\end{equation}
where we implicitly set gauge-transformations to zero. The last of these can be quickly established for all fields but $\psi_+$, as
\begin{equation}
    [ \delta_{+,1}, \delta_{+,2}] V = \delta_{G,\alpha} V \ ,
\end{equation}
where $\delta_{G,\alpha}$ is a gauge transformation with parameter
\begin{equation}
    \alpha = 2 \sqrt{2} i \epsilon_{+,1}^T \epsilon_{+,2} X \ .
\end{equation}
However, for $\psi_+$ we have
\begin{equation}
    \delta_{+,1} \delta_{+,2} \psi_+ =  \brac{ i \Gamma_{ij} \epsilon_{+,2}\,  \epsilon_{+,1}^T \Gamma_{0i} D_j \psi_- + \sqrt{2} \epsilon_{+,2} \, \epsilon_{+,1}^T [X , \psi_+] } \ ,
\end{equation}
and so after using Fierz identities we find
\begin{equation}
    \delta_{+,1} \delta_{+,2} \psi_+ = \inv{8} \sum_M \brac{-i \Gamma_{ij} \Gamma^M \Gamma_{0i} D_j \psi_- - \sqrt{2} \Gamma^M [X , \psi_+]} \epsilon_{+,1}^T \Gamma_M \epsilon_{+,2} \ , 
\end{equation}
where $\{ \Gamma^M \}$ are a basis for the positive-chirality (with respect to $\Gamma_{0X}$) projection of the Clifford algebra. Antisymmetrising in the variations, the only surviving terms are when $\Gamma^M = \bbm{1}$ and $\Gamma^M = \Gamma_{ijkl}$\footnote{We must include an extra factor of $1/2$ to account for the duality constraint on $\Gamma_{ijkl}$ for the positive-chirality projection.}, and so
\begin{align} \nonumber
    [\delta_{+,1}, \delta_{+,2}] \psi_+ &= \inv{4} \brac{ 7 i \Gamma_{0i} D_i \psi_- - \sqrt{2} [X , \psi_+]} \epsilon_{+,1}^T \epsilon_{+,2} \\ \nonumber
    & \qquad - \frac{1}{16} \Gamma_{j_1 ... j_4} \brac{ i \Gamma_{0i} D_i \psi_- + \sqrt{2} [X , \psi_+]} \epsilon_{+,1}^T \Gamma_{j_1 ... j_4} \epsilon_{+,2} \\ \nonumber
    &= \delta_{G,\alpha} \psi_+ +  \brac{i \Gamma_{0i} D_i \psi_- + \sqrt{2} [X , \psi_+]} \bigg( \frac{7}{4} \epsilon_{+,1}^T \epsilon_{+,2} \\
    &\qquad - \inv{16} \Gamma_{j_1 ... j_4} \epsilon_{+,1}^T \Gamma_{j_1 ... j_4} \epsilon_{+,2} \bigg) \ .
\end{align}
This recovers the required gauge transformation only if we impose the constraint
\begin{equation}
    i \Gamma_{0i} D_i \psi_- + \sqrt{2} [X , \psi_+] = 0 \ :
\end{equation}
however, this is exactly the (nine-dimensional) equation of motion for the non-dynamical field $\psi_+$. By acting on this with $\delta_+$ and again using Fierz identities we recover the $A_0$ equation of motion
\begin{equation}
    D_i D_i X + i [X , D_0 X] - \inv{\sqrt{2}} \{ \psi_- , \psi_- \} = 0 \ ,
\end{equation}
where we use the notation $\{\xi , \zeta \}$ for anticommuting spinors $\xi$ and $\zeta$ by
\begin{equation}
    \{ \xi , \zeta \} = \xi^T \zeta + \zeta^T \xi \ .
\end{equation}
As the spatial translations can be extended to arbitrary functions of time at the level of the action, one would imagine (as was done for other supersymmetric non-Lorentzian theories in \cite{Lambert:2024yjk, Lambert:2024ncn}) that $\epsilon_+$ can similarly be promoted to have time-dependence; however, given that the bosonic symmetry is broken by boundary conditions the same will occur here and so these are not of much interest for our current purposes.

As was the case for the bosonic symmetries, additional structure arises in two spatial dimensions. The non-relativistic theory is the null reduction of four-dimensional $\cal{N}=4$ SYM which possesses the additional superconformal symmetry
\begin{equation}
    \epsilon = x^{\mu} \Gamma_{\mu} \eta \ ,
\end{equation}
with $\eta$ an arbitrary ten-dimensional MW spinor. This symmetry is spacetime-dependent, and so only survives the null reduction if we take it to be independent of $x^+$. To do this we must impose the constraint
\begin{equation} \label{eq: superconformal spinor constraint}
    \Gamma_+ \eta = 0 \ ,
\end{equation}
or equivalently $P_+ \eta = \eta$, on the spinor. Upon relabelling $x^- \Rightarrow t$ we find
\begin{equation}
    \epsilon = \brac{\sqrt{2} t \Gamma_0 + x^i \Gamma_i} \eta \ .
\end{equation}
Reducing the supersymmetry transformations \eqref{eq: 8d SGYM susy} to $d$-dimensions using the condensed notation $A_I = (A_i, Y^A)$ and $\partial_I = (\partial_i , 0)$, the proposed superconformal transformations are then
\begin{subequations}
\begin{align}
    \delta X &= 2 i t \eta^T \Gamma_0 \psi_- \ , \\
    \delta A_0 &= - \sqrt{2} i x^i \eta^T \Gamma_i \psi_+ \ , \\
    \delta A_I &= i \eta^T \brac{ x^j \Gamma_j \Gamma_{0I} \psi_- + \sqrt{2} t \Gamma_I \psi_+  } \ , \\
    \delta \psi_+^T &=  2 t \eta^T  \Gamma_I F_{0I} + x^i \eta^T \Gamma_i \brac{\inv{2} F_{JK} \Gamma_{JK} + D_0 X} \ , \\
    \delta \psi_-^T &= - \sqrt{2} \eta^T\brac{ x^i \Gamma_i \Gamma_{0J} D_J X + t \Gamma_0 \brac{\inv{2} F_{JK} \Gamma_{JK} - D_0 X} } \ ,
\end{align}
\end{subequations}
and lead to the one-fermion terms
\begin{align}
    \delta S_{\text{SGYM}}^{(1f)} = \inv{2g^2} \tr \int dt d^d x \bigg( 
    2\sqrt{2}(d-2) i \eta^T \Gamma_J D_J X \psi_+ - (d-2) i \eta^T F_{IJ} \Gamma_{IJ} \Gamma_0 \psi_- \bigg) \ ,
\end{align}
in the transformation of the $d$-dimensional action. We therefore see that if $d=2$ the action is invariant at the one-fermion level; as the cubic fermion terms do not involve derivatives of any fields the same computation as for $\epsilon$ shows that they vanish here too. As we had to impose the constraint \eqref{eq: superconformal spinor constraint} on $\eta$ it has eight independent components, leading to a total of 24 supercharges in the theory.

We close this section by briefly clarifying our nomenclature. While we have referred to the theory \eqref{eq: sgym action} simply as `supersymmetric Galilean Yang-Mills', there are of course many supersymmetric extensions of Galilean Yang-Mills that one can obtain via e.g. null reduction of Lorentzian gauge theories with a lesser number of supercharges. Its relation to maximally supersymmetric Lorentzian gauge theory makes it reasonable to propose that this is the maximal amount of supersymmetry a Galilean gauge theory can have, but the question of whether this theory is unique is not as obvious: for this reason, we shall be somewhat nondescript when referring to the theory.

\section{\texorpdfstring{Background Field Quantization of \newline Three-Dimensional SGYM}{Background Field Quantization of Three-Dimensional SGYM}} \label{sect: quantization}

\subsection{Setting up the Expansion} \label{sect: background field setup}

With our understanding of the classical SGYM in hand, let us move on to its quantization. As discussed above, the interesting case to consider is the three-dimensional theory. In this case the coupling is classically marginal and we have (classical) superconformal invariance; it is then interesting to ask whether this is preserved once one-loop corrections are included. Quantization of Abelian Galilean gauge theories was previously studied in \cite{Chapman:2020vtn, Baiguera:2022cbp}. As we work on the Coulomb branch, there will be some similarities between the structure of the theory studied here and those considered there.

We will quantize the theory using the background field method. To do this we expand our fields in terms of fixed classical backgrounds and fluctuating perturbative quantum fields,
\begin{subequations} \label{eq: bf expansion}
\begin{align}
    X &= \Tilde{X} + g \phi \ , \\
    A_0 &= \Tilde{A}_0 + g a_0 \ , \\
    A_i &= \Tilde{A}_i + g a_i \ , \\
    Y^A &= \Tilde{Y}^A + g y^A \ ,
\end{align}
\end{subequations}
where we will only consider bosonic background fields (in other words, we only need to rescale the fermions by a factor of $g$). Note that our conventions throughout will be to denote fluctuating fields by lower-case variables. The original fields' gauge transformations split into gauge transformations of the background fields (under which the fluctuating fields transform in the adjoint representation) and gauge transformations of the fluctuating fields, which are given in their infinitesimal form by
\begin{subequations} \label{eq: fluctuating gauge trans}
\begin{align}
    \delta a_{\mu} &= \bD_{\mu} \alpha - i g [a_{\mu}, \alpha] \ , \\
    \delta \phi &= i [\alpha, \bX] + i g [\alpha, \phi] \ , \\
    \delta y^A &= i [\alpha, \bY^A] + i g [\alpha, y^A] \ , \\
    \delta \psi_{\pm} &= i g [\alpha, \psi_{\pm}] \ .
\end{align}
\end{subequations}
Using \eqref{eq: bf expansion} we can expand the classical action in $g$,
\begin{equation}
    g^{-2} S_{\text{SGYM}}[\Tilde{\Lambda} + g \lambda] = g^{-2} S_{\text{SGYM}}[\Tilde{\Lambda}] + g^{-1} S^{(1)}[\lambda; \Tilde{\Lambda}] + S^{(2)}[\lambda ; \Tilde{\Lambda}] + O(g) \ ,
\end{equation}
where we are using $\Tilde{\Lambda}$ and $\lambda$ to collectively denote the background and fluctuating fields respectively. By taking the background fields to satisfy the classical equations of motion we see that the second term in the expansion vanishes, and a standard argument \cite{Abbott:1981ke} tells us that the one-loop correction to the quantum effective action is then given by
\begin{equation}
    e^{i \Gamma_1[\Tilde{\Lambda}]} = \bigintssss \frac{D\lambda}{\text{Vol}(\cal{G})} \, e^{i S^{(2)}[\lambda ; \Tilde{\Lambda}]} \ ,
\end{equation}
where $\text{Vol}(\cal{G})$ is the (infinite) gauge-group volume that must be cancelled by gauge-fixing the transformations \eqref{eq: fluctuating gauge trans}. We will compute $S_2$ for non-vanishing gauge-sector background and non-vanishing scalar background separately in order to keep all expressions manageable; as we shall only be interesting in computing the 1PI correlation functions of fields within each sector there is no issue with doing this.

Let us begin with the gauge sector. Expanding the action gives
\begin{align} \nonumber
    S^{(2)} = \inv{2} \tr \int dt d^2 x \, \bigg(&
    (\bD_0 \phi - i [\bX , a_0] - \bD_i a_i)^2 + 4 i \bD_0 \phi [\bX, a_0] \\ \nonumber
    &+ 2 \bD_i \phi \bD_i a_0 - 2 i \bD_0 a_i [\bX, a_i] - \bD_i a_j \bD_i a_j \\ \nonumber
    &- 2 i \bD_0 y^A [\bX, y^A] - \bD_i y^A \bD_i y^A - 4 i \bF_{0i} [\phi, a_i] \\ \nonumber
    &+ 2i \bD_0 \bX [\phi, a_0] + 4i [a_0, a_i] \bD_i \bX + 2i \bF_{ij} [a_i, a_j] \\
    &-\sqrt{2} i \psi_-^T  \Tilde{D}_0 \psi_- - 2i \psi_-^T \Gamma_{0i} \Tilde{D}_i \psi_+ - \sqrt{2} \psi_+^T [\bX, \psi_+]  
    \bigg) \ .
\end{align}
\label{eq: bfm quad action 1}
We must, of course, fix the gauge symmetry of the fluctuating fields. A convenient choice of gauge-fixing function is
\begin{equation}
    G = \bD_0 \phi - i [\bX, a_0] - \bD_i a_i
\end{equation}
due to its appearance in the action. The function also arises as the null reduction of the function defining Lorenz gauge in the background field quantization of (Lorentzian) (d+2)-dimensional Yang-Mills. We note in passing that upon setting $\bA_0 = \bA_i = 0$ and $\bX = v \sigma$ we recover \eqref{eq: gauge-fixing 1}, retroactively justifying choice of gauge-fixing in the free theory. The gauge-fixed path integral is then
\begin{equation}
    e^{i \Gamma_1[\Tilde{\Lambda}]} = \int D\lambda D\omega \, \det\brac{\frac{\delta G}{\delta \alpha}}\delta[G - \omega] \, e^{i S^{(2)}[\lambda ; \Lambda] - i S_{\xi}[\omega]} \ ,
\end{equation}
where we are including the Gaussian gauge-fixing term
\begin{equation}
    S_{\xi} = \inv{2\xi} \tr \int dt d^2x \, \omega^2
\end{equation}
in terms of an auxiliary field $\omega$. We will work exclusively in 't Hooft-Feynman gauge and set $\xi = 1$ to cancel the term quadratic in $G$ in $S^{(2)}$.

Before we can compute anything we must rewrite the functional determinant using Faddeev-Popov ghosts. The transformation of the gauge-fixing condition under \eqref{eq: fluctuating gauge trans} is
\begin{align} \nonumber
    \delta G = - 2 i& [\bX, \bD_0 \alpha] - i [\bD_0 \bX, \alpha] - \bD_i \bD_i \alpha \\
    &+ g \brac{i \bD_0 [\alpha, \phi] - [\bX, [a_0, \alpha]] + i \bD_i [a_i, \alpha]} \ .
\end{align}
The Faddeev-Popov ghosts are already fluctuating fields, meaning the $O(g)$ term in $\delta G$ can be neglected in one-loop computations. We are therefore left with the one-loop path integral
\begin{equation}
    e^{i \Gamma_1[\Tilde{\Lambda}]} = \int D\lambda D\bar{c} D c \, e^{i S_{\text{1-loop}}} \ ,
\end{equation}
with action
\begin{align} \nonumber
    S_{\text{1-loop}} = \inv{2} \tr \int dt d^2 x \, \bigg(&
    4 i \bD_0 \phi [\bX, a_0] + 2 \bD_i \phi \bD_i a_0 - 2 i \bD_0 a_i [\bX, a_i] \\ \nonumber
    &- \bD_i a_j \bD_i a_j - 2 i \bD_0 y^A [\bX, y^A] - \bD_i y^A \bD_i y^A \\ \nonumber
    &- 4 i \bF_{0i} [\phi, a_i] + 2i \bD_0 \bX [\phi, a_0] + 4i [a_0, a_i] \bD_i \bX \\ \nonumber
    &+ 2i \bF_{ij} [a_i, a_j] +2i [\bar{c}, \bX] \bD_0 c - 2i \bD_0 \bar{c} [\bX, c] - 2\bD_i \bar{c} \bD_i c \\ \label{eq: gauge sector}
    &- \sqrt{2} i \psi_-^T  \Tilde{D}_0 \psi_- - 2i \psi_-^T \Gamma_{0i} \Tilde{D}_i \psi_+ - \sqrt{2} \psi_+^T [\bX, \psi_+]  \bigg) \ .
\end{align}
To keep the action real we will take $c$ to be Hermitian and $\bar{c}$ anti-Hermitian. As we shall be expanding the path integral in the background fields to compute the one-loop 1PI correlation functions, we must separate out the constant asymptotic value of $\bX$ from the spatially-varying part,
\begin{equation}
    \bX = v \sigma + \Tilde{\Phi} \ ,
\end{equation}
and expand in $\Tilde{\Phi}$. We shall do this explicitly in section \ref{sect: gauge sector}. If we instead keep the scalar background fields\footnote{Note that when referring to scalar fields here we are excluding $\Tilde{X}$.} and set the others to zero (where we keep $\bX = v \sigma$) an almost identical procedure gives the one-loop action
\begin{align} \nonumber
    S_{\text{1-loop}} = \tr \int dt d^2 x \bigg(& 
    2i v \partial_0 \phi [\sigma, a_0] + \partial_i\phi \partial_i a_0 - i v \partial_0 a_i [\sigma, a_i] - \inv{2} \partial_i a_j \partial_i a_j \\ \nonumber
    &- i v \partial_0 y^A [\sigma, y^A] - \inv{2} \partial_i y^A \partial_i y^A - \frac{i}{\sqrt{2}}  \psi_-^T \partial_0 \psi_- \\ \nonumber
    &-  i \psi_-^T \Gamma_{0i} \partial_i \psi_+ - \frac{v}{\sqrt{2}}  \psi_+^T [\sigma, \psi_+]
    - i \partial_0 \bY^A [\phi, y^A] \\ \nonumber
    &- i \partial_0 y^A [\phi, \bY^A] - v [a_0 , y^A] [\sigma, \bY^A] - v [a_0, \bY^A] [\sigma, y^A] \\ \nonumber
    &+ i \partial_i y^A [a_i , \bY^A] + i \partial_i \bY^A [a_i , y^A] - \psi_+^{\alpha} (\Gamma_{0A})_{\alpha \dalpha} [\bY^A, \psi_-^{\dalpha}] 
    \\ \nonumber
    &- [a_0 , \bY^A] [\phi, \bY^A] + \inv{2} [a_i , \bY^A] [a_i , \bY^A] + \inv{2} [y^A , y^B] [\bY^A , \bY^B] \\ \label{eq: scalar sector}
    &+ \inv{2} [y^A , \bY^B] [y^A , \bY^B] - \inv{2} [y^A , \bY^B] [y^B , \bY^A]  \bigg) \ .
\end{align}
where we have ignored the role of the ghosts as they do not couple to $\bY^A$ at one-loop.

We will work with the $\frak{g} = \frak{su}(2)$ theory from here onwards. The background-independent parts of both actions take the same form as \eqref{eq: free action}, with the addition of ghosts in the gauge sector; using the expansions
\begin{subequations} \label{eq: field expansions}
\begin{align}
    a_{\mu} &= \sqrt{2} a_{\mu}^{(0)} \sigma + b_{\mu} \sigma^+ + \bar{b}_{\mu} \sigma^- \ , \\
    \phi &= \sqrt{2} \phi^{(0)} \sigma + \varphi \sigma^+ + \bar{\varphi} \sigma^- \ , \\
    y^A &= \sqrt{2} y^A_{(\sigma)} \sigma + \omega^A \sigma^+ + \bar{\omega}^A \sigma^- \ , \\
    c &= \sqrt{2} c^{(\sigma)} \sigma + d \sigma^+ + d^* \sigma^- \ , \\
    \bar{c} &= \sqrt{2}i \bar{c}^{(\sigma)} \sigma + \bar{d} \sigma^+ - \bar{d}^* \sigma^- \ , \\
    \psi_{\pm} &= \sqrt{2} \psi_{\pm}^{(\sigma)} \sigma + \rho_{\pm} \sigma^{+} + \bar{\rho}_{\pm} \sigma^- \ ,
\end{align}
\end{subequations}
for the fluctuating fields we therefore have the action
\begin{align} \nonumber
    S_{1-loop}^{(0)} = \int dt d^2 x \bigg(&
    2 i v \bar{b}_i \partial_0 b_i - \partial_i \bar{b}_j \partial_i b_j + 2 i v \bar{\omega}^A \partial_0 \omega^A - \partial_i \bar{\omega}^A \partial_i \omega^A \\ \nonumber
    &+ 2 i v \brac{b_0 \partial_0 \bar{\varphi} - \bar{b}_0 \partial_0 \varphi} + \partial_i \varphi \partial_i \bar{b}_0 + \partial_i \bar{\varphi} \partial_i b_0 \\ \nonumber
    &+ i v \brac{ \partial_0 \bar{d}^* d - \bar{d}^* \partial_0 d + \partial_0 \bar{d} d^* - \bar{d} \partial_0 d^*} - \partial_i \bar{d} \partial_i d^* \\ \nonumber
    &+ \partial_i \bar{d}^* \partial_i d - \inv{2} \partial_i a_j^{(\sigma)} \partial_i a_{j}^{(\sigma)} - \inv{2} \partial_i y^A_{(\sigma)} \partial_i y^A_{(\sigma)} + \partial_i \phi^{(\sigma)} \partial_i a_0^{(\sigma)} \\ \nonumber
    &- i \partial_i \bar{c}^{(\sigma)} \partial_i c^{(\sigma)} - \sqrt{2} \bar{\rho}_-^T \partial_0 \rho_- - i \bar{\rho}_-^T \Gamma_{0i} \partial_i \rho_+ - i \bar{\rho}_+^T \Gamma_{0i} \partial_i \rho_- \\ \label{eq: free perturbative theory}
    &- \sqrt{2} v \bar{\rho}_+^T \rho_+ - \inv{\sqrt{2}} \psi_{-}^{(\sigma) \, T} \partial_0 \psi_{-}^{(\sigma)} - i \psi_{-}^{(\sigma) \, T} \Gamma_{0i} \partial_i \psi_{+}^{(\sigma)}
    \bigg) \ ,
\end{align}
defining the propagators of the theory. With this choice $c^{(\sigma)}$ and $\bar{c}^{(\sigma)}$ are real Grassmann-value fields, and $d$ and $\bar{d}$ are complex Grassmann-valued fields.

With this in hand, we can compute the free theory's propagators. Let us start with the dynamical fields. The only non-trivial two-point functions are between unconjugated and conjugated fields, which we write in the form
\begin{equation}
    \bra{0} \cal{T}\{ \cal{A}(t_1,x_1) \bar{\cal{B}}(t_2,x_2) \} \ket{0} = \bigintssss \frac{dE d^2p}{(2\pi)^3} \, e^{-i E(t_1 - t_2) + i p_i (x_1 - x_2)^i} \langle \cal{A}(E,p) \bar{\cal{B}}(E,p) \rangle \ ,
\end{equation}
using the Fourier transform conventions
\begin{subequations}
\begin{align} \label{eq: fourier convention 1}
    \cal{A}(t,x) = \bigintssss\frac{dE d^2p}{(2\pi)^3} e^{-i E t + i p_i x^i} \cal{A}(E,p) \ , \\
    \bar{\cal{A}}(t,x) = \bigintssss\frac{dE d^2p}{(2\pi)^3} e^{i E t - i p_i x^i} \bar{\cal{A}}(E,p) \ .
\end{align}
\end{subequations}
Defining the momentum-space Schr\"odinger propagator
\begin{equation}
    G(E,p) = \frac{i}{2vE - p^2 + i\varepsilon} \ ,
\end{equation}
we find
\begin{subequations}
\begin{align}
    \langle b_i(E,p) \bar{b}_j(E,p) \rangle &= \delta_{ij} G(E,p) \ , \\
    \langle \omega^A(E,p) \bar{\omega}^B(E,p) \rangle &= \delta^{AB} G(E,p) \ , \\
    \langle b_0(E,p) \bar{\varphi}(E,p) \rangle &= - G(E,p) \ , \\
    \langle \varphi(E,p) \bar{b_0}(E,p) \rangle &= - G(E,p) \ , \\
    \langle d(E,p) \bar{d}^*(E,p) \rangle &= - G(E,p) \ , \\
    \langle \bar{d}(E,p) d^*(E,p) \rangle &= - G(E,p) \ , \\
    \langle \rho_-^{\dalpha}(E,p) \brho_-^{\dbeta}(E,p) \rangle &= \sqrt{2} v\, \delta^{\dalpha\dbeta} G(E,p) \ , \\ \nonumber
    \langle \rho_+^{\alpha}(E,p) \brho_-^{\dalpha}(E,p) \rangle &= -p_i (\Gamma_{0i})_{\dalpha \alpha} G(E,p) \ , \\
    \langle \rho_-^{\dalpha}(E,p) \brho_+^{\alpha}(E,p) \rangle &= - p_i (\Gamma_{0i})_{\dalpha \alpha} G(E,p) \ , \\ \nonumber
    \langle \rho_+^{\alpha}(E,p) \brho_+^{\beta}(E,p) \rangle &= \sqrt{2} \delta^{\alpha\beta} E \,G(E,p) \ ,
\end{align}
\end{subequations}
where we use $\{\alpha,\beta,...\}$ as spinor indices for the subspace defined by $P_-$ and $\{\dalpha, \dbeta, ... \}$ for that defined by $P_+$. To keep expressions compact we shall often denote the Schr\"odinger propagator simply by $G(p)$ when the two arguments take the same form, e.g. $G(E + E_i, p + p_i) \equiv G(p + p_i)$.

There is, however, an ambiguity here that we must deal with. We will illustrate this using one of the complex scalars that we will denote simply by $\omega$. Performing the integral over $E$, the real-time momentum space two-point function is
\begin{equation} \label{eq: equal time correlator 1}
    \bra{0} \cal{T} \{ \omega(t_1,p_1) \bar{\omega}(t_2,p_2)  \} \ket{0} = \frac{(2\pi)^2}{2v} \delta^{(2)}(p_1 - p_2) \Theta(t_1 - t_2) e^{- \frac{i p_1^2 (t_1 - t_2)}{2v}} \ .
\end{equation}
This is well-defined for $\abs{t_1 - t_2} > 0$, but is ambiguous when $t_1 = t_2$ due to the presence of $\Theta(0)$. This can be understood as a consequence of the system being first-order in time derivatives, which has two important implications for the theory within canonical quantization. The first is that $\omega(t,x)$ only has a single oscillator mode in its Fourier transform\footnote{Which is equivalent to the familiar statement that the non-relativistic theory has no antiparticles in its spectrum.} and so can be taken to annihilate the vacuum,
\begin{equation}
    \omega(t,x) \ket{0} = 0 \ .
\end{equation}
The second is that (after canonically normalising the fields) $i\bar{\omega}(t,x)$ is the conjugate momentum of $\omega(t,x)$ and the two obey the equal-time commutation relation
\begin{equation}
    [\omega(t,x) , \bar{\omega}(t,y)] = \delta^{(2)}(x-y) \ .
\end{equation}
As the fields commute under the time-ordering symbol a naive evaluation of the time-ordered two-point function can either yield
\begin{align} \nonumber
    \bra{0} \cal{T} \{ \omega(t,x) \bar{\omega}(t,y) \} \ket{0} &= \bra{0}  \omega(t,x) \bar{\omega}(t,y) \ket{0} \\ \nonumber
    &= \bra{0}  [\omega(t,x) , \bar{\omega}(t,y)]\ket{0} \\
    &= \delta^{(2)}(x-y) \ ,
\end{align}
or
\begin{align} \nonumber
    \bra{0} \cal{T} \{ \omega(t,x) \bar{\omega}(t,y) \} \ket{0} &= \bra{0}   \bar{\omega}(t,y) \omega(t,x) \ket{0} \\
    &= 0 \ .
\end{align}
These are clearly inconsistent, and we need to specify an ordering procedure for the equal-time product of operators. A commonly used prescription in the quantum many body systems literature\footnote{See \cite{Shankar:2017zag} for a textbook account of this.} is to shift $t_2$ by a positive infinitesimal value $\eta$. This has no effect when $\abs{t_1 - t_2} > 0$ but shifts the argument of the equal time $\Theta$-function to $-\eta$, causing it to vanish. This can be more easily implemented by a redefinition of the Schr\"odinger propagator,
\begin{equation} \label{eq: propagator time splitting}
    \frac{i}{2vE - p^2 + i \varepsilon} \Rightarrow \frac{i e^{i \eta E}}{2vE - p^2 + i \varepsilon} \equiv G(E,p) \ ,
\end{equation}
which we shall take as part of the definition of $G(E,p)$ for the rest of this work. Since \eqref{eq: equal time correlator 1} is the inverse Fourier transform of $G(E,p)$ with respect to time, a consequence of this definition is that
\begin{equation} \label{eq: single G vanishing}
    \bigintssss \frac{dE}{2\pi} \, G(E,p) = 0 \ .
\end{equation}
This initially seems somewhat surprising, as the integral is naively logarithmically divergent via a power counting argument: however, a quick computation in any regularisation shows that the divergence cancels, leaving behind a finite regulator-dependent piece. We note here that a different regularisation that assigns this integral a finite value can induce additional divergences through the integral over loop momenta.

It was pointed out in \cite{Auzzi:2019kdd, Chapman:2020vtn} that presence of only a single (complex) pole in the Schr\"odinger propagator immediately gives the result
\begin{equation} \label{eq: multi G vanishing}
    \bigintssss \frac{dE}{2\pi} \, \prod_{i=1}^n G(E + E_i , p + p_i) = 0
\end{equation}
for $n\geq2$, as the integrand is sufficiently convergent for $\abs{E}\to\infty$ for us to close the integration contour in the upper-half complex plane (where $G(E + E_i , p+ p_i)$ has no poles) at no additional cost. Such integrals correspond to one-loop Feynman diagrams in which all propagators of dynamical fields are oriented along the direction of loop momentum flow. As we have interactions containing derivatives of fields this result must be generalised to include insertions of loop energy; by rewriting $E$ in terms of inverse powers of $G$ and energy-independent pieces we can use the results \eqref{eq: single G vanishing} and \eqref{eq: multi G vanishing} to get
\begin{equation} \label{eq: vanishing with additional E}
    \bigintssss \frac{dE}{2\pi} \, E^m \prod_{i=1}^n G(E + E_i , p + p_i) = 0
\end{equation}
for $n-m\geq1$. This will prove to be extraordinarily useful when we begin computing the contributions to $\Gamma_{1}$.

Finally, the momentum-space correlation functions of the non-dynamical fields are
\begin{subequations}
\begin{align}
    \bra{0} a_i^{(\sigma)}(E_1 ,p_1) a_j^{(\sigma)}(E_2 ,p_2) \ket{0} &= - (2\pi)^3 \delta(E_1 + E_2) \delta^{(2)}(p_1 + p_2) \delta_{ij} \frac{i}{p_1^2} \ , \\
    \bra{0} y^A_{(\sigma)} (E_1 , p_1) y^B_{(\sigma)} (E_2 , p_2) \ket{0} &= - (2\pi)^3 \delta(E_1 + E_2) \delta^{(2)}(p_1 + p_2) \delta^{AB} \frac{i}{p_1^2} \ , \\
    \bra{0} \phi^{(\sigma)}(E_1 , p_1) a_0^{(\sigma)}(E_2,p_2) \ket{0} &= (2\pi)^3 \delta(E_1 + E_2) \delta^{(2)}(p_1 + p_2) \frac{i}{p_1^2} \ , \\
    \bra{0} c^{(\sigma)}(E_1 , p_1) \bar{c}^{(\sigma)}(E_2 , p_2) \ket{0} &= - (2\pi)^3 \delta(E_1 + E_2) \delta^{(2)}(p_1 + p_2) \inv{p_1^2} \ , \\
    \bra{0} \gamma_+^{\alpha}(E_1,p_1) \gamma_+^{\beta}(E_2,p_2) \ket{0} &= -  (2\pi)^3 \delta(E_1 + E_2) \delta^{(2)}(p_1 + p_2) \delta^{\alpha\beta} \frac{\sqrt{2} i E_1}{p_1^2} \ , \\
    \bra{0} \gamma_+^{\alpha}(E_1 , p_1) \gamma_-^{\dalpha}(E_2 , p_2)  \ket{0} &= -   (2\pi)^3 \delta(E_1 + E_2) \delta^{(2)}(p_1 + p_2) \frac{i (\Gamma_{0i})_{\dalpha \alpha} p_{1,i}}{p_1^2} \ , \\
    \bra{0} \gamma_-^{\alpha}(E_1 , p_1) \gamma_+^{\dalpha}(E_2 , p_2)  \ket{0} &= -   (2\pi)^3 \delta(E_1 + E_2) \delta^{(2)}(p_1 + p_2) \frac{i (\Gamma_{0i})_{\dalpha \alpha} p_{1,i}}{p_1^2} \ , \\
    \bra{0}  \gamma_-^{\dalpha}(E_1, p_1) \gamma_-^{\dbeta}(E_2,p_2) \ket{0} &= 0 \ ,
\end{align}
\end{subequations}
 where we use the Fourier transform convention \eqref{eq: fourier convention 1} for the real fields. As these are independent of energy the real-space functions we get after inverting the Fourier transform are proportional to temporal delta functions, meaning the fields mediate instantaneous forces as we previously asserted. These propagators agree with those found in previous work on quantizing Galilean electrodynamics \cite{Chapman:2020vtn, Baiguera:2022cbp}\footnote{An alternative quantization scheme for the Galilean electromagnetism multiplet was recently proposed in \cite{Hernandez:2026stx}, and it would be interesting to better understand the physical content of the difference in the two approaches.}.

\subsection{Quantizing the Gauge Sector} \label{sect: gauge sector}

We are now ready to begin computing the theory's 1PI correlation functions, starting with the gauge sector. As the original fields will not play a role in the remainder of this section we will drop the tildes on the background fields to help simplify our notation. We will also use the same notation for the expansion of the background fields into $\frak{su}(2)$ components as we used for the full fields in \eqref{eq: matrix expansions}. In particular, $\Tilde{Y}^A$ and $\Tilde{\Phi}$ are now expanded as
\begin{subequations} \label{eq: background field defs}
\begin{align}
   \Tilde{Y}^A \to Y^A &= \sqrt{2} Y^A_{(\sigma)} \sigma + \Omega^A \sigma^+ + \bar{\Omega}^A \sigma^- \ , \\ 
   \Tilde{\Phi} \to \Phi &= \sqrt{2} \Phi_{(\sigma)} \sigma + {\cal X} \sigma^+ + \bar{\cal X} \sigma^- \ .
\end{align}
\end{subequations}

Let us first consider only taking $\Phi$ to be non-vanishing. The terms involving both $\Phi$ and the fluctuating fields are
\begin{align} \nonumber
    S_{\Phi} = \tr \int dt d^2x \bigg( &
    2i \partial_0 \phi [\Phi, a_0] - i \partial_0 a_i [\Phi, a_i] - i \partial_0 y^A [\Phi, y^A] \\ \nonumber
    &+ i [\bar{c}, \Phi] \partial_0 c - i \partial_0 \bar{c} [\Phi, c] + 2i \partial_i \Phi [a_0 , a_i]  \\
    &+ i \partial_0 \Phi [\phi, a_0] - \inv{\sqrt{2}} \psi_+^T [ \Phi, \psi_+]  \bigg) \ .
\end{align}
Expanding both the background and fluctuating fields in $\frak{su}(2)$ components (using \eqref{eq: field expansions} for the fluctuating fields) and Fourier transforming gives
\begin{align} \nonumber
    S_{\Phi} = \sqrt{2} \bigintsss  d\mu_1 d\mu_2 \Bigg(&
    \Phi^{(\sigma)}_{1} \bigg[
    -  (E_1+ 2E_3) b_{0,2} \bar{\varphi}_{3} - (2E_2-E_1) \varphi_{2} \bar{b}_{0,3} \\ \nonumber
    &+ (E_2 + E_3) \bigg( b_{i,2} \bar{b}_{i,3} +  \omega^A_{2} \bar{\omega}^A_{3} + \bar{d}_{2} d^*_{3} - \bar{d}^*_{3} d_{2} \bigg) \\ \nonumber
    &- 2p_1^i \brac{b_{0,2} \bar{b}_{i,3} - b_{i,2} \bar{b}_{0,3} } - \sqrt{2} \rho_{+,2}^{\alpha} \brho_{+,3}^{\alpha} \bigg]
    \\ \nonumber
    &+ \cal{X}_{1} \bigg[ 
    (2E_2-E_1) \phi^{(\sigma)}_2 \bar{b}_{0,3} + (2E_3 + E_1) a^{(\sigma)}_{0,2} \bar{\varphi}_{3}   \\ \nonumber
    &- (E_2+E_3) \bigg( a^{(\sigma)}_{i,2} \bar{b}_{i,3} + y^A_{(\sigma),2} \bar{\omega}^A_{3} + i  \bar{c}^{(\sigma)}_{2} d^*_{3} -  \bar{d}^*_{3} c^{(\sigma)}_{2} \bigg) \\ \nonumber
    &- 2p_1^i  \brac{a^{(\sigma)}_{i,2} \bar{b}_{0,3} - a^{(\sigma)}_{0,2} \bar{b}_{i,3}} + \sqrt{2} \gamma_{+,2}^{\alpha} \brho^{\alpha}_{+,3} \bigg] \\ \nonumber
    &+ \bar{\cal{X}}_{3} \bigg[
    (2E_2+E_3) a^{(\sigma)}_{0,1} \varphi_{2} - (2E_2 + E_3) b_{0,1} \phi^{(\sigma)}_{2}   
    \\ \nonumber
    &+ (E_1 - E_2) \bigg( a^{(\sigma)}_{i,1} b_{i,2} + y^A_{(\sigma),1} \omega^A_{2} + i \bar{c}^{(\sigma)}_{1} d_{2} - \bar{d}_{1} c^{(\sigma)}_{2} \bigg) \\ 
    &+ 2p_3^i \brac{a^{(\sigma)}_{0,1} b_{i,2} - b_{0,1} a^{(\sigma)}_{i,2}} - \sqrt{2} \gamma_{+,1}^{\alpha} \rho_{+,2}^{\alpha} \bigg] \Bigg) \ .
\end{align}
Here we use the condensed notation $\cal{A}_i \equiv \cal{A}(E_i,p_i)$, $E_3 = E_1 + E_2$, $p^i_3 = p^i_1 + p^i_2$, and
\begin{equation}
    d\mu_i = \frac{d E_i d^2 p_i}{(2\pi)^3} \ .
\end{equation}
The one-loop effective action of $\Phi$ has the expansion 
\begin{equation}
    e^{i \Gamma_1[\Phi]} = \cal{Z}_0 \brac{
    1 + i \langle S_{\Phi} \rangle - \inv{2} \langle S^2_{\Phi} \rangle - \frac{i}{6} \langle S^3_{\Phi} \rangle + \inv{24} \langle S^4_{\Phi} \rangle } + O(\Phi^5) \ ,
\end{equation}
where we compute expectation values using \eqref{eq: free perturbative theory}. As it will play no further role we will formally set $\cal{Z}_0 = 1$. 

Our strategy in performing the perturbative calculations will be to perform the integral over the loop energy before that of the loop momentum in order to exploit the result \eqref{eq: vanishing with additional E}. Diagrammatically, the result means that (for a choice of loop momentum orientation) any graph whose internal Schr\"odinger propagators are all aligned\footnote{Or, equivalently, all anti-aligned.} with the loop momentum either have vanishing or power-law divergent loop energy integral. As such, there is no physical information contained in these diagrams, and one may hope that they cancel internally: in this section we shall show that (at least for the examples we consider) this is indeed the case. It is straightforward to see that all diagrams that can contribute to the one-point, two-point, and three-point functions are of this form. The first genuine quantum corrections to the theory therefore lie in the four-point 1PI correlation function. 

Before tackling this, however, we will show that the power-law divergences in the loop energy integrals cancel in the lower-point cases. The one-point function of $S_{\Phi}$ is given by
\begin{align} \nonumber
    \langle S_{\Phi} \rangle &= 2 \sqrt{2} \int d\mu \,\Phi^{(\sigma)}_p \,(2\pi)^3 \delta^{(3)}(p) \int d\Tilde{\mu} \, \Tilde{E}\, G(\Tilde{p}) \brac{\delta_{ii} + \delta_{AA} - \delta_{\alpha \alpha}} \\
    &= 0 \ .
\end{align}
As in Lorentzian supersymmetric gauge theories, matching between bosonic and fermionic degrees of freedom ensures that this vanishes and keeps the VEV stable at one-loop. There are two potential contributions to $\langle S_{\Phi}^2 \rangle$ allowed by $U(1)$ invariance of the effective action: we either have two copies of the interaction terms proportional to $\Phi^{(\sigma)}$ in the correlation function, or a pair proportional to $\cal{X}$ and $\bcalX$ respectively. The first of these is\footnote{As the one-point function vanishes we can focus on the connected contribution.}
\begin{align} \nonumber
    \langle S_{\Phi}^2 \rangle^{(1)} = 2 \bigintssss &d \mu_1  d\mu_2 d\mu_4 d\mu_5 \Phi_1^{(\sigma)} \Phi^{(\sigma)}_4 \bigg[ 
    (E_1 + 2E_3) (E_4 + 2 E_6) \langle b_{0,5} \bar{\varphi}_3 \rangle \langle b_{0,2} \bar{\varphi}_6 \rangle \\ \nonumber
    &+ (2E_2 - E_1) (2 E_5 - E_4) \langle \varphi_2 \bar{b}_{0,5} \rangle \langle \varphi_5 \bar{b}_{0,3} \rangle + 2 \langle \rho^{\alpha}_{+,2} \bar{\rho}^{\beta}_{+,6} \rangle \langle  \bar{\rho}^{\alpha}_{+,3} \rho^{\beta}_{+,5} \rangle \\ \nonumber
    &+ (E_2 + E_3) (E_5 + E_6) \Big( 
    \langle b_{i,2} \bar{b}_{j,6} \rangle \langle b_{j,5} \bar{b}_{i,3} \rangle + \langle \omega^A_2 \bar{\omega}^B_6 \rangle \langle \omega^B_5 \bar{\omega}^A_3 \rangle \\
    &+ \langle \bar{d}_2 d_6^* \rangle \langle d_3^* \bar{d}_5 \rangle + \langle 
    \bar{d}_3^* d_5 \rangle \langle  d_2 \bar{d}_6^* \rangle \Big) \bigg] \ ,
\end{align}
which upon evaluating the correlation functions is
\begin{align} \nonumber
    \langle S_{\Phi}^2 \rangle^{(1)} &= 4 \int d\mu \,\Phi_p^{(\sigma)} \Phi_{-p}^{(\sigma)} \int d\Tilde{\mu} \, G(\Tilde{p}) G(p + \Tilde{p}) \Big[ 3 (2 \Tilde{E} + E)^2 \\ \nonumber
    &\qquad \qquad \qquad + (3 E + 2 \Tilde{E}) (2 \Tilde{E} - E) - 16 \Tilde{E} (E + \Tilde{E}) \Big] \\
    &= 0 \ ,
\end{align}
as hoped. Similarly, the $\cal{X}\bcalX$ term is
\begin{align} \nonumber
    \langle S^2_{\Phi} \rangle^{(2)} = 4 \bigintssss &d \mu_1 d\mu_2 d\mu_4 d\mu_5 \, \cal{X}_1 \bcalX_6 \bigg[ 
    (2 E_2 - E_1) (2 E_5 + E_6) \langle \phi_2^{(\sigma)} a_{0,4}^{(\sigma)} \rangle \langle \bar{b}_{0,3} \varphi_5 \rangle \\ \nonumber
    &- (2 E_3 + E_1) ( 2 E_5 + E_6) \langle b_{0,4} \bar{\varphi} \rangle \langle a_{0,2}^{(\sigma)} \phi^{(\sigma)}_5 \rangle - 2 \langle \gamma^{\alpha}_{+,2} \gamma^B_{+,4} \rangle \langle \rho^{\beta}_{+,5} \bar{\rho}^{\alpha}_{+,3} \rangle \\ \nonumber
    &- (E_2 + E_3) (E_4 - E_5) \Big( 
    \langle a_{i,2}^{(\sigma)} a_{j,4}^{(\sigma)} \rangle \langle \bar{b}_{i,3} b_{j,5} \rangle + \langle y^A_{(\sigma),2} y^B_{(\sigma),4 } \rangle \langle \bar{\omega}^A_3 \omega^B_5 \rangle \\
    &- i \langle c_5^{(\sigma)} \bar{c}^{(\sigma)}_2 \rangle \langle \bar{d}_4 d_3^* \rangle + i \langle c_2^{(\sigma)} \bar{c}_4^{(\sigma)} \rangle\langle d_5 \bar{d}_3^* \rangle \Big) \bigg] \ ,
\end{align}
which evaluates to
\begin{align}
    \langle S_{\Phi}^2 \rangle^{(1)} &= -8i \bigintssss d\mu \,\cal{X}_p \bcalX_p \bigintssss d\Tilde{\mu} \, \frac{G(p + \Tilde{p})}{\Tilde{p}^{2}} \Big[ 3 (2 \Tilde{E} + E)^2 \\ \nonumber
    &\qquad \qquad \qquad + (3 E + 2 \Tilde{E}) (2 \Tilde{E} - E) - 16 \Tilde{E} (E + \Tilde{E}) \Big] \\
    &= 0 \ .
\end{align}
Finally, we must do the same for $\langle S_{\Phi}^3 \rangle$. There are two terms which can appear here, with both a single neutral and charged pair of background fields, and three neutral fields allowed by $U(1)$-invariance. A quick computation shows that the first of these is given by the fairly lengthy expression
\begin{subequations}
\begin{align}
    \langle S_{\Phi}^3 \rangle^{(1)} &= 12 \sqrt{2} \int d\mu_1 d\mu_2 d\mu_4 d\mu_5 d\mu_7 d\mu_8 \, \Phi^{(\sigma)}_1 \cal{X}_4 \bcalX_9 \, \cal{I}_1 \ , \\ \nonumber
    \cal{I}_1 &= (E_1 + 2 E_3) (2 E_6 + E_4) (2 E_8 + E_9) \langle b_{0,2} \bar{\varphi}_{6} \rangle \langle b_{0,7} \bar{\varphi}_{3} \rangle \langle a_{0,5}^{(\sigma)} \phi_8^{(\sigma)} \rangle \\ \nonumber
    &\qquad - (2 E_2 - E_1) (2 E_5 - E_4) (2 E_8 + E_9) \langle \varphi_2 \bar{b}_{0,6} \rangle \langle \varphi_8 \bar{b}_{0,3} \rangle \langle \phi_5^{(\sigma)} a_{0,7}^{(\sigma)} \rangle \\ \nonumber
    &\qquad- (E_2 + E_3) (E_5 + E_6) (E_7 - E_8) \Big( 
    \langle b_{i,2} \bar{b}_{j,6} \rangle \langle b_{k,8} \bar{b}_{i,3} \rangle \langle a_{j,5}^{(\sigma)} a_{k,7}^{(\sigma)} \rangle \\ \nonumber
    &\qquad+ \langle \omega_2^A \bar{\omega}_6^B \rangle \langle \omega_8^C \bar{\omega}_3^A \rangle \langle y^B_{(\sigma),5} y^C_{(\sigma),7} \rangle + i \langle \bar{d}_2 d_6^* \rangle \langle \bar{d}_7 d_3^* \rangle \langle c_8^{(\sigma)} \bar{c}_5^{(\sigma)} \rangle \\
    &\qquad- i \langle d_2 \bar{d}^*_6 \rangle \langle d_8 \bar{d}_3^* \rangle \langle c_5^{(\sigma)} \bar{c}^{(\sigma)}_7 \rangle \Big) - 2 \sqrt{2} \langle \rho^{\alpha}_{+,2} \bar{\rho}^{\beta}_{+,6} \rangle \langle \rho^{\gamma}_{+,8} \brho^{\alpha}_{+,3} \rangle \langle \gamma^{\beta}_{+,5} \gamma^{\gamma}_{+,7} \rangle  \ .
\end{align}
\end{subequations}
However, upon evaluating this we find
\begin{align} \nonumber
    \langle S_{\Phi}^3 \rangle^{(1)} &= - 12\sqrt{2} i \bigintssss d\mu_1 d\mu_2 \, \Phi_1^{(\sigma)} \cal{X}_2 \bcalX_3 \bigintssss d\Tilde{\mu} \,
    \frac{G(\Tilde{p}) G(p_1 + \Tilde{p})}{(\Tilde{p} - p_2)^2} \\ \nonumber
    & \hspace{2.0cm}\times \bigg[ 
    (2 \Tilde{E} + 3 E_1) (2 \Tilde{E} + E_2 ) (2 \Tilde{E} - E_1 - 3 E_2) \\ \nonumber
    &\hspace{2.7cm}+ (2 \Tilde{E} - 3 E_2) (2 \Tilde{E} - E_1) ( 2 \Tilde{E} + 3 E_1 + E_2) \\ \nonumber
    &\hspace{2.7cm}+ 6 (2 \Tilde{E} + E_1) (2 \Tilde{E} - E_2) (2 \Tilde{E} + E_1 - E_2) \\ \nonumber
    &\hspace{2.7cm}- 64 \Tilde{E} (\Tilde{E} + E_1) (\Tilde{E} - E_2) \bigg] \\
    &= 0 \ .
\end{align}
The term with three neutral background fields is
\begin{subequations}
\begin{align} 
    \langle S_{\Phi}^3 \rangle^{(2)} &= 2\sqrt{2} \bigintssss d\mu_1 d\mu_2 d\mu_4 d\mu_5 d\mu_7 d\mu_8 \, \Phi_1^{(\sigma)} \Phi^{(\sigma)}_4 \Phi_7^{(\sigma)} \,\cal{I}_2 \ , \\ \nonumber
    \cal{I}_2 &= -(E_1 + 2 E_3) (E_4 + 2E_6) (E_7 + 2 E_9) \langle b_{0,2} \bar{\varphi}_3 b_{0,5} \bar{\varphi}_6 b_{0,8} \bar{\varphi}_9 \rangle \\ \nonumber
    & - (2 E_2 - E_1 )( 2 E_5 - E_4) (2E_8 - E_7) \langle \varphi_2 \bar{b}_{0,3} \varphi_5 \bar{b}_{0,6} \varphi_8 \bar{b}_{0,9} \\ \nonumber
    & + (E_2 + E_3) (E_5 + E_6) (E_8 + E_9) \bigg( 
    \langle b_{i,2} \bar{b}_{i,3} b_{j,5} \bar{b}_{j,6} b_{k,8} \bar{b}_{k,9} \rangle \\ \nonumber
    &+ \langle \omega^A_{2} \bar{\omega}^A_{3} \omega^B_{5} \bar{\omega}^B_{6} \omega^C_{8} \bar{\omega}^C_{9} \rangle + \Big\langle 
    \Big( \bar{d}_2 d^*_3 - \bar{d}^*_3 d_2 \Big) \Big( \bar{d}_5 d^*_6 - \bar{d}^*_6 d_5 \Big) \Big( \bar{d}_8 d^*_9 - \bar{d}^*_9 d_8 \Big) \Big\rangle 
    \bigg) \\
    & - 2 \sqrt{2} \langle \rho^{\alpha}_{+,2} \bar{\rho}^{\alpha}_{+,3} \rho^{\beta}_{+,5} \bar{\rho}^{\beta}_{+,6} \rho^{\gamma}_{+,8} \bar{\rho}^{\gamma}_{+,9} \rangle \ ,
\end{align}   
\end{subequations}
which evaluates to 
\begin{align} \nonumber
    \langle S_{\Phi}^3 \rangle^{(2)} &= 4 \sqrt{2} \bigintssss d\mu_1 d\mu_2 \Phi_1^{(\sigma)} \Phi^{(\sigma)}_2 \Phi^{(\sigma)}_{-3} \bigintssss d\Tilde{\mu} \, G(\Tilde{p}) G(\Tilde{p} + p_1) G(\Tilde{p} + p_1 + p_2) \\ \nonumber
    & \qquad \qquad\times  \bigg[ 
    (2\Tilde{E} + 3 E_1) (2\Tilde{E} + 2 E_1 + 3 E_2) (2\Tilde{E} - E_1 - E_2) \\ \nonumber
    & \qquad \qquad \hspace{0.7cm} + (2 \Tilde{E} - E_1) ( 2 \Tilde{E} + 2 E_1 - E_2) (2\Tilde{E} + 3 E_1 + 3 E_2) \\ \nonumber
    &\qquad \qquad \hspace{0.7cm}  + 6 (2 \Tilde{E} + E_1 ) ( 2 \Tilde{E} + 2 E_1 + E_2 )( 2 \Tilde{E} + E_1 + E_2) \\ \nonumber
    &\qquad \qquad \hspace{0.7cm} - 64 \Tilde{E} ( \Tilde{E} + E_1) ( \Tilde{E} + E_1 + E_2) \bigg] \\
    &= 0 \ ,
\end{align}
and so the power-law divergences in the $\Tilde{E}$ integral cancel within the one, two, and three-point functions as hoped. As mentioned above, it is essential that we have a supersymmetric theory in order to obtain these cancellations.

With these out of the way, let us turn our attention to the four-point function. If we think about the allowed diagrams for each set of possible background fields, it is straightforward to see that we can only construct terms whose Schr\"odinger propagators have opposing orientation with respect to the loop momentum if the external fields are two copies of $\cal{X}$ and two copies of $\bcalX$. For this reason we shall focus our attention here and set $\Phi^{(\sigma)} = 0$. After a lengthy computation one finds the surprisingly simple result
\begin{align} \nonumber
    \inv{24} \langle S_{\Phi}^4 \rangle &= -12 \bigintssss  \prod_{i=1}^3 d\mu_i \, \cal{X}_1 \cal{X}_2 \bcalX_3 \bcalX_4 \bigintssss d\Tilde{\mu} \, E_1 E_2 E_3 E_4 \, \cal{J} \ , \\ \nonumber
    \cal{J} &= \frac{G(p_1 + \Tilde{p})G( p_2 - \Tilde{p})}{\Tilde{p}^2} \brac{\inv{ (p_1 - p_3 + \Tilde{p})^2} + \inv{ (p_3 - p_2 + \Tilde{p})^2}}  \\ 
    &\hspace{0.5cm}+ \frac{G(p_1 + \Tilde{p})G(p_3 + \Tilde{p})}{\Tilde{p}^2 } \brac{\inv{(p_3 - p_2 + \Tilde{p})^2} + \inv{(p_1 - p_4 + \Tilde{p})^2}} \ ,
\end{align}
where the external energies and momenta obey the constraints $E_1 + E_2 = E_3 + E_4$, $p_1^i + p_2^i = p_3^i + p_4^i$. The final line is zero using \eqref{eq: multi G vanishing}; evaluating the first then gives the one-loop vertex function
\begin{align} \nonumber
    \inv{24} \langle S_{\Phi}^4 \rangle &= - \bigintssss  \prod_{i=1}^4 d\mu_i \, \cal{X}_1 \cal{X}_2 \bcalX_3 \bcalX_4 \, V_4^{(\Phi)}(E_i,p_i) , \\ \nonumber
    V_4^{(\Phi)}(E_i,p_i) &= 12\bigintssss \frac{d^2 \Tilde{p}}{(2\pi)^2} \, \frac{i E_1 E_2 E_3 E_4 }{2v\Tilde{p}^2\brac{2v (E_1 + E_2) - (p_1 + \Tilde{p})^2 - (p_2 - \Tilde{p})^2 + 2 i \epsilon}} \\
    &\qquad \qquad \qquad \times \brac{\inv{ (p_1 - p_3 + \Tilde{p})^2} + \inv{ (p_3 - p_2 + \Tilde{p})^2}} \ .
\end{align}
This is IR-divergent, as may be expected from the perturbative expansion of a scaleless theory, but is perfectly well-defined for large loop momentum. 

The next set of correlation functions to consider are those formed from the spatial components of the gauge field. We shall restrict our interest to the case of charged external fields,
\begin{equation}
    A_i = B_i \sigma^+ + \bar{B}_i \sigma^- \ ,
\end{equation}
since (as we saw above) this allows for diagrams contributing to four-point functions which don't trivially vanish or contribute a power-law divergence. Expanding \eqref{eq: gauge sector} with $A_0 = 0$ and $X = v \sigma$ gives the interaction terms
\begin{align} \nonumber
    S_{B,1} = \sqrt{2} \bigintssss d\mu_1 d\mu_2 \bigg[& 
    B_{i,1} \bigg( 
    (p_2 + p_3)_i \Big( a_{0,2}^{(\sigma)} \bar{\varphi}_3 + \phi^{(\sigma)}_2 \bar{b}_{0,3} - a_{j,2}^{(\sigma)} \bar{b}_{j,3} - y^A_{(\sigma),2} \bar{\omega}_3^A \\ \nonumber
    &\hspace{0.9cm} - c_2^{(\sigma)} \bar{d}^*_3 - i \bar{c}^{(\sigma)}_2 d_3^* \Big) - 2 E_1 \big( a_{i,2}^{(\sigma)} \bar{\varphi}_3 - \phi_2^{(\sigma)} \bar{b}_{i,3} \big) \\ \nonumber
    &\hspace{0.9cm}+ 2 p_{1,j} \big( a_{j,2}^{(\sigma)} \bar{b}_{i,3} - a_{i,2}^{(\sigma)} \bar{b}_{j,3} \big) - 2v \big( a_{i,2}^{(\sigma)} \bar{b}_{0,3} - a_{0,2}^{(\sigma)} \bar{b}_{i,3} \big) \\ \nonumber
    &\hspace{0.9cm} - (\Gamma_{0i})_{\alpha \dalpha} \big( 
    \gamma_{-,2}^{\dalpha} \bar{\rho}_{+,3}^{\alpha} + \gamma_{+,2}^{\alpha} \bar{\rho}_{-,3}^{\dalpha} \big)
    \bigg) \\ \nonumber
    + &\bar{B}_{i,3} \bigg(
    (p_1 - p_2)_i \Big( 
    \varphi_1 a_{0,2}^{(\sigma)} + b_{0,1} \phi_2^{(\sigma)} - b_{j,1} a_{j,2}^{(\sigma)} - \omega^A_1 y^A_{(\sigma),2} \\ \nonumber
    &\hspace{0.9cm} - \bar{d}_1 c_2^{(\sigma)} + i d_1 \bar{c}_2^{(\sigma)} \Big) - 2 E_3 \big( \varphi_1 a_{i,2}^{(\sigma)} - b_{i,1} \phi_2^{(\sigma)} \big) \\ \nonumber
    &\hspace{0.9cm}+ 2 p_{3,j} \big( b_{i,1} a_{j,2}^{(\sigma)} - b_{j,1} a_{i,2}^{(\sigma)} \big) - 2 v \big( b_{0,1} a_{i,2}^{(\sigma)} - b_{i,1} a_{0,2}^{(\sigma)} \big) \\ \label{eq: B interactions 1}
    &\hspace{0.9cm}- (\Gamma_{0i})_{\alpha \dalpha} \big( \rho_{-,1}^{\dalpha} \gamma_{+,2}^{\alpha} + \rho_{+,1}^{\alpha} \gamma_{-,2}^{\dalpha} \big) \bigg) \bigg] \ ,
\end{align}
and
\begin{align} \nonumber
    S_{B,2} = \bigintssss d\mu_1 d\mu_2 d\mu_3 \bigg[& 
    B_{i,1} B_{i,2} \Big( \bar{b}_{j,3} \bar{b}_{j,4} + \bar{\omega}^A_3 \bar{\omega}^A_4 - 2 \bar{b}_{0,3} \bar{\varphi}_4 + 2 d^*_3 \bar{d}^*_4  \Big) \\ \nonumber
    +& \bar{B}_{i,3} \bar{B}_{i,4} \Big( 
    b_{j,1} b_{j,2} + \omega^A_1 \omega^A_2 - 2 b_{0,1} \varphi_2 - 2 d_1 \bar{d}_2 \Big) \\ \nonumber
    + &2 B_{i,1} \bar{B}_{i,3} \Big( 
    2 \phi^{(\sigma)}_2 a_{0,-4}^{(\sigma)} + b_{0,2} \bar{\varphi}_4 + \varphi_2 \bar{b}_{0,4}  - y^A_{(\sigma),2} y^A_{(\sigma),-4} \\ \nonumber
    &\hspace{1.8cm}- \omega^A_2 \bar{\omega}^A_4 + 2 i c^{(\sigma)}_2 \bar{c}^{(\sigma)}_{-4} - \bar{d}_2 d^*_4 - d_2 \bar{d}^*_4
    \Big) \\
    +& 2 B_{i,1} \bar{B}_{j,3} \Big( b_{i,2} \bar{b}_{j,4} - b_{j,2} \bar{b}_{i,4} - \delta_{ij} \big( a_{k,2}^{(\sigma)} a_{k,-4}^{(\sigma)} + b_{k,2} \bar{b}_{k,4} \big) \Big)
    \bigg] \ .
\end{align}
The one-loop effective action for $B_i$ is then given by the expansion
\begin{subequations} \label{eq: Gamma one-loop B}
\begin{align} 
    e^{i\Gamma_1[B]} &= 1 + i \Gamma_1^{(2)}[B] + \bigg( i\Gamma_1^{(4)}[B] - \inv{2} \Gamma_1^{(2)}[B]^2 \bigg) + O(B^6) \ , \\
    \Gamma_1^{(2)}[B] &= \langle S_{B,2} \rangle + \frac{i}{2} \langle S_{B,1}^2 \rangle \ , \\ \label{eq: Gamma1 A4}
    \Gamma_1^{(4)}[B] &= \frac{i}{2} \langle S_{B,2}^2 \rangle^{(\text{conn.})} - \inv{2} \langle S_{B,1}^2 S_{B,2} \rangle^{(\text{conn.})} - \frac{i}{24} \langle S_{B,1}^4 \rangle^{(\text{conn.})} \ ,
\end{align}
\end{subequations}
in increasing powers of the background fields. It is straightforward to see that the non-vanishing contributions to the two-point function are
\begin{align}
    \langle S_{B,2} \rangle = 16 i \bigintssss d\mu \, B_i \bar{B}_i \bigintssss d\Tilde{\mu} \, \inv{\Tilde{p}^2} \ ,
\end{align}
and
\begin{align}
    \langle S_{B,1}^2 \rangle = 32 i \bigintssss d\mu \, B_i \bar{B}_i \bigintssss d\Tilde{\mu} \, \frac{2v \Tilde{E}}{\Tilde{p}^2} G(p + \Tilde{p})  \ .
\end{align}
However, as
\begin{equation} \label{eq: S_1^2 identity}
    2v \Tilde{E} \, G(\Tilde{p} + p) = i e^{i\eta (\Tilde{E} + E)} - \big( 2vE - (p + \Tilde{p})^2 + i \varepsilon \big) G(\Tilde{p} + p) \ ,
\end{equation}
(where we will neglect infinitesimal quantities in what follows) the integral is just
\begin{align}
    \langle S_{B,1}^2 \rangle = - 32  \bigintssss d\mu \, B_i \bar{B}_i \bigintssss d\Tilde{\mu} \, \inv{\Tilde{p}^2} \ ,
\end{align}
exactly cancelling the contribution from $\langle S_{B,2} \rangle$ in $\Gamma_1^{(2)}[B]$, meaning
\begin{equation}
    \Gamma_1^{(2)}[B] = 0 \ .
\end{equation}

\begin{figure}[!ht]
\centering
\includegraphics[width=1\textwidth]{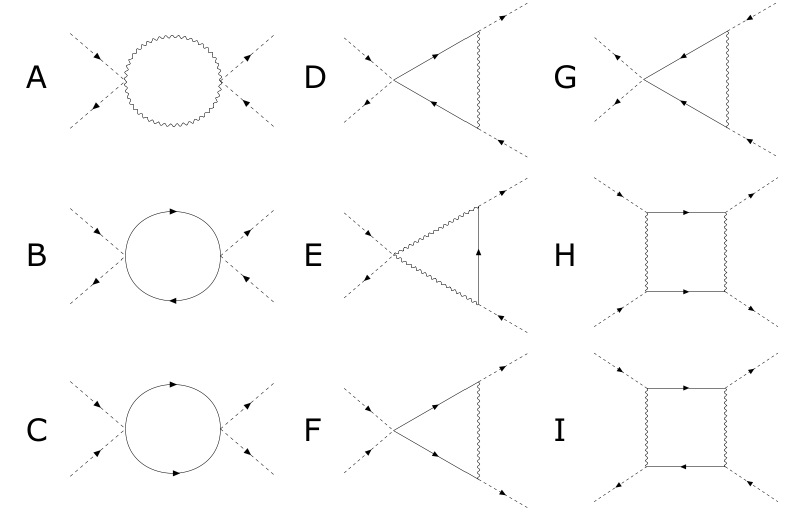}
    \caption{The diagrams contributing to the one-loop 1PI four-point correlation functions of charged fields with both cubic and quartic interaction terms. A solid line with an arrow indicates a dynamical field, while wavy lines are non-dynamical fields. Dashed lines indicate background fields. Diagrams C, F, G and H have Schr\"odinger propagators anti-aligned with respect to any choice of loop energy and hence give rise to poles in the upper half plane. The remaining diagrams lead to integrals of the form (\ref{eq: vanishing with additional E}) and diverge only when $m\ge n$. }
    \label{fig:diagrams}
\end{figure}

Now let us compute the four-point function. Note that as the interactions in \eqref{eq: B interactions 1} only include factors of external energies, so any diagram containing internal bosonic lines that fully align with the loop momentum vanish automatically. The different diagrams contributing to $\langle B^2 \bar{B}^2 \rangle^{(\mathrm{1PI})}$ at one-loop are shown in figure \ref{fig:diagrams}, with the diagrams A to C arising from $\langle S_{B,2}^2 \rangle$, diagrams D to G from $\langle S_{B,1}^2 S_{B,2} \rangle$, and diagrams H and I from $\langle S_{B,1}^4\rangle$.

Before continuing to the full calculation it is insightful to look at what kinds of divergences can arise from these diagrams. For simplicity we set all external momenta and energy to zero. Then it is not too hard   to see that the diagrams give integrals of the form, ignoring any tensor structures, 
\begin{subequations}
    \begin{align}
        A&\sim 	\int \frac{dEd^2p}{(2\pi)^3}\left(\frac{ 1}{p^2}\right)^2\ ,\\
        B&\sim \int \frac{dEd^2p}{(2\pi)^3}\left(G(E)\right)^2 \ ,\\
         C&\sim \int \frac{dEd^2p}{(2\pi)^3}G(E)G(-E)\ ,\\
          D&\sim\int \frac{dEd^2p}{(2\pi)^3} (E+C_Dp^2)(G(E))^2\left(\frac{1}{p^2}\right) \ , \\
           E&\sim\int \frac{dEd^2p}{(2\pi)^3}(E+C_Ep^2)G(E)\left(\frac{ 1}{p^2}\right)^2\ , \\
            F&\sim \int \frac{dEd^2p}{(2\pi)^3}(E+C_Fp^2)G(E)G(-E)\left(\frac{ 1}{p^2}\right)\ ,\\
             G&\sim\int \frac{dEd^2p}{(2\pi)^3}(E+C_Gp^2)G(E)G(-E)\left(\frac{ 1}{p^2}\right)\ ,\\
             H& \sim \int \frac{dEd^2p}{(2\pi)^3}(E^2+C_HEp^2 + C'_Hp^4)G(E)G(-E)\left(\frac{ 1}{p^2}\right)^2 ,\\
              I&\sim \int \frac{dEd^2p}{(2\pi)^3}(E^2+C_IEp^2 + C'_Ip^4)(G(E))^2\left(\frac{ 1}{p^2}\right)^2\ \ ,
    \end{align}
\end{subequations}
where $C_D,C_E,C_F,C_G,C_H,C_H',C_I,C_I'$ are tensor-valued constants. In particular the factors of $G(E), G(-E)$ and $1/p^2$ come from propagators whereas the powers of $E$ and $p^2$ come from interaction vertices. Note that the diagrams $B$ and $D$, as well as the terms involving $C_E,C_I,C_I'$, vanish as a consequence of (\ref{eq: multi G vanishing}).   Ultimately all the remaining  integrals reduce to just two types of divergence:
\begin{subequations}
    \begin{align}
    \int \frac{dEd^2p}{(2\pi)^3} \frac{ 1}{p^4} \qquad {\rm and}\qquad
         \frac{i}{ v}\int \frac{d^2p}{(2\pi)^2} \frac{ 1}{p^2}\ .
    \end{align}
\end{subequations}
In particular the first divergence gives a real contribution to $i\Gamma_{1}$ and hence would indicate an instability. However we find that all these divergences cancel, though this is not obvious at this stage. 

Let us now start the full calculation. Using the notation
\begin{equation}
    \Gamma_1^{(4)}[B] = \bigintssss d\mu_1 d\mu_2 d\mu_3 \, B_{i,1} B_{j,2} \bar{B}_{k,3} \bar{B}_{l,4} \, \cal{U}_{ijkl}(p_i) \ ,
\end{equation}
we find the contribution
\begin{align}
    \cal{U}_{ijkl}^{(1)} = 16i \bigintssss d\Tilde{\mu} \bigg( 
    \delta_{ij} \delta_{kl} G(p_1 + \Tilde{p}) G(p_2 - \Tilde{p}) - \frac{2 \delta_{ik} \delta_{jl}}{\Tilde{p}^2 (p_1 - p_3 + \Tilde{p})^2}
    \bigg) \ ,
\end{align}
to $\cal{U}$ from the first term of \eqref{eq: Gamma1 A4}. The second term gives
\begin{align} \nonumber
    \cal{U}_{ijkl}^{(2)} = - 8 i \bigintssss d\Tilde{\mu} \,\bigg(& 
    \delta_{kl} \frac{G(p_1 + \Tilde{p}) G(p_2 - \Tilde{p})}{\Tilde{p}^2} \Big( 
    2v (E_1 + E_2) \delta_{ij} - 2 p_1 \cdot p_2 \delta_{ij} \\ \nonumber
    &\qquad+ p_{1,i} p_{2,j} + p_{2,i} p_{1,j} - 2 (p_1 + 2 \Tilde{p})_i (p_2 - 2\Tilde{p})_j \Big) \\ \nonumber
    + &\delta_{ij} \frac{G(p_3 + \Tilde{p}) G(p_4 - \Tilde{p})}{\Tilde{p}^2} \Big( 
    2v (E_3 + E_4) \delta_{kl} - 2 p_3 \cdot p_4 \delta_{kl} \\
    &\qquad + p_{3,k} p_{4,l} + p_{3,l} p_{4,k} - 2 (p_3 + 2\Tilde{p})_k (p_4 - 2 \Tilde{p})_l \Big) \bigg) \ .
\end{align}
The terms from $\langle S_{B,1}^4 \rangle$ are somewhat less clean, and we shall only evaluate them at vanishing external energies and momenta as this is all one needs to identify divergences in this case. Using the identities
\begin{subequations}
\begin{align}
    2v \bigintssss \frac{d\Tilde{E}}{2\pi} \, \Tilde{E} G(\Tilde{p}) G(-\Tilde{p}) &= 0 \ , \\
    4v^2 \bigintssss \frac{d\Tilde{E}}{2\pi} \, \Tilde{E}^2 G(\Tilde{p}) G(-\Tilde{p}) &= \bigintssss \frac{d\Tilde{E}}{2\pi} \bigg( 1 + \Tilde{p}^4 G(\Tilde{p}) G(-\Tilde{p}) \bigg) \ , \\
    \bigintssss\frac{d^2 \Tilde{p}}{(2\pi)^2} \, \Tilde{p}_i \Tilde{p}_j f(\Tilde{p}^2) &= \inv{2} \delta_{ij} \bigintssss\frac{d^2 \Tilde{p}}{(2\pi)^2} \, \Tilde{p}^2 f(\Tilde{p}^2) \ , \\
    \bigintssss\frac{d^2 \Tilde{p}}{(2\pi)^2} \, \Tilde{p}_i \Tilde{p}_j \Tilde{p}_k \Tilde{p}_l f(\Tilde{p}^2) &= \inv{8} \brac{ \delta_{ij} \delta_{kl} + \delta_{ik} \delta_{jl} + \delta_{il} \delta_{jk} } \bigintssss\frac{d^2 \Tilde{p}}{(2\pi)^2} \, \Tilde{p}^4 f(\Tilde{p}^2) \ ,
\end{align}
\end{subequations}
for any function $f(\Tilde{p}^2)$, we find
\begin{align}
    \cal{U}^{(3)}_{ijkl}(0) = 16 i \bigintssss d\Tilde{\mu} \bigg( (3 \delta_{ij} \delta_{kl} + 4 \delta_{ik} \delta_{jl}) G(\Tilde{p}) G(-\Tilde{p}) + \frac{2 \delta_{ik} \delta_{jl}}{\Tilde{p}^4} \bigg) \ .
\end{align}
Summing the three contributions gives
\begin{align} \nonumber
    \cal{U}_{ijkl}(0) &= 64i \, \delta_{ik} \delta_{jl} \bigintssss d \Tilde{\mu} \, G(\Tilde{p}) G(- \Tilde{p}) \\ \label{eq: Bi div}
    &= \frac{16 \delta_{ik} \delta_{jl}}{v} \bigintssss \frac{d^2 \Tilde{p}}{(2\pi)^2} \, \inv{\Tilde{p}^2} \ ,
\end{align}
where we have integrated over loop energy to reach the final form. While the linear divergences in the loop energy cancel, we still find a logarithmic divergence arising from the integral over loop momentum. 

The index structure of \eqref{eq: Bi div} means the divergence cannot be absorbed by renormalizing any existing quartic gauge-field terms in the action \eqref{eq: sgym action}. We must therefore add additional marginal terms to the action to obtain a consistent action: a gauge-invariant term which would allow the divergence to be absorbed via counterterms is
\begin{equation}
    S_{DX} = - \alpha^2 \tr \bigintssss dt d^2 x \, D_i X D_i X D_j X D_j X \ ,
\end{equation}
though obviously we may find more by inserting additional factors of the dimensionless scalar $X$. Put another way, the action \eqref{eq: sgym action} is not renormalizable without the addition of further (classically) marginal terms; the RG flow one gets from once these are included will therefore fundamentally alter the structure of the theory. As we have used the gauge-invariance violating temporal point-splitting regularization scheme \eqref{eq: propagator time splitting} for the Schr\"odinger propagators, one may wonder to what extent the result we have obtained here is regulator-dependent. It would therefore be illuminating to compute the four-point function in a different scheme, which we leave to future work.

\subsection{Quantizing the Scalar Sector} \label{sect: scalar sector}

Let us now move on to the case of background scalar fields. As with the gauge sector, the interesting physics is contained within correlation functions of external fields charged under the remnant $U(1)$ gauge symmetry and so we shall work with $Y_{(\sigma)}^A=0$ and non-zero $\Omega^A$, where the background field $\Omega^A$ is defined in \eqref{eq: background field defs}. The terms coupling $\Omega^A$ to the fluctuating fields are, upon Fourier transforming,
\begin{align} \nonumber
    S_{\Omega,1} = \sqrt{2} \bigintsss d\mu_1 d\mu_2 \Bigg[& 
    \Omega^A_1 \bigg( 
    (E_1 + E_3) \phi_2^{(\sigma)} \bar{\omega}^A_3 - (E_1 - E_2) y^A_{(\sigma),2} \bar{\varphi}_3 \\ \nonumber
    &+ (p_1 + p_3)^i a_{i,2}^{(\sigma)} \bar{\omega}^A_3 - (p_1 - p_2)^i y^A_{(\sigma),2} \bar{b}_{i,3} - v \Big(y^A_{(\sigma),2} \bar{b}_{0,3} \\ \nonumber
    &- 2 a^{(\sigma)}_{0,2} \bar{\omega}^A_3\Big) - (\Gamma_{0A})_{\alpha\dalpha} \Big( \gamma^{\alpha}_{+,2} \brho_{-,3}^{\dalpha} - \brho_{+,3}^{\alpha} \gamma_{-,2}^{\dalpha} \Big)
    \bigg) \\ \nonumber
    & + \bar{\Omega}^A_3 \bigg(
    (E_1 + E_3) \omega^A_1 \phi^{(\sigma)}_2 - (E_2 + E_3 ) \varphi_1 y^A_{(\sigma),2} \\ \nonumber
    &+ (p_1 + p_3)^i \omega^A_1 a^{(\sigma)}_{i,2} - (p_2 + p_3)^i b_{i,1} y^A_{(\sigma),2} - v\Big( b_{0,1} y^A_{(\sigma),2} \\ 
    &- 2 \omega^A_1 a^{(\sigma)}_{0,2} \Big) - (\Gamma_{0A})_{\alpha\dalpha} \Big( \rho^{\alpha}_{+,1} \gamma^{\dalpha}_{-,2} - \gamma^{\alpha}_{+,2} \rho_{-,1}^{\dalpha} \Big)
    \bigg)
    \Bigg] \ ,
\end{align}
and
\begin{align} \nonumber
    S_{\Omega,2} = \bigintsss d\mu_1 d\mu_2 d\mu_3 \Bigg[ &
    \Omega^A_1 \Omega^B_2 \bigg( 
    \delta_{AB} \Big( \bar{b}_{i,3} \bar{b}_{i,4} + \bar{\omega}^C_3 \bar{\omega}^C_4 - 2 \bar{b}_{0,3} \bar{\varphi}_4 \Big) - \bar{\omega}_3^A \bar{\omega}^B_4 \bigg)
    \\ \nonumber
    &+ \bar{\Omega}^A_3 \bar{\Omega}^B_4 \bigg(
    \delta_{AB} \Big( b_{i,1} b_{i,2} + \omega^C_1 \omega^C_2 - 2 b_{0,1} \varphi_2 \Big) - \omega^A_1 \omega^B_2 \bigg)
    \\ \nonumber
    &+ 2 \Omega^A_{1} \bar{\Omega}^B_3 \bigg( 
    \delta_{AB} \Big( 2 a_{0,2}^{(\sigma)} \phi_{(\sigma),-4} +  b_{0,2} \bar{\varphi}_4 + \bar{b}_{0,4} \varphi_2 \\ \nonumber
    &- a_{i,2}^{(\sigma)} a_{i,-4}^{(\sigma)}- b_{i,2} \bar{b}_{i,4} - y^C_{(\sigma),2} y^C_{(\sigma),-4} - \omega^C_2 \bar{\omega}^C_4 \Big) \\
    &+  2\omega^A_2 \bar{\omega}^B_4  - \omega^B_2 \bar{\omega}^A_{4}+  y^A_{(\sigma),2} y^B_{(\sigma),-4} \bigg) \Bigg] \ ,
\end{align}
where we are using the same conventions as in the gauge sector. The one-loop effective action for $\Omega^A$ is determined in the same way as for $B_i$, using \eqref{eq: Gamma one-loop B} with the substitution $B_i \to \Omega^A$.

Using \eqref{eq: single G vanishing} we see that a single insertion of $S_{\Omega,2}$ gives
\begin{equation} \label{eq: S-omega2 correlation}
    \langle S_{\Omega,2} \rangle = 18 i \bigintssss  d\mu \, \Omega^A(E,p) \bar{\Omega}(E,p) \bigintssss d\Tilde{\mu} \, \inv{\Tilde{p}^2} \ .
\end{equation}
The other term in $\Gamma_1^{(2)}[\Omega]$ is
\begin{equation}
    \langle S_{\Omega,1}^2 \rangle = 36 i \bigintssss d\mu \, \Omega^A(E,p) \bar{\Omega}^A(E,p) \bigintssss d\Tilde{\mu} \, \frac{2v \Tilde{E} \, G(\Tilde{p} + p)}{\Tilde{p}^2} \ ,
\end{equation}
so using \eqref{eq: S_1^2 identity} this is
\begin{equation}
    \langle S_{\Omega,1}^2 \rangle = - 36 \bigintssss d\mu \, \Omega^A(E,p) \bar{\Omega}^A(E,p) \bigintssss d\Tilde{\mu} \, \inv{\Tilde{p}^2} \ ,
\end{equation}
and so upon combining with \eqref{eq: S-omega2 correlation} we find
\begin{equation}
    \Gamma_1^{(2)}[\Omega] = 0 \ .
\end{equation}
The more complicated calculation is that of $\Gamma_1^{(4)}[\Omega]$. Writing it as
\begin{equation}
    \Gamma_1^{(4)}[\Omega] = \int d\mu_1 d\mu_2 d\mu_3 \, \Omega_1^A \Omega^B_2 \bar{\Omega}^C_3 \bar{\Omega}^D_4 \, \cal{V}_{ABCD}(p_i) \ ,
\end{equation}
the non-vanishing contributions from the $S_{\Omega,2}^2$ term in $\Gamma_1^{(4)}$ are
\begin{align} \nonumber
    \cal{V}_{ABCD}^{(1)} = 2i \bigintssss d\Tilde{\mu} \bigg[&  G(\Tilde{p} + p_1) G(p_2 - \Tilde{p}) \Big( 8 \delta_{AB} \delta_{CD} + \delta_{AC} \delta_{BD} \Big) \\ \label{eq: V1}
    &- \frac{1}{\Tilde{p}^2 (p_1 - p_3 + \Tilde{p})^2} \Big( 16 \delta_{AC} \delta_{BD} + \delta_{AD} \delta_{BC} + \delta_{AB} \delta_{CD} \Big) \bigg] \ ,
\end{align}
where we have used the contraction of $\cal{V}_{ABCD}$ with the factors of $\Omega$ in $\Gamma_1^{(4)}$ to simplify the result. Similarly, the contribution from the $S_{\Omega,2} S_{\Omega,1}^2$ term is
\begin{align} \nonumber
    \cal{V}^{(2)}_{ABCD} = -  2i\bigintsss d\Tilde{\mu} \, \Bigg[& 
    \frac{G(p_1 + \Tilde{p}) G(p_2 - \Tilde{p})}{\Tilde{p}^2}  \Big(  \delta_{AB} \delta_{CD} \big( 3v (E_1 + E_2) \\ \nonumber
    &- (2 p_1 + \Tilde{p})\cdot (2p_2 - \Tilde{p}) - (p_1 - \Tilde{p}) \cdot (p_2 + \Tilde{p}) \big) \\ \nonumber
    &- \delta_{AC} \delta_{BD} \big( 4v (E_1 + E_2) - (2 p_1 + \Tilde{p}) \cdot (2p_2 - \Tilde{p}) \big) \Big)
    \\ \nonumber
    &+ \frac{G(p_3 + \Tilde{p}) G(p_4 - \Tilde{p})} {\Tilde{p}^2}  \Big(  \delta_{AB} \delta_{CD} \big( 3v (E_3 + E_4) \\ \nonumber
    &- (2 p_3 + \Tilde{p})\cdot (2p_4 - \Tilde{p}) - (p_3 - \Tilde{p}) \cdot (p_4 + \Tilde{p}) \big) \\ \nonumber
    &- \delta_{AC} \delta_{BD} \big( 4v (E_3 + E_4) - (2 p_3 + \Tilde{p}) \cdot (2p_4 - \Tilde{p}) \big) \Big) \\ \label{eq: V2}
    &- \frac{2}{\Tilde{p}^2 (p_1 - p_3 + \Tilde{p})^2} \Big(\delta_{AB} \delta_{CD} + 2 \delta_{AC} \delta_{BD} + \delta_{AD} \delta_{BC} \Big) \Bigg] \ .
\end{align}
Finally, we have the contribution from $S_{\Omega,1}^4$. Using the substitutions
\begin{subequations} 
\begin{align} \nonumber 
    2v \bigintssss \frac{d\Tilde{E}}{2\pi} \,\Tilde{E} \,G(p_1 + \Tilde{p}) G(p_2 - \Tilde{p}) &=  - \inv{2} \bigg( 2v(E_1 - E_2) - 2 \Tilde{p}\cdot(p_1 + p_2) + p_2^2 - p_1^2 \bigg) \\
    &\hspace{0.8cm} \times\bigintssss \frac{d\Tilde{E}}{2\pi} \,G(p_1 + \Tilde{p}) G(p_2 - \Tilde{p}) \ , \\ \nonumber
    4v^2\bigintssss \frac{d\Tilde{E}}{2\pi} \,\Tilde{E}^2 G(p_1 + \Tilde{p}) G(p_2 - \Tilde{p}) &= \bigintssss \frac{d\Tilde{E}}{2\pi} \bigg(1 + \inv{2} \bigg((2v E_1 - (p_1 + \Tilde{p})^2)^2 + (2v E_2 \\
    &\hspace{0.5cm} - (p_2 - \Tilde{p})^2 )^2 \bigg) G(p_1 + \Tilde{p}) G(p_2 - \Tilde{p}) \bigg) \ ,
\end{align}
\end{subequations}
within the $\Tilde{E}$ integral, we obtain
\begin{align} \nonumber
    \cal{V}^{(3)}_{ABCD} = 2i \bigintsss d\Tilde{\mu} \Bigg[& 
    \frac{G(p_1 + \Tilde{p}) G(p_2 - \Tilde{p})}{\Tilde{p}^2 (p_1 - p_3 + \Tilde{p})^2} \bigg( 
    4 \Big( \delta_{AB} \delta_{CD} - \delta_{AC} \delta_{BD} + \delta_{AD} \delta_{BC} \Big) \\ \nonumber
    &\times\Big( 
    (2v E_1 - (p_1 + \Tilde{p})^2)^2 + (2v E_2 - (p_2 - \Tilde{p})^2 )^2  - \big( 2v( E_1 - E_2) \\ \nonumber
    &- 2 \Tilde{p} \cdot (p_1 + p_2) + p_2^2 - p_1^2 \big) \big( 2v(E_1 - E_3) + 2 \Tilde{p} \cdot (p_1 + p_2) \\ \nonumber
    &+ (p_1 + p_2) \cdot (p_1 - p_3) \big) + 4 \big( 
    v (E_1 + E_2) \,\Tilde{p}\cdot (p_1 - p_3 + \Tilde{p}) \\ \nonumber
    &+ v (E_1 - E_3)\, \Tilde{p}\cdot (p_1 + p_2) - p_2 \cdot p_3 \, \Tilde{p} \cdot (\Tilde{p} + p_1) \\ \nonumber
    &+ \Tilde{p}\cdot p_3 \,(p_1 + \Tilde{p})\cdot p_2 + (p_1 + \Tilde{p})\cdot (p_1 - p_3 + \Tilde{p}) \,\Tilde{p} \cdot (p_2 - \Tilde{p})
    \big) \Big) \\ \nonumber
    &- \inv{2} \delta_{AB} \delta_{CD} \Big( 
    (2v E_1 - (p_1 + \Tilde{p})^2)^2 + (2v E_2 - (p_2 - \Tilde{p})^2 )^2 \\ \nonumber
    &+ \big( 2v( E_1 - E_2) - 2 \Tilde{p} \cdot (p_1 + p_2) + p_2^2 - p_1^2 \big) \big( 2v (E_3 - E_4) \\ \nonumber
    &- ( p_1 - \Tilde{p}) \cdot (2 p_3 - p_1 - \Tilde{p}) + (p_2 + \Tilde{p}) \cdot (2 p_4 - p_2 + \Tilde{p}) \big) \\ \nonumber
    &- 2 \big( 2v E_3 - (p_1 - \Tilde{p}) \cdot (2p_3 - p_1 - \Tilde{p}) \big) \big( 2v E_4 \\ \nonumber
    &- (p_2 + \Tilde{p}) \cdot (2p_4 - p_2 + \Tilde{p}) \big)
    \Big) + \delta_{AC} \delta_{BD} \Big( 4v (E_1 + E_2) \\ \nonumber
    &- (2p_1 + \Tilde{p}) \cdot (2p_2 - \Tilde{p}) \Big) \Big( 4v (E_1 + E_2) - (p_1 + p_3 + \Tilde{p}) \cdot (p_2 \\ \label{eq: V3}
    &+ p_4 - \Tilde{p}) \Big) \bigg)
    - \frac{\delta_{AB} \delta_{CD} + \delta_{AD}\delta_{BC} - 12 \delta_{AC} \delta_{BD}}{\Tilde{p}^2 (p_1 - p_3 + \Tilde{p})^2}
    \Bigg] \ .
\end{align}
By summing \eqref{eq: V1}, \eqref{eq: V2}, and \eqref{eq: V3} we see that the power-law divergences cancel. However, we still have to worry about logarithmic divergences arising from the loop momentum integral; setting the external energies and momenta to zero, we find
\begin{align} \nonumber
    \cal{V}_{ABCD}(0) &= - 8 i (\delta_{AB} \delta_{CD} - \delta_{AC} \delta_{BD} ) \bigintssss d\Tilde{\mu} \, G(p) G(-p) \\ \label{eq: divergent result}
    &= - \frac{2(\delta_{AB} \delta_{CD} - \delta_{AC} \delta_{BD})}{v} \bigintssss \frac{d^2 \Tilde{p}}{(2\pi)^2} \, \inv{\Tilde{p}^2} \ ,
\end{align}
and so as with the gauge field's four-point function we have a log divergence at one-loop. In contrast to the four-point function of $B_i$, the index structure of $\cal{V}_{ABCD}$ is precisely what one finds from expansion of the commutator terms involving $Y^A$ in the initial action. This means that, unlike in the previous case, the divergences here are proportional to terms already included in the theory.

To deal with the divergence we must renormalise the $\Omega^4$ term in the action. As the divergent terms appear in the initial action \eqref{eq: sgym action}, one may hope that the divergence can be resolved using counterterms for terms already within the theory. To see why this cannot be the case, we observe that the terms we are interested in come from the square of the commutator $[Y^A, Y^B]$. This means their coefficient is proportional to the analogous terms involving two copies of the neutral component of the scalars along with a complex pair. However, it is straightforward to see that these cannot receive any quantum corrections; any diagram with these external legs cannot posses a Schr\"odinger propagator oriented against the direction of loop momentum at one-loop, and so we will only find power-law divergences that we expect to cancel due to supersymmetry. Thus we cannot cancel \eqref{eq: divergent result} against existing terms in the action while preserving $\frak{su}(2)$-invariance, and must add new terms to achieve this. The conclusions reached for the gauge field hold in this case also, and the theory is non-renormalizable without the addition of these terms. To generate the tensorial structure of \eqref{eq: divergent result} we can consider adding commutators of $X$ into the existing commutator terms involving $Y^A$, which preserves gauge-invariance. The original action's interaction terms will then acquire a coupling to $\phi^{(\sigma)}$, with a similar structure observed for Galilean electrodynamics in \cite{Chapman:2020vtn, Baiguera:2022cbp}.

We can compute the integral \eqref{eq: divergent result} using dimensional regularisation in the spatial dimensions. As usual, this requires us to introduce a renormalisation scale $\mu$ and allow the coupling to run in order for the final result to be independent of $\mu$. However, as we discussed above we cannot allow $g(\mu)$ to run. Resolving this fully requires deforming the action in a way consistent with supersymmetry and the non-Abelian gauge group, and then promoting the coupling of the deformation to be $\mu$-dependent; here we will work directly with the $U(1)$ effective field theory (EFT) and neglect the non-Abelian supersymmetric origin of each term, allowing us to alter the coefficients of each term by hand and introduce a running coupling $\Tilde{g}(\mu)$ to parametrise the complex scalar's quartic potential within the EFT. We take this to agree with the original coupling $g$ at tree-level (meaning our previous one-loop computation is still valid). Introducing an IR regulator $\Delta$, this is
\begin{align} \nonumber
    \cal{V}_{ABCD}(0) &= - \frac{2\brac{ \delta_{AB} \delta_{CD} - \delta_{AC} \delta_{BD} }}{v}  \mu^{\varepsilon} \bigintssss \frac{d^{2 - \varepsilon} \Tilde{p}}{(2\pi)^{2-\varepsilon}} \, \inv{\Tilde{p}^2 + \Delta^2} \\
    &= - \frac{2 \brac{ \delta_{AB} \delta_{CD} - \delta_{AC} \delta_{BD} }}{v} \brac{ \inv{\pi \varepsilon} + \inv{2\pi} \brac{ \ln \brac{ \frac{\pi \mu^2}{\Delta^2} } - \gamma } + O(\varepsilon) } \ .
\end{align}
To deal with the divergence we add the counterterms
\begin{align}
    \Gamma^{(4)}_{1,\mathrm{ct}}[\Omega] = \alpha \brac{ \delta_{AB} \delta_{CD} - \delta_{AC} \delta_{BD} } \bigintssss dt d^2 x \, \Omega^A \Omega^B \bar{\Omega}^C \bar{\Omega}^D \ .
\end{align}
We use the MS-bar renormalisation scheme 
\begin{align}
    \alpha = \frac{2}{v} \brac{\inv{\pi \varepsilon} + \frac{\ln \pi - \gamma}{2\pi}}
\end{align}
to remove the pole and additional constants, allowing us to take the $\varepsilon\to 0$ limit. This gives the one-loop momentum-independent quartic scalar terms
\begin{subequations}
\begin{align} \nonumber
    \Gamma^{(4)}[\Omega] &= \Gamma_0^{(4)}[\Omega] + \Gamma_1^{(4)}[\Omega] + \Gamma^{(4)}_{1,\mathrm{ct}}[\Omega] \\
    &= \bigintssss d\mu_1 d\mu_2 d\mu_3 \, \Omega^A_1 \Omega^B_2 \bar{\Omega}^C_3 \bar{\Omega}^D_4 \, \brac{ \delta_{AB} \delta_{CD} - \delta_{AC} \delta_{BD} }  \Tilde{\cal{V}} \ , \\
    \Tilde{\cal{V}} &= \inv{\Tilde{g}^2(\mu)} - \frac{1}{\pi v} \ln \brac{\frac{\mu^2}{\Delta^2}} + \dots  \ ,
\end{align}
\end{subequations}
where the ellipsis represents the $\mu$-independent momentum terms one would find in the full computation. In order for correlation functions to be independent of $\mu$ we must have
\begin{equation}
    0 = \mu \frac{d}{d \mu} \brac{ \inv{\Tilde{g}^2(\mu)} - \frac{1}{\pi v} \ln \brac{\frac{\mu^2}{\Delta^2}} } \ ;
\end{equation}
defining the rescaling-invariant coupling $\Tilde{e}(\mu)$ for $\Tilde{g}(\mu)$ as in \eqref{eq: true coupling}, we find the one-loop beta-function
\begin{equation}
    \mu \frac{d \Tilde{e}(\mu)}{d\mu} = - \frac{\Tilde{e}^3(\mu)}{\pi} \ .
\end{equation}
The coupling is asymptotically free, tending to zero at small length-scales, and the theory seems to be a good starting for a non-relativistic UV-fixed point. However, requiring that the modification to the complex scalar's quartic terms comes from a supersymmetric non-Abelian action means there will be a plethora of additional terms we must also include, and we must check whether any new running couplings these terms define are also well-behaved in the UV before such a claim can be properly made (though one may hope this is guaranteed by supersymmetry).

\section{Marginal Deformations and Supersymmetry} \label{sect: deformation}

As the field $X$ does not scale under the Lifshitz scaling symmetry there are an infinite number of classically marginal terms that one could add to the action \eqref{eq: sgym action}. While our one-loop four-point function calculations show that higher-order terms in $X$ must be added to the theory, one would need to compute a large number of higher-point functions to determine exactly what must be added. We will therefore proceed in a different way and ask whether we can add terms in such a way that the one-loop corrections are UV-finite. In addition to the infinite number of operators involving higher powers of $X$ one may turn on, in three spacetime dimensions we could also consider adding in Chern-Simons terms for the gauge field. To simplify the discussion here we shall only consider deformations consistent with T-duality and gauge invariance, allowing us to view the theory as a reduction from nine-dimensions. Given the difference in the structure of the logarithmic divergence for the $B_i$ four-point function in \eqref{eq: Bi div} and the scalar four-point function in \eqref{eq: divergent result}, one may suspect that this approach is not sufficient to accommodate the full set of counterterms required for quantization of the initial action; however, in the absence of a concrete proposal for a three-dimensional action of interest we shall take this as a starting point. In particular, we do not propose a fully supersymmetric set of deformations.

To diagnose the effects of adding higher-order terms we shall focus on the four-point function of $\Omega^A$. Schematically, the possible interaction terms at one-loop using the background field method take the form (where we suppress momentum indices for now)
\begin{subequations}
\begin{align}
    S_{\Omega,1} &= \bigintssss d\mu_1 d\mu_2 \, \Omega^A \Big( \bar{\cal{A}} y^A_{(\sigma)}  + \cal{B} \, \bar{\omega}^A + \cal{C} \gamma^T \Gamma_{0A} \bar{\rho}  \Big) + \mathrm{h.c.} \ , \\ \nonumber
    S_{\Omega,2} &= \bigintssss d\mu_1 d\mu_2 d\mu_3 \, \bigg[ \Omega^A \Omega^B \Big( \cal{D} \, \bar{\omega}^A \bar{\omega}^B + \cal{E} \delta_{AB}  + \cal{F} \bar{\rho}_-^T \Gamma_{AB} \bar{\rho}_-  \Big) + \mathrm{h.c.} \\
    &\hspace{3.3cm}+ \Omega^A \bar{\Omega}^B \cal{G}_{AB} \bigg] \ .
\end{align}
\end{subequations}
The only assumption we have made to reach this form is that none of the coefficients have negative scaling dimension (which is guaranteed by locality). While we will also find that terms cubic and quartic in the background fields are added by the deformations, these cannot contribute to the 1PI four-point function's log divergence on kinematic grounds. Similarly, if we focus our attention on the log divergence of the $\brac{\Omega^A \bar{\Omega}^A}^2$ term in $\Gamma_1^{(4)}$ it is straightforward to see that the only terms that contribute are $\cal{A}, \cal{B},$ and $\cal{D}$. While this greatly simplifies the form of the interaction terms needed to compute the bosonic contribution, it crucially means that this divergence is independent of the theory's fermionic terms. We can therefore use it to probe the structure of bosonic deformations of SGYM while remaining agnostic as to their supersymmetric completion, if one exists.

The bosonic extension that we shall be interested in here is
\begin{align} \nonumber
    S_B = \inv{2g^2} \tr \bigintssss dt d^8 x \bigg(& (D_{0}X)^2 - 2 D_i X F_{0i} - \inv{2} F_{ij} F_{ij} - \kappa \Big( [X , D_0 X]^2 \\ \nonumber
    &- 2 [X , F_{0i}] [X , D_i X] - \inv{2} [X , F_{ij}]^2 \Big) \\ \label{eq: bosonic deformation 1}
    &+ i \lambda F_{ij} [D_i X , D_j X] + \frac{\lambda^2}{2} [D_i X , D_j X]^2 \bigg) \ .
\end{align}
When $\kappa$ is positive this is manifestly positive-definite for any arbitrary coupling $\lambda$, and so seems a sensible theory to consider. Reducing this to three dimensions gives
\begin{align} \nonumber
    S_B = \inv{2g^2} \tr \bigintssss dt d^2 x \bigg(& (D_{0}X)^2 - 2 D_i X F_{0i} - \inv{2} F_{ij} F_{ij} - 2i [X , Y^A] D_0 Y^A \\ \nonumber
    &- D_i Y^A D_i Y^A + \inv{2} [Y^A , Y^B]^2 - \kappa \Big( [X , D_0 X]^2 \\ \nonumber
    &- 2 [X , F_{0i}] [X , D_i X] - \inv{2} [X , F_{ij}]^2 - [X , D_i Y^A]^2 \\ \nonumber
    &- 2 i [X , [X , Y^A]][X , D_0 Y^A] + \inv{2} [X , [Y^A , Y^B]]^2 \Big)  \\ \nonumber
    & + i \lambda \Big( F_{ij} [D_i X , D_j X] + 2 i D_i Y^A [D_i X , [ X , Y^A]] \\ \nonumber
    & + i [Y^A , Y^B][[X , Y^A] , [X , Y^B]] \Big)+ \frac{\lambda^2}{2} \Big( [D_i X , D_j X]^2 \\ 
    &- 2 [[X , Y^A ], D_i X]^2 + [[X , Y^A], [X , Y^B]]^2 \Big) \bigg) \ .
\end{align}
In general, quantising this action is a tall order. However, we can simplify matters tremendously by focusing our attention on the projection of the scalar four-point function discussed above (in other words, only keeping $\cal{A}$, $\cal{B}$, and $\cal{D}$) and working with $\lambda = \kappa$. The details of this are given in appendix \ref{sect: deformed quant}, with the result being the logarithmic divergence 
\begin{equation} \label{eq: modified divergence 1}
    \cal{V} = \frac{2}{v} \brac{1 - \frac{2 v^2 \lambda}{1 + v^2 \lambda}}^2 \bigintssss \frac{d^2 \Tilde{p}}{(2\pi)^2} \, \inv{\Tilde{p}^2} \ ,
\end{equation}
in place of \eqref{eq: divergent result}. This vanishes for
\begin{equation}
    \frac{2 v^2 \lambda}{1 + v^2 \lambda} = 1 \ ,
\end{equation}
meaning
\begin{equation}
    \lambda = v^{-2} \ .
\end{equation}
This is a fairly striking result, and clearly warrants further investigation. The other correlation functions receive contributions from the theory's fermions; in the undeformed theory cancellation of the non-logarithmic divergences required supersymmetry, and it seems reasonable to believe that the same will hold true here. It is therefore important to ask whether the action \eqref{eq: bosonic deformation 1} for $\lambda = \kappa$ admits a supersymmetric extension. While we shall not resolve this question, we will find an interesting structure emerges by requiring consistency of the theory.

A convenient way to approach the construction of the fermionic action is to note that the $O(\lambda)$ terms in \eqref{eq: bosonic deformation 1} can be constructed from the undeformed action by making the field redefinitions
\begin{subequations}
\begin{align}
    \hat{A}_0 &= A_0 - \frac{i \lambda}{2} [X , D_0 X] \ , \\
    \hat{A}_i &= A_i - \frac{i \lambda}{2} [X , D_i X] \ ,
\end{align}
\end{subequations}
where the hatted variables are those of the undeformed theory. Hence, by making the corresponding fermionic redefinitions
\begin{subequations}
\begin{align}
    \hat{\psi}_- &= \psi_- + \frac{\lambda}{2} [X , [X , \psi_-]] \ , \\ \label{eq: field redef psi plus}
    \hat{\psi}_+ &= \psi_+ + \frac{\lambda}{2} [X , [X , \psi_+]] \ ,
\end{align}
\end{subequations}
we find that invariance of the undeformed action under the supersymmetry transformations \eqref{eq: 8d SGYM susy} implies the fermionic action
\begin{align} \nonumber
    S_{F} = \inv{2g^2} \tr \bigintssss dt d^8 x \bigg(& 
    - \sqrt{2} i \psi_-^T D_0 \psi_- - 2i \psi_-^T \Gamma_{0i} D_i \psi_+ - \sqrt{2} \psi_+^T [X , \psi_+] \\ \nonumber
    &+ \lambda \Big(
    \sqrt{2} i [X , \psi_-^T] [X , D_0 \psi_-] - \sqrt{2} i [X , \psi_-^T] [D_0 X , \psi_-] \\ \nonumber
    &+ i [X , \psi_-^T] \Gamma_{0i} [X , D_i \psi_+] + i [X , \psi_+^T] \Gamma_{0i} [X , D_i \psi_-] \\ \nonumber
    &- i [X , \psi_+^T] \Gamma_{0i} [D_i X , \psi_-] - i [X , \psi_-^T] \Gamma_{0i} [D_i X , \psi_+] \\ 
    & + \sqrt{2} [X , \psi_+^T] [X , [X ,\psi_+] ]]
    \Big) \bigg) \ ,
\end{align}
cancels the $O(\lambda)$ terms in the supersymmetry transformation of \eqref{eq: bosonic deformation 1}. However, as we have redefined our fields the supersymmetry transformations are altered; focusing on the $\epsilon_+$ supercharge, we find
\begin{subequations} \label{eq: deformed susy 1}
\begin{align}
    \delta A_0 &= - \sqrt{2} i \epsilon_+^T \psi_+ \ , \\
    \delta A_i &= i \epsilon_+^T \Gamma_{0i} \psi_- \ , \\ 
    \delta \psi_+^T &= \epsilon_+^T \brac{D_0 X + \inv{2} \Gamma_{ij} F_{ij}  - \frac{i \lambda}{2} \Gamma_{ij} [D_i X , D_j X] }\ , \\
    \delta \psi_-^T &= - \sqrt{2} \epsilon_+^T \Gamma_{0i} D_i X \ .
\end{align}
\end{subequations}
It does not appear to be possible to find an extension that does not require altering the supersymmetry transformations, though it would be worthwhile to further explore if this can be done.

A straightforward check as to whether such an approach is consistent is to ask whether the fermionic constraint imposed by demanding the on-shell condition
\begin{equation}
    [\delta_{+,1} , \delta_{+,2}] \psi_+ = 0 \ ,
\end{equation}
as well as the corresponding bosonic constraint obtained by acting with supersymmetry, match the equations of motion of $\psi_+$ and $A_0$. As the modified supersymmetry transformations introduce new terms in $\delta \psi_+$, a similar computation to that performed in section \ref{sect: GYM review} fixes the $A_0$ and $\psi_+$ constraints to be
\begin{subequations}
\begin{align} \nonumber
    0 &= i [X, D_0 X] + D_i D_i X + \lambda [D_i X , [X , D_i X]] \\ \label{eq: deformed Gauss law}
    &\qquad - \inv{\sqrt{2}} \{ \psi_- , \psi_- \} + \frac{\lambda}{\sqrt{2}}  \{ [X , \psi_-] , [X , \psi_-] \} \ , \\ \nonumber
    0 &= i \Gamma_{0i} D_i \psi_- + \sqrt{2} [ X, \psi_+ ] + i \lambda \Gamma_{0i} [D_i X , [X , \psi_- ] ] \ ,
\end{align}
\end{subequations}
where we use the same notation as in section \ref{sect: GYM review}. The deformed theory in its current form does not reproduce these as the equations of motion for $A_0$ and $\psi_+$ and therefore cannot be made supersymmetric. This is remedied through the inclusion of
\begin{align} \nonumber
    S' =  \frac{ \lambda^2}{2g^2} \tr &\bigintssss dt d^8x \bigg(2 i [X , D_0 X ] [D_i X , [X , D_i X]] + 2 D_j D_j X [D_i X , [X , D_i X]]  \\ \label{eq: S prime}
    &-2i [X , [X , \psi_+^T]] \Gamma_{0i}  [D_i X , [X , \psi_-]] - \sqrt{2} i [X, \psi_-^T] [[X, D_0 X], [X , \psi_-]] \bigg) \ ;
\end{align}
the second term, while not strictly necessary to match the constraints with the equations of motion, appears to be required if one searches for a supersymmetric completion of the remaining terms at the one-fermion level using \eqref{eq: deformed susy 1} and so we include it in our discussion here. Using the notation $\cal{E}_A$ and $\cal{E}_{\psi}$ for the constraints derived from closure of the supersymmetry algebra, a quick calculation shows that the $A_0$ and $\psi_+$ equations of motion are now
\begin{subequations}
\begin{align}
    0 &= \cal{E}_A + \lambda [X , [X , \cal{E}_A]] \ , \\
    0 &= \cal{E}_{\psi} + \lambda [X , [X , \cal{E}_{\psi}]] \ ,
\end{align}
\end{subequations}
and so as hoped the two approaches match. One may also consider adding higher-order terms in $\lambda$ to \eqref{eq: deformed susy 1}, which would require the introduction of further terms to match the new constraint equations with the equations of motion.

The conserved energy associated with the bosonic action given by the sum of \eqref{eq: bosonic deformation 1} and first two terms in \eqref{eq: S prime} is
\begin{align} \nonumber
    E_B = \inv{2g^2} \tr \bigintssss d^8 x \bigg(& (D_0 X)^2 + \inv{2} \big(F_{ij} - i \lambda [D_i X , D_j X]\big)^2 - \lambda [X , D_0 X]^2 \\
    &- \frac{\lambda}{2} [X , F_{ij}]^2 - 2\lambda^2 D_j D_j X [D_i X , [X , D_i X]] \bigg) \ ,
\end{align}
which is not manifestly positive-definite. However, with $\psi_- = 0$ the Gauss law constraint \eqref{eq: deformed Gauss law} implies
\begin{equation}
    - \tr [X , D_0 X]^2 = \tr \big( D_i D_i X + \lambda [D_i X , [X , D_i X]] \big)^2 \ ,
\end{equation}
meaning the energy evaluated on the constraint becomes
\begin{align} \nonumber
    E_B' = \inv{2g^2} \tr \bigintssss d^8 x \bigg[& 
    (D_0 X)^2 + \lambda (D_i D_i X)^2 + \inv{2} \Big(F_{ij} - i \lambda [D_i X, D_j X]\Big)^2 \\ \label{eq: energy on constraint}
    &- \frac{\lambda}{2} [X , F_{ij}]^2 + \lambda^3 [D_i X , [X , D_i X]]^2 \bigg] \ ;
\end{align}
the indefinite term cancels, leaving behind a sum of positive-definite terms. The additional terms in the bosonic action will further alter the UV behaviour of the $(\Omega^A \bar{\Omega}^A)^2$ correlation function; in appendix \ref{sect: deformed quant} we show that the effect is to change \eqref{eq: modified divergence 1} to
\begin{equation} \label{eq: modified V}
    \Tilde{\cal{V}} = \frac{2}{v} \brac{1 - \frac{2v^2 \lambda}{1 + v^2 \lambda} + \frac{4 v^6 \lambda^3}{(1 + v^2 \lambda)^2}}^2 \bigintssss \frac{d^2 \Tilde{p}}{(2\pi)^2} \, \inv{\Tilde{p}^2} \ .
\end{equation}
which has no roots for positive $\lambda$. However, importantly the terms arrange themselves into a squared quantity, maintaining the sign of the beta-function discussed in section \ref{sect: scalar sector}.

It is amusing to note that the additional terms which prevent a conformal fixed-point at one-loop can be remedied through the inclusion of the three-dimensional potential terms
\begin{equation} \label{eq: new potential terms}
    S_{\mathrm{pot.}} = \inv{2g^2} \tr \bigintssss dt d^2 x \bigg( \lambda [Y^A , [X , Y^A]]^2 + \lambda^3 [[X , Y^A] , [X , [X , Y^A]]]^2 \bigg) \ ,
\end{equation}
which contributes to the quadratic terms in $\Omega$,
\begin{equation}
    \frac{8 v^4 \lambda^2}{(1 + v^2 \lambda)^2} \Rightarrow 4 v^2 \lambda \ .
\end{equation}
At $\lambda = v^{-2}$ this cancels the additional contribution from $\Tilde{\beta}$. The terms in \eqref{eq: new potential terms} are negative-definite, but appear in the energy with the exact coefficients to ensure their cancellation against the reduction of the second and final terms in \eqref{eq: energy on constraint} to three dimensions, preserving the positive-definiteness of the theory. While the $O(\lambda^3)$ term can be lifted to the nine-dimensional theory and given a supersymmetric extension at the one-fermion level, the same is not true for the additional $O(\lambda)$ term without adding a term quadratic in $D_i D_i X$.
Including this appears to fundamentally alter the theory, as it would change the propagator structure of $X$ and $A_0$ while leaving the others unchanged: it seems likely that such a term is therefore not consistent with supersymmetry. It is also important to note that attempting to cancel the final term in \eqref{eq: modified V} by altering the coefficients of terms already included will lead to an action that is unbounded from below. We observe that there is an interesting tension between supersymmetry, UV-finiteness, and bounded energies in theories obtained from the reduction of nine-dimensional QFTs, where demanding that any two are satisfied appears to disallow the third.

\section{Conclusion and Outlook} \label{sect: conclusion}

In this paper we have studied non-Abelian supersymmetric Galilean Yang-Mills. We discussed its perturbative dynamics, which are obtained by expanding about a point on the Coulomb branch in order for the dynamical fields to have propagating degrees of freedom, and the structure of its (super)symmetries before moving on to the computation one-loop effects for two, three, and four-point functions in the three-dimensional theory using the background field method. While there are non-trivial cancellations due to supersymmetry that prevent power-law divergences in integrals over loop energy and momentum, as with previously studied Abelian Galilean gauge-theories \cite{Chapman:2020vtn, Baiguera:2022cbp} we found logarithmic divergences in the 1PI four-point functions of charged Coulomb branch fields. These divergences cannot be cancelled using counterterms that modify the coefficients of terms in the bare action while preserving gauge-invariance and require us to introduce additional marginal deformations of the theory. We considered a particular class of deformations obtained from the nine-dimensional theory, where it was seen that supersymmetry, the requirement that the theory's energy is bounded from below, and UV-finiteness of a particular projection of the charged scalar four-point function appear to be in tension.

As was also reported in previous work on Abelian non-relativistic three-dimensional gauge theories, the quantum features that we have observed here are reminiscent of two-dimensional sigma-models where the scalar fields are also dimensionless, leading to a rich class of marginal deformations related to the geometry of a target space. In string theory, sigma-models with a fixed geometry define  consistent backgrounds only if conformal invariance is preserved, leading to the supergravity equations of motion on the metric and other fields.   It is natural to wonder if a similar approach could be considered for SGYM, where background field configurations for which  the loop divergences cancel play a preferred role within the string theoretic embedding of the QFT. 

To simplify the computations performed here we only studied the $\mathfrak{su}(2)$ SGYM theory. It is desirable to go beyond this and study the theory at arbitrary $N$, particularly if one is interested in potential large-$N$ holographic applications. It is possible that the divergences we have found are subleading in $N$ and hence are not visible at leading order in the supergravity dual which admits a scaling symmetry. It is also important to determine whether the structure of the theory found here extends to an arbitrary gauge group: while our experience with relativistic theories may make one assume this is the case, it is of course not guaranteed. The brute force approach taken in calculations here (namely, explicitly expanding the fields in components before taking contractions) would make calculations for a general gauge group incredibly convoluted, and it seems a worthwhile endeavour to find a more covariant way to describe the perturbative dynamics on the Coulomb branch to make such an extension tractable.

It would be interesting to apply the approach developed here to other non-Lorentzian non-Abelian gauge theories. An example of particular note is the theory obtained via the M2-brane limit of BLG/ABJM \cite{Lambert:2019nti}. This is a supersymmetric Chern-Simons matter theory which (like three-dimensional SGYM) is classically invariant under the Schr\"odinger algebra. All 24 generators of the original theory's superconformal transformations\footnote{Enhanced to 32 supercharges for the non-Lorentzian limit of BLG.} are also (classically) preserved by the limit, and it would be fascinating to see whether this holds once quantum corrections are included. Additionally, it is believed that relativistic three-dimensional $U(N)$ $\cal{N}=8$ SYM flows to $U(N)_1 \times U(N)_{-1}$ ABJM in the IR, which for the $N=2$ case considered in this paper is the BLG theory. Whether such a connection holds between the non-Lorentzian limits of the two theories is an intriguing one. Our results suggest that the SGYM theory should flow to the M2-brane limit of BLG/ABJM in the IR. This  is also supported from the spacetime embedding of the set-up; in appendix \ref{sect: IIA NRST} we show that the non-Lorentzian limit associated with an intersecting M2-brane configuration (which was found in \cite{Lambert:2024uue} to be the spacetime configuration from which the M2-brane limit of the field theory arises) descends to the non-relativistic string limit of the D2-brane solution through a circle-reduction on one of the compact directions. This perhaps hints at a corresponding link between the Coulomb branch dynamics of the two field theories, as one may expect from the relation between the relativistic theories.

A question that should be dealt with is the fate of supersymmetry-preserving marginal deformations in the three-dimensional theory that render the theory renormalizable. As mentioned previously, it does not appear that one can find a well-defined nine-dimensional supersymmetric extension whose reduction to three dimensions is free of logarithmic divergences, and the structure of divergences means such an approach will not encompass all required counterterms. A more complete analysis of the possible three-dimensional terms that can be added is therefore desirable, though obviously far more complex; the hope would be that within this set lies a theory in which the Schr\"odinger symmetry is preserved at the quantum level. A related question is whether any deformation also preserves the $\epsilon_-$-supersymmetry. In \cite{Blair:2025ewa} it was shown that the backreacting M2-brane solution \eqref{eq: m2-brane solution} of eleven-dimensional supergravity's M2-brane limit only preserves 8 Killing spinors; as the D2-brane solution \eqref{eq: NH NL D2 Solution} of type IIA supergravity's non-relativistic string limit is obtained from this via reduction on a circle, one would imagine its supersymmetry structure is the same. One may therefore imagine that the deformation of the QFT required to relate the theory to the non-Lorentzian gravitational dual solution must break the $\epsilon_-$-supersymmetry. More work should be done on both sides of this problem to investigate this further.

\section*{Acknowledgements}

We would like to thank Hyungrok Kim and Ibrahima Bah for helpful conversations during the completion of this work. N.L. was supported in part by the STFC grant ST/X000753/1.

\appendix

\section{The D2-Brane Solution of Type IIA NRST} \label{sect: IIA NRST}

In this appendix we shall briefly review the non-relativistic string limit of type IIA supergravity bosonic sector and study its D2-brane solution, first presented in \cite{Harmark:2025ikv} following the construction of the D4-brane solution in \cite{Lambert:2024ncn}. The theory was constructed in \cite{Blair:2021waq} by reducing the M2-brane limit of eleven-dimensional supergravity; more directly, it can be found by making the string-frame field reparameterisations
\begin{subequations} \label{eq: iia reparam}
\begin{align}
    G_{\mn} &= c^2 \cal{T}_{\mn} + \cal{E}_{\mn} \ , \\
    G^{\mn} &= \cal{E}^{\mn} + c^{-2} \cal{T}^{\mn} \ , \\
    e^{\phi} &= c e^{\varphi} \ , \\
    B_2 &= - c^2 \tau^0 \wedge \tau^1 + b_2 \ , \\
    C_1 &= c_1 \ , \\
    C_3 &= c^2 \tau^0 \wedge \tau^1 \wedge c_1 + c_3 \ ,
\end{align}
\end{subequations}
before taking the $c\to\infty$ limit. This parameterisation of the NS-NS sector is identical to the NRST limit of type IIB supergravity \cite{Bergshoeff:2023ogz}. Here we define a pair of one-forms $\tau^A_{\mu}$ and eight orthogonal vector fields $e^{\mu}_a$, so $e^{\mu}_a \tau_{\mu}^A = 0$, by
\begin{subequations}
\begin{align}
    \cal{T}_{\mn} &= \eta_{AB} \tau^A_{\mu} \tau^B_{\nu} \ , \\
    \cal{E}^{\mn} &= e_a^{\mu} e^{\nu}_a \ ,
\end{align}
\end{subequations}
with their projective inverses $\{ \tau^{\mu}_A , e^a_{\mu} \}$ then defined by the conditions
\begin{subequations}
\begin{align}
    \tau^A_{\mu} \tau^{\mu}_B &= \delta^A_B \ , \\
    e^a_{\mu} e^{\mu}_b &= \delta^a_b \ , \\
    \tau^A_{\mu} \tau^{\nu}_A + e^a_{\mu} e^{\nu}_a &= \delta^{\nu}_{\mu} \ .
\end{align}
\end{subequations}
Note that this does not uniquely define $\{ \tau^{\mu}_A , e^a_{\mu} \}$. Upon substituting this into the type IIA action with the definitions
\begin{subequations}
\begin{align}
    f_{p+1} &= d c_p \ , \\
    h_3 &= d b_2 \ , \\
    \Tilde{f}_4 &= f_4 - c_1 \wedge h_3 \ , \\
    \Omega \,dx^0 \wedge dx^1 \wedge ... \wedge dx^9 &= \tau^0 \wedge \tau^1 \wedge e^1 \wedge ... \wedge e^8 \ ,
\end{align}
\end{subequations}
we find
\begin{align} \nonumber
    S_{IIA} = \inv{4\kappa^2} \bigintsss \Bigg[& 
    d^{10} x \, \Omega \bigg( 
    - \frac{c^2}{2 \cdot 4! } \brac{ \Tilde{f}_{a_1 ... a_4} - \inv{4!} \epsilon_{a_1 ... a_4 b_1 ... b_4} \Tilde{f}_{b_1 ... b_4} }^2 + 2 e^{-2\varphi} \Big( R^{(0)} \\ \nonumber
    &+ 4 \cal{E}^{\mn} \partial_{\mu} \varphi \partial_{\nu} \varphi - \inv{12} h_{abc} h_{abc}  + \inv{2} \epsilon_{AB} T^A_{ab} h^{Bab} \Big) - f_{Aa} f^{Aa} \\
    &- \inv{3!} \Tilde{f}_{Aabc} \Tilde{f}^{Aabc} - \inv{2} \epsilon_{AB} f_{ab} \Tilde{f}^{ABab} \bigg) - b_2 \wedge f_4 \wedge f_4 \Bigg] +O(c^{-2}) \ .
\end{align}
Here we simply denote the finite terms in the expansion of the relativistic Ricci scalar by $R^{(0)}$ (leaving the details to the previous references) and freely contract our form-fields with $e^{\mu}_a$ and $\tau^{\mu}_A$, with the local indices then raised and lowered with $\delta_{ab}$ and $\eta_{AB}$. The divergent term can be regulated by rewriting it using a Hubbard-Stratonovich field $\Tilde{g}_{a_1 ... a_4}$,
\begin{equation}
    S_{\Tilde{g}} = \inv{4\kappa^2} \int d^{10} x \,\Omega \brac{ \Tilde{g}_{a_1 ... a_4} \brac{ \Tilde{f}_{a_1 ... a_4} - \inv{4!} \epsilon_{a_1 ... a_4 b_1 ... b_4} \Tilde{f}_{b_1 ... b_4} } + \frac{4!}{2 c^2} \Tilde{g}_{a_1 ... a_4} \Tilde{g}_{a_1 ... a_4}} \ ,
\end{equation}
which upon imposing the algebraic equation of motion
\begin{equation} \label{eq: tilde g eom}
    \Tilde{g}_{a_1 ... a_4} = - 4! \, c^{-2} \brac{ \Tilde{f}_{a_1 ... a_4} - \inv{4!} \epsilon_{a_1 ... a_4 b_1 ... b_4} \Tilde{f}_{b_1 ... b_4} } \ ,
\end{equation}
recovers the appropriate term in the initial action. This allows the $c\to\infty$ limit to be taken, giving
\begin{align} \nonumber
    S_{IIA,NR} = \inv{4\kappa^2} \bigintsss \Bigg[& 
    d^{10} x \, \Omega \bigg( 2 e^{-2\varphi} \Big( R^{(0)} + 4 \cal{E}^{\mn} \partial_{\mu} \varphi \partial_{\nu} \varphi - \inv{12} h_{abc} h_{abc}  + \inv{2} \epsilon_{AB} T^A_{ab} h^{Bab} \Big) \\ \nonumber
    &- f_{Aa} f^{Aa} - \inv{3!} \Tilde{f}_{Aabc} \Tilde{f}^{Aabc} - \inv{2} \epsilon_{AB} f_{ab} \Tilde{f}^{ABab} \\ \label{eq: NR IIA action}
    &+ \Tilde{g}_{a_1 ... a_4} \brac{ \Tilde{f}_{a_1 ... a_4} - \inv{4!} \epsilon_{a_1 ... a_4 b_1 ... b_4} \Tilde{f}_{b_1 ... b_4} } \bigg) - b_2 \wedge f_4 \wedge f_4 \Bigg] \ .
\end{align}
As $\Tilde{g}$ appears linearly it imposes the self-duality constraint
\begin{equation} \label{eq: NRIIA constraint}
    \Tilde{f}_{a_1 ... a_4} = \inv{4!} \epsilon_{a_1 ... a_4 b_1 ... b_4} \Tilde{f}_{b_1 ... b_4} \ ,
\end{equation}
on the totally transverse components of the D2-brane flux.

Aside from diffeomorphism invariance, there are two important gauge symmetries of this action: the local dilatation transformations
\begin{subequations} \label{eq: local dilatation symmetry}
\begin{align}
    \hat{\tau}^A_{\mu} &= e^{\sigma} \tau^A_{\mu} \ , \\
    \hat{\varphi} &= \varphi + \sigma \ , \\
    \hat{\Tilde{g}}_{a_1 ... a_4} &= e^{-2\sigma} 
\end{align}
\end{subequations}
which we see are in practice equivalent to a local rescaling of $c$ (where the scaling of $\Tilde{g}$ follows from the equation of motion \eqref{eq: tilde g eom}), and the local 1-brane Galilean boosts
\begin{subequations}
\begin{align}
    \delta e^a_{\mu} &= \tensor{\lambda}{^a_A} \tau^A_{\mu} \ , \\
    \delta \tau^{\mu}_A &= - \tensor{\lambda}{^a_A} e^{\mu}_a \ , \\ 
    \delta b_2 &= - \epsilon_{AB} \tensor{\lambda}{_a^A} e^a \wedge \tau^B \ , \\
    \delta c_3 &= \epsilon_{AB} \tensor{\lambda}{_a^A} e^a \wedge \tau^B \wedge c_1 \ , \\
    \delta \Tilde{g}_{a_1 ... a_4} &= \inv{3} \Tilde{f}_{A [a_1 a_2 a_3} \tensor{\lambda}{_{a_4]} ^A} \ .
\end{align}
\end{subequations}
These transformations are also present in the limit of type IIB supergravity, with the transformations of the NS-NS fields identical in both cases; this means that we can use the results of \cite{Bergshoeff:2023ogz} for invariance of this sector and focus on the R-R fields. As none of the form fields scale with $\sigma$, there are no subtleties surrounding derivatives of $\sigma$ and it is straightforward to see that the scaling of each terms cancels that of $\Omega$. The theory is therefore invariant under \eqref{eq: local dilatation symmetry} and is missing an equation of motion present in the relativistic theory, which is the Poisson equation computed in \cite{Blair:2021waq}. The calculation for local 1-brane Galilean boosts is slightly more involved, but by using the result
\begin{equation}
    \delta \Tilde{f}_4 = \epsilon_{AB} \tensor{\lambda}{_a^A} e^a \wedge \tau^B \wedge f_2 \ , 
\end{equation}
it is straightforward to see that the transformation of the R-R terms cancel between themselves. Due to the local dilatation symmetry \eqref{eq: NR IIA action} is really a psuedo-action and must be augmented with the Poisson equation to get the full set of equations of motion. We should then check this is also invariant under the two symmetries; however, we shall just assume this property holds here.

With an understanding of the non-Lorentzian supergravity theory in hand, let us move on to a discussion of the D2-brane solution. We shall obtain this through a reparameterisation of the relativistic solution into the form \eqref{eq: iia reparam}. The most straightforward way to do this is to consider the F1-D2 solution, which has the non-trivial fields
\begin{subequations}
\begin{align}
    G_{10} &= \cal{H}_s^{-1} \Big( - \cal{H}^{-1/2} dt^2 + \cal{H}^{1/2} dX^2 \Big) + \cal{H}^{-1/2} dx^i dx^i + \cal{H}^{1/2} dY^A dY^A \ , \\
    e^{\phi} &= \cal{H}_s^{-1/2} \cal{H}^{1/4} \ , \\
    B_2 &= - \cal{H}_s^{-1} dt \wedge dX \ , \\
    C_3 &= \brac{\cal{H}^{-1} - 1} dt \wedge dx^1 \wedge dx^2 \ ,
\end{align}
\end{subequations}
where $\cal{H}_s$ and $\cal{H}$ are the string and D2-brane harmonic functions respectively. Taking 
\begin{equation}
    \cal{H}_2 = c^{-2} \ ,
\end{equation}
and choosing to smear $\cal{H}$ along $X$ (which we therefore compactify, $X \sim X + 2 \pi R_X$) then gives 
\begin{subequations} \label{eq: NL D2brane solution}
\begin{align}
    G_{10} &= c^2 \Big( - \cal{H}^{-1/2} dt^2 + \cal{H}^{1/2} dX^2 \Big) + \cal{H}^{-1/2} dx^i dx^i + \cal{H}^{1/2} dY^A dY^A \ , \\
    e^{\phi} &= c \cal{H}^{1/4} \ , \\
    B_2 &= - c^2 dt \wedge d X \ , \\
    C_3 &= \brac{ \cal{H}^{-1} - 1 } dt \wedge dx^1 \wedge dx^2 \ ,
\end{align}
\end{subequations}
with the harmonic function
\begin{equation}
    \cal{H} = 1 + \frac{R^4}{r^4} \ ,
\end{equation}
where we are using the notation $r^2 = Y^A Y^A$. This is of the correct form for the non-Lorentzian limit; furthermore, as the transverse projection of $F_4$ vanishes the constraint \eqref{eq: NRIIA constraint} is trivial, and so \eqref{eq: NL D2brane solution} is a good solution of the non-Lorentzian theory. It is also straightforward to check that with the distributions of $c$ given above the total D2-brane charge of the solution is $c$-independent and is therefore preserved by the limit. As the Hubbard-Stratonovich field obeys \eqref{eq: tilde g eom} at finite $c$ it vanishes for this solution. We shall be somewhat loose with our notation here and keep factors of $c$ around, though it should obviously be understand that when we're discussing the non-Lorentzian theory we take $c\to\infty$.

We can use the local dilatation symmetry to trivialise the spatial variation of the dilaton; in practical terms this is just the substitution
\begin{equation}
    c \cal{H}^{1/4} = \Tilde{c} \ ,
\end{equation}
and so we get
\begin{subequations}
\begin{align}
    G_{10} &= \Tilde{c}^2 \big( - \cal{H}^{-1} dt^2 + dX^2 \big) + \cal{H}^{-1/2} dx^i dx^i + \cal{H}^{1/2} dY^A dY^A \ , \\
    e^{\phi} &= \Tilde{c} \ , \\
    B_2 &= - \Tilde{c}^2 \cal{H}^{-1/2} \, dt \wedge d X \ , \\
    C_3 &= \brac{ \cal{H}^{-1} - 1 } dt \wedge dx^1 \wedge dx^2 \ .
\end{align}
\end{subequations}
The solution simplifies in the near-horizon limit, and we find the IIA NRST spacetime with non-trivial fields
\begin{subequations} \label{eq: NH NL D2 Solution}
\begin{align}
    G_{10} &= \Tilde{c}^2 \bigg( - \frac{r^4}{R^4} dt^2 + dX^2 \bigg) + \frac{r^2}{R^2} dx^i dx^i + \frac{R^2}{r^2} dr^2 + R^2 d\Omega_5^2 \ , \\
    C_3 &= \frac{r^4}{R^4} dt \wedge dx^1 \wedge dx^2 \ .
\end{align}
\end{subequations}
Let us compute the symmetries of the solution. To do this we will need a vielbein, for which we take the obvious diagonal choice
\begin{subequations}
\begin{align}
    \tau^t &= \frac{r^2}{R^2} dt \ , \\
    \tau^X &= dX \ , \\
    e^i &= \frac{r}{R} dx^i \ , \\
    e^r &= \frac{R}{r} dr \, \\
    e^I &= R \, e_{S^5}^I \ .
\end{align}
\end{subequations}
It is clear that the $\frak{iso}(2)$ transformations
\begin{equation}
    \delta x^i = \xi^i + \tensor{\omega}{^i_j} x^j \ ,
\end{equation}
with $\omega_{(ij)}=0$, are symmetries of $\cal{E}$. Using the local 1-brane Galilean boost this is still the case if can extend $\xi^i$ and $r_{ij}$ to be functions of $t$ and $X$, where the boost parameters are
\begin{subequations}
\begin{align}
    \tensor{\lambda}{^i_t} = - \frac{R}{r} \brac{\partial_t \xi^i + \partial_t\tensor{\omega}{^i_j} x^j} \ , \\
    \tensor{\lambda}{^i_X} = - \frac{r}{R} \brac{\partial_X \xi^i + \partial_X\tensor{\omega}{^i_j} x^j} \ .
\end{align}
\end{subequations}
However, these act non-trivially on $b_2$, with
\begin{equation}
    \delta b_2 = - \partial_t \brac{\xi_i + \tensor{\omega}{_{ij}} x^j } dx^i \wedge dX - \frac{r^4}{R^4} \partial_X \brac{\xi_i + \tensor{\omega}{_{ij}} x^j } dx^i \wedge dt \ .
\end{equation}
This is therefore only a symmetry of the full solution if $\delta b_2$ is a gauge transformation, or equivalently $d\brac{\delta b_2} = 0$. This restricts us to translations, rotations, and Galilean boosts,
\begin{equation}
    \delta x^i = \xi^i + \zeta^i t + \tensor{\omega}{^i_j} x^j \ .
\end{equation}
Similarly, the $SO(6)$ symmetries of the $S^5$ cannot be made $t$ or $X$-dependent without inducing a non-zero value of $b_2$. We also have translations of $X$, which form a $U(1)$ symmetry of \eqref{eq: NH NL D2 Solution}.

The more interesting of transformations are those that act non-trivially on $t$. The coordinate transformation
\begin{subequations}
\begin{align}
    \delta t &= 2 F(t) \ , \\
    \delta x^i &= \dot{F}(t) x^i \ , \\
    \delta r &= - \dot{F}(t) r \ ,
\end{align}
\end{subequations}
combined with the 1-brane boosts
\begin{subequations}
\begin{align}
    \tensor{\lambda}{^i_t} &= - \frac{R}{r} x^i \Ddot{F} \ , \\
    \tensor{\lambda}{^r_t} &= \frac{R^3}{r^2} \Ddot{F} \ ,
\end{align}
\end{subequations}
are symmetries of $\cal{\tau}_{\mn}$ and $\cal{E}_{\mn}$\footnote{Or equivalently $\cal{E}^{\mn}$ and $\cal{\tau}^{\mn}$.}. However, the boosts transform $b_2$ by
\begin{equation}
    \delta b_2 = - x^i \Ddot{F} dx^i \wedge dX + \frac{R^4}{r^3} \Ddot{F} dr \wedge dX \ ,
\end{equation}
meaning we must restrict to
\begin{equation}
    F(t) = a + b t + c t^2 \ .
\end{equation}
These possess non-trivial commutation relations with the transformations of $x^i$, and with a bit of thought we see that together they form the two-dimensional Schr\"odinger algebra $\frak{schr}(2)$. Altogether, the symmetries of \eqref{eq: NH NL D2 Solution} are then
\begin{equation} \label{eq: NH NL D2 sym}
    \frak{g} = \frak{schr}(2) \oplus \frak{u}(1) \oplus \frak{so}(6) \ .
\end{equation}

The discussion above can be lifted to the M2-brane limit of M-theory. The relevant solution, which arises from two stacks of intersecting M2-branes, is \cite{Blair:2025ewa}
\begin{subequations}
\begin{align}
    G_{11} &= c^2 \brac{ - \cal{H}^{-2/3} dt^2 + \cal{H}^{1/3} dX^a dX^a  } + c^{-1} \brac{\cal{H}^{-2/3} dx^i dx^i + \cal{H}^{1/3} dY^A dY^A} \ , \\
    \Tilde{C}_3 &= - c^3 dt \wedge dX^1 \wedge dX^2 + \brac{\cal{H}^{-1} - 1} dt \wedge dx^1 \wedge dx^2 \ ,
\end{align}
\end{subequations}
where the harmonic function $\cal{H}$ is
\begin{equation}
    \cal{H} = 1 + \frac{R^4}{r^4} \ ,
\end{equation}
with $r^2 = Y^A Y^A$, and we smear over the compact $X^a$ directions (with $X^a \sim X^a + 2\pi R_a$). The M2-brane limit also possesses a local dilatation symmetry which acts as a local rescaling of $c$; making the redefinition $c = \cal{H}^{-1/6} \hat{c}$ and taking the near-horizon limit gives
\begin{subequations} \label{eq: m2-brane solution}
\begin{align}
    G_{11} &= \hat{c}^2 \brac{ - \frac{r^4}{R^4} dt^2 + dX^a dX^a} + \hat{c}^{-1} \brac{ \frac{r^2}{R^2} dx^i dx^i + \frac{R^2}{r^2} dr^2 + R^2 d\Omega_5 } \ , \\
    \Tilde{C}_3 &= - \hat{c}^3 \frac{r^2}{R^2} dt \wedge dX^1 \wedge dX^2 + \frac{r^4}{R^4} dt \wedge dx^1 \wedge dx^2 \ .
\end{align}
\end{subequations}
The symmetries of this solution are \cite{SmithThesis}
\begin{equation}
    \frak{g} = \frak{schr}(2) \oplus \frak{u}(1) \oplus \frak{u}(1) \oplus \frak{so}(6) \ ,
\end{equation}
which differs from \eqref{eq: NH NL D2 sym} by an additional $\frak{u}(1)$ factor corresponding to translations along the additional compact longitudinal direction. Using the reduction ansatz
\begin{subequations}
\begin{align}
    G_{11} &= e^{-2\phi/3} G_{10} + e^{4\phi/3} (dX^2)^2 \ , \\
    \Tilde{C}_3 &= B_2 \wedge dX^2 + C_3 \ ,
\end{align}
\end{subequations}
for a reduction along $X^2$, we find that defining $\hat{c}^3 = \Tilde{c}^2$ and relabelling $X^1$ to $X$ exactly reproduces \eqref{eq: NH NL D2 Solution}, as one would expect.

\section{Quantization of Deformed Action} \label{sect: deformed quant}

Let us detail the calculation of the logarithmic divergence in the one-loop contribution to the correlation function
\begin{equation}
    \langle \Omega^A \Omega^B \bar{\Omega}^A \bar{\Omega}^B \rangle^{(\mathrm{1PI})} \ ,
\end{equation}
in the deformed action \eqref{eq: bosonic deformation 1} introduced in section \ref{sect: deformation}. In the background field expansion 
\begin{subequations}
\begin{align}
    X &= v \sigma + g \varphi \ , \\
    A_{\mu} &= g a_{\mu} \ , \\
    Y^A &= \Tilde{Y}^A + g y^A \ ,
\end{align}
\end{subequations}
the relevant $O(1)$ action is
\begin{subequations}
\begin{align} \nonumber
    S_{0} = \inv{2} \tr \bigintssss dt d^2 x  \bigg(& 
    (\partial_0 \varphi)^2 - 2 (\partial_i \varphi + i v [\sigma, a_i]) (\partial_0 a_i - \partial_i a_0) - (\partial_i a_j)^2 \\ \nonumber
    &+ (\partial_i a_i)^2 - 2 i v [\sigma, y^A] \partial_0 y^A - (\partial_i y^A)^2 - v^2 \kappa \Big( [\sigma, \partial_0 \varphi]^2 \\ \nonumber
    &- 2 [\sigma, \partial_i \varphi + i v [\sigma, a_i]] [\sigma, \partial_0 a_i - \partial_i a_0] - [\sigma, \partial_i a_j]^2 \\ 
    &+ [\sigma, \partial_i a_i]^2 - 2 i v [\sigma, [\sigma, y^A]] [\sigma, \partial_0 y^A] - [\sigma, \partial_i y^A]^2 \Big)
    \bigg) \ , 
\end{align}
\begin{align} \nonumber
    S_{\Omega,1} = \tr \bigintssss dt d^2 x \bigg( &
    -i \Big( [\varphi, y^A] \partial_0 \Tilde{Y}^A - i v [\sigma, \Tilde{Y}^A] [a_0, y^A] + \partial_0 y^A [\varphi, \Tilde{Y}^A] \\ \nonumber
    &- i v [\sigma, y^A] [a_0 , \Tilde{Y}^A] \Big) +  i \partial_i y^A [a_i , \Tilde{Y}^A] +  i \partial_i \Tilde{Y}^A [a_i , y^A]  \\ \nonumber
    &+  i v^2 \kappa \Big( [\sigma, [\sigma, \Tilde{Y}^A]] \big( [\varphi, \partial_0 y^A] - i v [\sigma, [a_0 , y^A]] \big) \\ \nonumber
    &+ [\sigma, \partial_0 \Tilde{Y}^A] \big( [\varphi, [\sigma, y^A]] + [\sigma, [\varphi, y^A]] \big) + [\sigma, \partial_0 y^A] \big( [\varphi, [\sigma, \Tilde{Y}^A]] \\ \nonumber
    &+ [\sigma, [\varphi, \Tilde{Y}^A]] \big) + [\sigma, [\sigma, y^A]] \big( [\varphi, \partial_0 \Tilde{Y}^A] - i v [\sigma, [a_0 , \Tilde{Y}^A]] \big) \Big) \\ \nonumber
    &+ v\kappa \Big(  [\sigma, \partial_i \Tilde{Y}^A] \big( [\varphi, \partial_i y^A] - i v [\sigma, [a_i , y^A]] \big) \\ \nonumber
    &+  [\sigma, \partial_i y^A] \big( [\varphi, \partial_i \Tilde{Y}^A]  - i v [\sigma, [a_i, \Tilde{Y}^A]] \big) \Big) \\ \nonumber
    &-  \lambda v \Big( \partial_i \Tilde{Y}^A \big( [\partial_i \varphi , [\sigma, y^A]] + i v [[\sigma, a_i], [\sigma, y^A]] \big) \\ 
    &+ \partial_i y^A \big( [\partial_i \varphi , [\sigma, \Tilde{Y}^A]] + i v [ [\sigma, a_i], [\sigma, \Tilde{Y}^A]] \big) \Big) \bigg) \ , \label{eq: deformed interactions 1}
\end{align}
\begin{align} \nonumber
    S_{\Omega,2} = \inv{2} \tr \bigintssss dt d^2 x &\bigg( 
    [\Tilde{Y}^A ,\Tilde{Y}^B] [y^A , y^B] + [y^A , \Tilde{Y}^B] [y^A, \Tilde{Y}^B] - [y^A , \Tilde{Y}^B] [y^B, \Tilde{Y}^A] \\ \nonumber
    &- v^2 \kappa \Big(
    [\sigma, [\Tilde{Y}^A ,\Tilde{Y}^B]] [\sigma, [y^A , y^B]] + [\sigma, [y^A , \Tilde{Y}^B] ] [\sigma, [y^A , \Tilde{Y}^B] ] \\ \nonumber
    &- [\sigma, [y^A , \Tilde{Y}^B] ][\sigma, [y^B , \Tilde{Y}^A] ] \Big) - v^2 \lambda \Big( [\Tilde{Y}^A , \Tilde{Y}^B] [[\sigma, y^A], [\sigma, y^B]] \\ \nonumber
    &+ [y^A , y^B] [[\sigma, \Tilde{Y}^A], [\sigma, \Tilde{Y}^B]] + 2 [y^A , \Tilde{Y}^B] [[\sigma, y^A], [\sigma, \Tilde{Y}^B]] \\ \nonumber
    &- 2 [y^A , \Tilde{Y}^B] [[\sigma, y^B], [\sigma, \Tilde{Y}^A]] \Big) + v^4 \lambda^2  \Big( [[\sigma, \Tilde{Y}^A], [\sigma, \Tilde{Y}^B]] \\ \nonumber
    &\times [[\sigma, y^A],[\sigma, y^B]]] + [[\sigma, y^A],[\sigma, \Tilde{Y}^B]][[\sigma, y^A],[\sigma, \Tilde{Y}^B]] \\
    &- [[\sigma, y^A],[\sigma, \Tilde{Y}^B]][[\sigma, y^B],[\sigma, \Tilde{Y}^A]] \Big)
    \bigg)\ . \label{eq: deformed interactions 2}
\end{align}
\end{subequations}
While this adds a multitude of interaction terms, we see that the quadratic terms of the original theory are also augmented. These additions are all of the form
\begin{equation}
    \tr MN \Rightarrow \tr \big( MN - v^2 \kappa [\sigma, M] [\sigma, N] \big) \ ,
\end{equation}
so by expanding both fields as in \eqref{eq: field expansions} we see that this is
\begin{align}
    \tr \big( MN - v^2 \kappa [\sigma, M] [\sigma, N] \big) = M^{(\sigma)} N^{(\sigma)} + (1 + v^2 \kappa) ( M_+ N_- + M_- N_+) \ .
\end{align}
If we therefore rescale the complex components of our background and fluctuating field expansions to be of the form
\begin{equation}
    V = \sqrt{2} V^{(\sigma)} \sigma + \brac{1 + v^2 \kappa}^{-1/2} ( V_+ \sigma^+ + V_- \sigma_-)
\end{equation}
then $S_{B,0}$ is equivalent to the bosonic part of \eqref{eq: free perturbative theory} and the propagators obtained previously remain unchanged, with the effect shifted to the scaling of terms in \eqref{eq: deformed interactions 1} and \eqref{eq: deformed interactions 2}. Evaluating the traces in these actions (and again only keeping the terms that contribute to $\cal{A}$, $\cal{B}$, and $\cal{D}$), we find
\begin{subequations}
\begin{align} \nonumber
    S_{\Omega,1} = \sqrt{2} \bigintssss dt d^2 x \Bigg[&
    \Omega^A \bigg( 
    -\frac{i(1 + 3 v^2 \kappa)}{1 + v^2 \kappa} \phi^{(\sigma)} \partial_0 \bar{\omega}^A + i \bar{\varphi} \partial_0 y^A_{(\sigma)} + v ( 2 a_0^{(\sigma)} \bar{\omega}^A - y^A_{(\sigma)} \bar{b}_0 ) \\ \nonumber
    &+ i a_i^{(\sigma)} \partial_i \bar{\omega}^A - \frac{i(1 + v^2 \lambda)}{1 + v^2 \kappa} \bar{b}_i \partial_i y^A_{(\sigma)} - \frac{v \lambda}{1 + v^2 \kappa} ( \partial_i \phi^{(\sigma)} \partial_i \bar{\omega}^A \\ \nonumber
    &- \partial_i y^A_{(\sigma)} \partial_i \bar{\varphi} ) \bigg) - i \partial_0 \Omega^A \bigg(
    y^A_{(\sigma)} \bar{\varphi} + \frac{1 - v^2 \kappa}{1 + v^2 \kappa} \phi^{(\sigma)} \bar{\omega}^A \bigg)
    \\ \nonumber
    &+ \partial_i \Omega^A \bigg(
    \frac{i(1 - v^2 \kappa)}{1 + v^2 \kappa} ( a_i^{(\sigma)} \bar{\omega}^A - y^A_{(\sigma)} \bar{b}_i ) + \frac{v \kappa}{1 + v^2 \kappa} ( \bar{\varphi} \partial_i y^A_{(\sigma)} \\  \label{eq: modified interactions 1 expanded}
    &- 2 \phi^{(\sigma)} \partial_i \bar{\omega}^A ) - 
    \frac{v \lambda}{1 + v^2 \kappa} \bar{\omega}^A \partial_i \phi^{(\sigma)} \bigg) \Bigg] + \mathrm{c.c.} \ , 
\end{align}
\begin{align}
    S_{\Omega,2} &= - \brac{\frac{1 + v^2 \lambda }{1 + v^2 \kappa}}^2 \bigintssss dt d^2 x \brac{ \Omega^A \Omega^B \bar{\omega}^A \bar{\omega}^B + \bar{\Omega}^A \bar{\Omega}^B \omega^A \omega^B} \ .
\end{align}
\end{subequations}
The most natural choice is to take $\lambda = \kappa$, which we shall do from here onwards. As our interest will be solely in the coefficient of the log divergence we will set all external momenta to zero in vertices, taking
\begin{equation}
    \partial_0 \Omega^A = \partial_i \Omega^A = 0 \ .
\end{equation}
The interactions are then almost identical to those of the initial theory, with the only differences being the modified coefficient of the first term in \eqref{eq: modified interactions 1 expanded} and the terms quadratic in external momentum. In momentum space we have
\begin{subequations}
\begin{align} \nonumber
    S_{\Omega,1} = \sqrt{2} \bigintssss d\Tilde{\mu} \bigg[& 
    \Omega^A \Big(
    \bar{\omega}^A_{\Tilde{p}} \big( (\alpha \Tilde{E} - \beta \Tilde{p}^2) \phi^{(\sigma)}_{\Tilde{p}} + 2v a^{(\sigma)}_{0, \Tilde{p}} + \Tilde{p}^i a_{i,\Tilde{p}}^{(\sigma)} \big) \\ \nonumber
    &+ y^A_{(\sigma), \Tilde{p}} \big( ( \Tilde{E} + \beta \Tilde{p}^2) \bar{\varphi}_{\Tilde{p}} - v \bar{b}_{0,\Tilde{p}} + \Tilde{p}^i \bar{b}_{i,\Tilde{p}} \big)
    \Big) \\ \nonumber
    &+ \bar{\Omega}^A \Big( 
    \omega^A_{\Tilde{p}} \big( (\alpha \Tilde{E} - \beta \Tilde{p}^2) \phi^{(\sigma)}_{-\Tilde{p}} + 2v a_{0, - \Tilde{p}}^{(\sigma)} + \Tilde{p}^i a^{(\sigma)}_{i,-\Tilde{p}} \big) \\ \nonumber
    &+ y^A_{(\sigma), -\Tilde{p}} \big( 
    ( \Tilde{E} + \beta \Tilde{p}^2) \varphi_{\Tilde{p}} - v b_{0,\Tilde{p}} + \Tilde{p}^i b_{i,\Tilde{p}} \big) \Big)
    \bigg] \ ,
\end{align}
\begin{align}
    S_{\Omega,2} = - \bigintssss d\Tilde{\mu} \bigg[ 
    \Omega^A \Omega^B \bar{\omega}^A_{\Tilde{p}} \bar{\omega}^B_{-\Tilde{p}} + \bar{\Omega}^A \bar{\Omega}^B \omega^A_{\Tilde{p}} \omega^B_{-\Tilde{p}} \bigg] \ ,
\end{align}
\end{subequations}
where all background fields are evaluated at zero momentum and we define
\begin{subequations}
\begin{align}
    \alpha &= \frac{1 + 3 v^2 \lambda}{1 + v^2 \lambda} \ , \\
    \beta &= \frac{v \lambda}{1 + v^2 \lambda} \ .
\end{align}
\end{subequations}
Restricting our attention only to the log divergence of interest, a computation analogous to the one performed in section \ref{sect: scalar sector} gives the result
\begin{equation} \label{eq: modified divergence 1 appendix}
    \cal{V} = \frac{2 (1 - 2 v \beta)^2}{v} \int \frac{d^2 \Tilde{p}}{(2\pi)^2} \, \inv{\Tilde{p}^2} \ ,
\end{equation}
in place of \eqref{eq: divergent result} for the projection along $\delta_{AC} \delta_{BD}$\footnote{ As a sanity check, setting $\beta = 0$ reproduces our earlier result.}.

We can also do the same for the additional bosonic terms in \eqref{eq: S prime}: a short calculation shows that the effect of this is to shift $\beta$ to
\begin{equation}
    \beta \Rightarrow \Tilde{\beta} = \beta - \frac{2 v^3 \lambda^2}{1 + v^2 \lambda} \ ,
\end{equation}
and scales the terms quadratic in the background fields to
\begin{equation}
    S_{\Omega,2} \Rightarrow \Tilde{S}_{\Omega,2} = \brac{ 1 - \frac{8 v^4 \lambda^2}{(1 + v^2 \lambda)^2} } S_{\Omega, 2} \ .
\end{equation}
The result of this is then
\begin{equation}
    \Tilde{\cal{V}} = \frac{2}{v} \brac{1 - \frac{2v^2 \lambda}{1 + v^2 \lambda} + \frac{4 v^6 \lambda^3}{(1 + v^2 \lambda)^2}}^2 \bigintssss \frac{d^2 \Tilde{p}}{(2\pi)^2} \, \inv{\Tilde{p}^2} \ ,
\end{equation}
in place of \eqref{eq: modified divergence 1 appendix}.

\printbibliography

@article{Chapman:2020vtn,
    author = "Chapman, Shira and Di Pietro, Lorenzo and Grosvenor, Kevin T. and Yan, Ziqi",
    title = "{Renormalization of Galilean Electrodynamics}",
    eprint = "2007.03033",
    archivePrefix = "arXiv",
    primaryClass = "hep-th",
    doi = "10.1007/JHEP10(2020)195",
    journal = "JHEP",
    volume = "10",
    pages = "195",
    year = "2020"
}

@article{Hernandez:2026stx,
    author = "Hernandez, Rafael and Nieto Garcia, Juan Miguel and Urtiaga, Ander",
    title = "{Quantization of Galilean Electrodynamics: a non-trivially trivial theory}",
    eprint = "2608.01372",
    archivePrefix = "arXiv",
    primaryClass = "hep-th",
    month = "8",
    year = "2026"
}

@article{Lambert:2019nti,
    author = "Lambert, Neil and Mouland, Rishi",
    title = "{Non-Lorentzian RG flows and Supersymmetry}",
    eprint = "1904.05071",
    archivePrefix = "arXiv",
    primaryClass = "hep-th",
    doi = "10.1007/JHEP06(2019)130",
    journal = "JHEP",
    volume = "06",
    pages = "130",
    year = "2019"
}

@article{Lambert:2024uue,
    author = "Lambert, Neil and Smith, Joseph",
    title = "{Non-Relativistic M2-Branes and the AdS/CFT Correspondence}",
    eprint = "2401.14955",
    archivePrefix = "arXiv",
    primaryClass = "hep-th",
    month = "1",
    year = "2024"
}

@article{Blair:2021waq,
    author = "Blair, Chris D. A. and Gallegos, Domingo and Zinnato, Natale",
    title = "{A non-relativistic limit of M-theory and 11-dimensional membrane Newton-Cartan geometry}",
    eprint = "2104.07579",
    archivePrefix = "arXiv",
    primaryClass = "hep-th",
    doi = "10.1007/JHEP10(2021)015",
    journal = "JHEP",
    volume = "10",
    pages = "015",
    year = "2021"
}

@article{Harmark:2014mpa,
    author = "Harmark, Troels and Orselli, Marta",
    title = "{Spin Matrix Theory: A quantum mechanical model of the AdS/CFT correspondence}",
    eprint = "1409.4417",
    archivePrefix = "arXiv",
    primaryClass = "hep-th",
    doi = "10.1007/JHEP11(2014)134",
    journal = "JHEP",
    volume = "11",
    pages = "134",
    year = "2014"
}

@article{Andringa:2012uz,
    author = "Andringa, Roel and Bergshoeff, Eric and Gomis, Joaquim and de Roo, Mees",
    title = "{'Stringy' Newton-Cartan Gravity}",
    eprint = "1206.5176",
    archivePrefix = "arXiv",
    primaryClass = "hep-th",
    reportNumber = "UG-12-01, ICCUB-12-307, UB-ECM-PF-12-75",
    doi = "10.1088/0264-9381/29/23/235020",
    journal = "Class. Quant. Grav.",
    volume = "29",
    pages = "235020",
    year = "2012"
}

@article{Bergshoeff:2019pij,
    author = "Bergshoeff, Eric A. and Gomis, Jaume and Rosseel, Jan and \c{S}im\c{s}ek, Ceyda and Yan, Ziqi",
    title = "{String Theory and String Newton-Cartan Geometry}",
    eprint = "1907.10668",
    archivePrefix = "arXiv",
    primaryClass = "hep-th",
    doi = "10.1088/1751-8121/ab56e9",
    journal = "J. Phys. A",
    volume = "53",
    number = "1",
    pages = "014001",
    year = "2020"
}

@article{Bergshoeff:2023ogz,
    author = "Bergshoeff, Eric A. and Grosvenor, Kevin T. and Lahnsteiner, Johannes and Yan, Ziqi and Zorba, Utku",
    title = "{Non-Lorentzian IIB supergravity from a polynomial realization of SL(2, \ensuremath{\mathbb{R}})}",
    eprint = "2306.04741",
    archivePrefix = "arXiv",
    primaryClass = "hep-th",
    reportNumber = "NORDITA 2023-030",
    doi = "10.1007/JHEP12(2023)022",
    journal = "JHEP",
    volume = "12",
    pages = "022",
    year = "2023"
}

@article{Baiguera:2023fus,
    author = "Baiguera, Stefano",
    title = "{Aspects of Non-Relativistic Quantum Field Theories}",
    eprint = "2311.00027",
    archivePrefix = "arXiv",
    primaryClass = "hep-th",
    month = "10",
    year = "2023"
}

@article{Blair:2023noj,
    author = "Blair, Chris D. A. and Lahnsteiner, Johannes and Obers, Niels A. J. and Yan, Ziqi",
    title = "{Unification of Decoupling Limits in String and M-theory}",
    eprint = "2311.10564",
    archivePrefix = "arXiv",
    primaryClass = "hep-th",
    reportNumber = "NORDITA-2023-071, IFT-UAM/CSIC-23-151",
    month = "11",
    year = "2023"
}

@article{Bagchi:2015qcw,
    author = "Bagchi, Arjun and Basu, Rudranil and Kakkar, Ashish and Mehra, Aditya",
    title = "{Galilean Yang-Mills Theory}",
    eprint = "1512.08375",
    archivePrefix = "arXiv",
    primaryClass = "hep-th",
    reportNumber = "MIT-CTP-4755",
    doi = "10.1007/JHEP04(2016)051",
    journal = "JHEP",
    volume = "04",
    pages = "051",
    year = "2016"
}

@article{Bagchi:2022twx,
    author = "Bagchi, Arjun and Basu, Rudranil and Islam, Minhajul and Kolekar, Kedar S. and Mehra, Aditya",
    title = "{Galilean gauge theories from null reductions}",
    eprint = "2201.12629",
    archivePrefix = "arXiv",
    primaryClass = "hep-th",
    doi = "10.1007/JHEP04(2022)176",
    journal = "JHEP",
    volume = "04",
    pages = "176",
    year = "2022"
}

@article{Baiguera:2022cbp,
    author = "Baiguera, Stefano and Cederle, Lorenzo and Penati, Silvia",
    title = "{Supersymmetric Galilean Electrodynamics}",
    eprint = "2207.06435",
    archivePrefix = "arXiv",
    primaryClass = "hep-th",
    doi = "10.1007/JHEP09(2022)237",
    journal = "JHEP",
    volume = "09",
    pages = "237",
    year = "2022"
}

@article{Fontanella:2024rvn,
    author = "Fontanella, Andrea and Nieto Garc\'\i{}a, Juan Miguel",
    title = "{Constructing Non-Relativistic AdS$_5$/CFT$_4$ Holography}",
    eprint = "2403.02379",
    archivePrefix = "arXiv",
    primaryClass = "hep-th",
    reportNumber = "ZMP-HH/24-4, TCDMATH-24-01",
    month = "3",
    year = "2024"
}

@article{Lambert:2024yjk,
    author = "Lambert, Neil and Smith, Joseph",
    title = "{Non-Relativistic Intersecting Branes, Newton-Cartan Geometry and AdS/CFT}",
    eprint = "2405.06552",
    archivePrefix = "arXiv",
    primaryClass = "hep-th",
    month = "5",
    year = "2024"
}

@article{Blair:2024aqz,
    author = "Blair, Chris D. A. and Lahnsteiner, Johannes and Obers, Niels A. and Yan, Ziqi",
    title = "{Matrix Theory Reloaded: A BPS Road to Holography}",
    eprint = "2410.03591",
    archivePrefix = "arXiv",
    primaryClass = "hep-th",
    reportNumber = "IFT-UAM/CSIC-24-139, NORDITA 2024-033",
    month = "10",
    year = "2024"
}

@article{Fontanella:2024hgv,
    author = "Fontanella, Andrea and Nieto Garc{\'\i}a, Juan Miguel",
    title = "{Revisiting the symmetries of Galilean Electrodynamics}",
    eprint = "2411.19217",
    archivePrefix = "arXiv",
    primaryClass = "hep-th",
    reportNumber = "ZMP-HH/24-29",
    doi = "10.1007/JHEP06(2025)058",
    journal = "JHEP",
    volume = "06",
    pages = "058",
    year = "2025"
}

@article{Harmark:2025ikv,
    author = "Harmark, Troels and Lahnsteiner, Johannes and Obers, Niels A.",
    title = "{Gravitational solitons and non-relativistic string theory}",
    eprint = "2501.10178",
    archivePrefix = "arXiv",
    primaryClass = "hep-th",
    reportNumber = "NORDITA 2024-050",
    doi = "10.1007/JHEP05(2025)199",
    journal = "JHEP",
    volume = "05",
    pages = "199",
    year = "2025"
}

@article{Lambert:2024ncn,
    author = "Lambert, Neil and Smith, Joseph",
    title = "{Reciprocal non-relativistic decoupling limits of String Theory and M-Theory}",
    eprint = "2410.17074",
    archivePrefix = "arXiv",
    primaryClass = "hep-th",
    doi = "10.1007/JHEP12(2024)094",
    journal = "JHEP",
    volume = "12",
    pages = "094",
    year = "2024"
}

@article{Taylor:1996ik,
    author = "Taylor, Washington",
    title = "{D-brane field theory on compact spaces}",
    eprint = "hep-th/9611042",
    archivePrefix = "arXiv",
    reportNumber = "PUPT-1659",
    doi = "10.1016/S0370-2693(97)00033-6",
    journal = "Phys. Lett. B",
    volume = "394",
    pages = "283--287",
    year = "1997"
}

@article{Blair:2025ewa,
    author = "Blair, Chris D. A.",
    title = "{Supersymmetric solutions of non-relativistic 11-dimensional supergravity}",
    eprint = "2503.21577",
    archivePrefix = "arXiv",
    primaryClass = "hep-th",
    reportNumber = "IFT-UAM/CSIC-25-26",
    doi = "10.1007/JHEP09(2025)125",
    journal = "JHEP",
    volume = "09",
    pages = "125",
    year = "2025"
}

@article{Guijosa:2025mwh,
    author = "Guijosa, Alberto",
    title = "{On the underlying nonrelativistic nature of relativistic holography}",
    eprint = "2502.03031",
    archivePrefix = "arXiv",
    primaryClass = "hep-th",
    doi = "10.1007/JHEP08(2025)139",
    journal = "JHEP",
    volume = "08",
    pages = "139",
    year = "2025"
}

@article{Maskalaniec:2026vlk,
    author = "Maskalaniec, Dawid and Yan, Ziqi and Zorba, Utku",
    title = "{Non-Lorentzian Supergravity from Matrix Theory}",
    eprint = "2603.10278",
    archivePrefix = "arXiv",
    primaryClass = "hep-th",
    reportNumber = "NORDITA 2026-032",
    month = "3",
    year = "2026"
}

@article{Bergshoeff:2025grj,
    author = "Bergshoeff, Eric and Lambert, Neil and Smith, Joseph",
    title = "{The M5-brane limit of eleven-dimensional supergravity}",
    eprint = "2502.07969",
    archivePrefix = "arXiv",
    primaryClass = "hep-th",
    doi = "10.1007/JHEP06(2025)060",
    journal = "JHEP",
    volume = "06",
    pages = "060",
    year = "2025"
}

@article{Bergshoeff:2018yvt,
    author = "Bergshoeff, Eric and Gomis, Jaume and Yan, Ziqi",
    title = "{Nonrelativistic String Theory and T-Duality}",
    eprint = "1806.06071",
    archivePrefix = "arXiv",
    primaryClass = "hep-th",
    doi = "10.1007/JHEP11(2018)133",
    journal = "JHEP",
    volume = "11",
    pages = "133",
    year = "2018"
}

@article{Son:2008ye,
    author = "Son, D. T.",
    title = "{Toward an AdS/cold atoms correspondence: A Geometric realization of the Schrodinger symmetry}",
    eprint = "0804.3972",
    archivePrefix = "arXiv",
    primaryClass = "hep-th",
    reportNumber = "INT-PUB-08-08",
    doi = "10.1103/PhysRevD.78.046003",
    journal = "Phys. Rev. D",
    volume = "78",
    pages = "046003",
    year = "2008"
}

@book{Shankar:2017zag,
    author = "Shankar, Ramamurti",
    title = "{Quantum Field Theory and Condensed Matter}",
    publisher = "Cambridge University Press",
    doi = "10.1017/9781139044349",
    year = "2017"
}

@article{Abbott:1981ke,
    author = "Abbott, L. F.",
    title = "{Introduction to the Background Field Method}",
    reportNumber = "CERN-TH-3113",
    journal = "Acta Phys. Polon. B",
    volume = "13",
    pages = "33",
    year = "1982"
}

@article{Festuccia:2016caf,
    author = "Festuccia, Guido and Hansen, Dennis and Hartong, Jelle and Obers, Niels A.",
    title = "{Symmetries and Couplings of Non-Relativistic Electrodynamics}",
    eprint = "1607.01753",
    archivePrefix = "arXiv",
    primaryClass = "hep-th",
    doi = "10.1007/JHEP11(2016)037",
    journal = "JHEP",
    volume = "11",
    pages = "037",
    year = "2016"
}

@article{Bagchi:2017yvj,
    author = "Bagchi, Arjun and Chakrabortty, Joydeep and Mehra, Aditya",
    title = "{Galilean Field Theories and Conformal Structure}",
    eprint = "1712.05631",
    archivePrefix = "arXiv",
    primaryClass = "hep-th",
    doi = "10.1007/JHEP04(2018)144",
    journal = "JHEP",
    volume = "04",
    pages = "144",
    year = "2018"
}

@article{Santos:2004pq,
    author = "Santos, E. S. and de Montigny, M. and Khanna, F. C. and Santana, A. E.",
    title = "{Galilean covariant Lagrangian models}",
    doi = "10.1088/0305-4470/37/41/011",
    journal = "J. Phys. A",
    volume = "37",
    pages = "9771--9789",
    year = "2004"
}

@article{LeBellac:1973unm,
    author = "Le Bellac, M. and L{\'e}vy-Leblond, J. M.",
    title = "{Galilean electromagnetism}",
    doi = "10.1007/BF02895715",
    journal = "Nuovo Cim. B",
    volume = "14",
    number = "2",
    pages = "217--234",
    year = "1973"
}

@article{Bagchi:2014ysa,
    author = "Bagchi, Arjun and Basu, Rudranil and Mehra, Aditya",
    title = "{Galilean Conformal Electrodynamics}",
    eprint = "1408.0810",
    archivePrefix = "arXiv",
    primaryClass = "hep-th",
    doi = "10.1007/JHEP11(2014)061",
    journal = "JHEP",
    volume = "11",
    pages = "061",
    year = "2014"
}

@article{Blair:2025prd,
    author = "Blair, Chris D. A. and Lahnsteiner, Johannes and Obers, Niels A. and Yan, Ziqi",
    title = "{Dual non-Lorentzian backgrounds for matrix theories}",
    eprint = "2502.20310",
    archivePrefix = "arXiv",
    primaryClass = "hep-th",
    reportNumber = "IFT-UAM/CSIC-25-18, NORDITA 2025-014",
    doi = "10.1007/JHEP05(2025)200",
    journal = "JHEP",
    volume = "05",
    pages = "200",
    year = "2025"
}

@article{Fontanella:2026gaq,
    author = "Fontanella, Andrea and Nieto Garc{\'\i}a, Juan Miguel",
    title = "{An Introduction to String Newton-Cartan Holography and Integrability}",
    eprint = "2603.24657",
    archivePrefix = "arXiv",
    primaryClass = "hep-th",
    month = "3",
    year = "2026"
}

@phdthesis{SmithThesis,
    author = "Smith, Joseph",
    title = "{BPS Decoupling Limits of String Theory and Non-Lorentzian Quantum Field Theory}",
    school = "King's College London",
    year = "2025"
}

@article{Kim:2026hpe,
    author = "Kim, Hyungrok and Smith, Joseph",
    title = "{Non-relativistic limits of $\mathcal N=4$ supersymmetric Yang-Mills theory and S-duality}",
    eprint = "2606.21494",
    archivePrefix = "arXiv",
    primaryClass = "hep-th",
    month = "6",
    year = "2026"
}

@article{Banerjee:2022uqj,
    author = "Banerjee, Kinjal and Sharma, Aditya",
    title = "{Quantization of interacting Galilean field theories}",
    eprint = "2205.01918",
    archivePrefix = "arXiv",
    primaryClass = "hep-th",
    doi = "10.1007/JHEP08(2022)066",
    journal = "JHEP",
    volume = "08",
    pages = "066",
    year = "2022"
}

@article{Auzzi:2026gan,
    author = "Auzzi, Roberto and Baiguera, Stefano and Guo, Lihan and Harmark, Troels and Nardelli, Giuseppe",
    title = "{Finite scalar field theory with SU(1,1) spacetime symmetry from near-BPS limits of $\mathcal{N}=4$ SYM}",
    eprint = "2605.24085",
    archivePrefix = "arXiv",
    primaryClass = "hep-th",
    month = "5",
    year = "2026"
}

@article{Duval:1993hs,
    author = "Duval, C. and Horvathy, P. A.",
    title = "{On Schrodinger superalgebras}",
    eprint = "hep-th/0508079",
    archivePrefix = "arXiv",
    reportNumber = "CPT-93-P-2912",
    doi = "10.1063/1.530521",
    journal = "J. Math. Phys.",
    volume = "35",
    pages = "2516--2538",
    year = "1994"
}

@article{Auzzi:2019kdd,
    author = "Auzzi, Roberto and Baiguera, Stefano and Nardelli, Giuseppe and Penati, Silvia",
    title = "{Renormalization properties of a Galilean Wess-Zumino model}",
    eprint = "1904.08404",
    archivePrefix = "arXiv",
    primaryClass = "hep-th",
    doi = "10.1007/JHEP06(2019)048",
    journal = "JHEP",
    volume = "06",
    pages = "048",
    year = "2019"
}

\end{document}